\documentclass{emulateapj-rtx4}
\usepackage{apjfonts}
\usepackage{amssymb}
\usepackage{natbib}
\usepackage{lscape}
\shorttitle{The Circum-Galactic Medium at $z=2-3$}
\shortauthors{Steidel et al.}
\newcommand{\msun}{\ensuremath{\rm M_\odot}}
\newcommand{\msunyr}{\ensuremath{\rm M_{\odot}\;{\rm yr}^{-1}}}
\newcommand{\Ha}{\ensuremath{\rm H\alpha}}
\newcommand{\Hb}{\ensuremath{\rm H\beta}}
\newcommand{\lya}{\ensuremath{\rm Ly\alpha}}
\newcommand{\zla}{\ensuremath{z_{\rm Ly\alpha}}}
\newcommand{\zis}{\ensuremath{z_{\rm IS}}}
\newcommand{\zha}{\ensuremath{z_{\rm H\alpha}}}

\newcommand{\kms}{\rm km~s\ensuremath{^{-1}\,}}
\newcommand{\dvla}{\ensuremath{\rm \Delta v_{\lya}}}
\newcommand{\dvis}{\ensuremath{\rm \Delta v_{IS}}}

\newcommand{\mbar}{\ensuremath{\rm M_{bar}}}
\newcommand{\ztwo}{\ensuremath{z\sim2}}

\newcommand{\lyb}{Ly$\beta$}

\newcommand{\secpoint}{\mbox{$''\mskip-7.6mu.\,$}}

\newcommand{\et}{{\rm et al.}~}
\def\ltsima{$\; \buildrel < \over \sim \;$}
\def\simlt{\lower.5ex\hbox{\ltsima}}
\def\gtsima{$\; \buildrel > \over \sim \;$}
\def\simgt{\lower.5ex\hbox{\gtsima}}
\def\arcs{$''~$}
\def\arcm{$'~$}

\begin{document}
\title{THE STRUCTURE AND KINEMATICS OF THE CIRCUM-GALACTIC MEDIUM FROM FAR-UV SPECTRA OF $Z \simeq 2-3$ GALAXIES\altaffilmark{1}}
\slugcomment{Received 2010 February 23; Accepted 2010 May 11}
\author{\sc Charles C. Steidel\altaffilmark{2}, Dawn K. Erb\altaffilmark{3,4}, Alice E. Shapley\altaffilmark{5,6,7}, Max Pettini\altaffilmark{8,9}, Naveen Reddy\altaffilmark{10,11}, Milan Bogosavljevi\'c\altaffilmark{2}, Gwen C. Rudie\altaffilmark{2}, \& 
Olivera Rakic\altaffilmark{12}}

\altaffiltext{1}{Based on data obtained at the 
W.M. Keck Observatory, which 
is operated as a scientific partnership among the California Institute of 
Technology, the
University of California, and NASA, and was made possible by the generous 
financial
support of the W.M. Keck Foundation. 
}
\altaffiltext{2}{California Institute of Technology, MS 249-17, Pasadena, CA 91125}
\altaffiltext{3}{Department of Physics, University of California, Santa Barbara, Santa Barbara,
CA 93106}
\altaffiltext{4}{Spitzer Fellow}
\altaffiltext{5}{Department of Physics and Astronomy, University of California, Los Angeles, 430 Portola
Plaza, Box 951547, Los Angeles, CA 90095}
\altaffiltext{6}{Alfred P. Sloan Fellow}
\altaffiltext{7}{Packard Fellow}
\altaffiltext{8}{Institute of Astronomy, Madingley Road, Cambridge CB3 OHA UK}
\altaffiltext{9}{International Centre for Radio Astronomy Research,
University of Western Australia, 35 Stirling Highway,
Crawley, WA 6009, Australia}
\altaffiltext{10}{National Optical Astronomy Observatories, 950 N. Cherry Ave., Tucson, AZ 85258}
\altaffiltext{11}{Hubble Fellow}
\altaffiltext{12}{Leiden Observatory, Leiden University, P.O. Box 9513, 2300 RA Leiden, The Netherlands}

\begin{abstract}
We present new results on the kinematics and spatial distribution of metal-enriched gas 
within $\sim 125$ kpc of star-forming (``Lyman Break'') galaxies at 
redshifts $2 \simlt z \simlt 3$. In particular, we focus on constraints
provided by the rest-frame far-UV spectra of faint galaxies-- and demonstrate 
how galaxy spectra can be used to obtain key spatial and spectral information more efficiently
than possible with QSO sightlines.
Using a sample of $89$ galaxies with $\langle z \rangle = 2.3\pm0.3$ and with 
both rest-frame far-UV and \Ha\ spectra, we re-calibrate  
the measurement of accurate galaxy systemic redshifts using only survey-quality
rest-UV spectra. We use the velocity-calibrated sample to investigate 
the kinematics of the galaxy-scale outflows via the strong interstellar (IS) 
absorption lines and Lyman $\alpha$ emission (when present), as well as their 
dependence on other physical properties of the galaxies. 
We construct a sample of 512 close ($1-15$\arcs) angular pairs of $z \sim 2-3$ galaxies
with redshift differences indicating a lack of physical association. Sightlines
to the background galaxies provide new information on the spatial distribution 
of circumgalactic gas surrounding the foreground galaxies. 
The close pairs sample 
galactocentric impact parameters 3-125 kpc (physical) at $\langle z \rangle =2.2$, providing for the first
time a robust map of cool gas as a function of galactocentric distance for a well-characterized
population of galaxies.  We propose a simple model   
of circumgalactic gas that simultaneously matches the kinematics, depth, and profile shape 
of IS absorption and \lya\ emission lines, as well as the observed variation of absorption line strength  
(\ion{H}{1} and several metallic species) versus galactocentric impact parameter.
Within the model, cool gas is distributed symmetrically around
every galaxy, accelerating radially outward with $v_{out}(r)$ increasing with $r$ (i.e., the highest velocities
are located at the largest galactocentric distances $r$). The inferred radial dependence of 
the covering fraction of cool gas (which modulates the absorption line strength) is 
$f_c (r) \propto r^{-\gamma}$ with $0.2 \simlt \gamma \simlt 0.6$ depending on transition. 
We discuss the results of the observations in the context of ``cold accretion'', in which cool gas
is accreting via filamentary streams directly onto the central regions of galaxies. 
At present, we find little observational evidence for cool infalling material, while evidence supporting the large-scale
effects of super-wind {\it outflows} is strong.  
This ``pilot'' study using faint galaxy
spectra demonstrates the potential of using galaxies to trace baryons within galaxies,
in the circumgalactic medium, and ultimately throughout the IGM.  
\end{abstract}

\section{Introduction}
\label{sec:intro}

Observational studies of the galaxy formation process are
reaching a critical juncture, where the accumulation rate of new
data may be overtaking our ability to reach new understanding.  
There are now many
large surveys designed to study galaxy formation and evolution
over ever-increasing volumes and redshift ranges, backed by the unquestionable power of multi-wavelength
observations from X-rays to the far-IR/sub-mm and radio.
Interpretation of the survey data is aided
by a multitude of theoretical results based on simulations and/or semi-analytic calculations.
However, many of the most fundamental remaining questions
in galaxy formation involve complex baryonic processes that are difficult to model,
are not well-constrained by current observations, or involve physics
that are not yet well understood.

While there is general agreement about
the development of structure in the dark matter component on scales larger
than that of galaxies, the astrophysics of the baryonic response to the
dark matter structure, the subsequent feedback of energy from
star formation, supernova explosions, and accretion energy from the
growth of supermassive black holes, and the flow of gas into and out of galaxies,
remain largely unconstrained-- and thus subject to substantial debate.
``Feedback'' has become a buzzword, universally acknowledged as something important to understand,
but there is little agreement about what it really means and how it
affects the ``big picture''. Nevertheless, some kind of feedback is invoked to explain many otherwise-inexplicable
observations: the cessation of star formation in massive galaxies at high redshift; the small number of low-mass
galaxies relative to the dark matter halo mass function predicted by the otherwise successful $\Lambda$CDM
cosmology; the correlation between galaxy spheroid mass and
the mass of central super-massive black holes; the general absence of cooling flows in clusters of galaxies; the
metal enrichment of intracluster gas, just to name a few.
The most obvious sources of the energy and/or momentum required to explain these phenomena are massive stars, supernovae,
and AGN activity, all of which must have exerted most of their influence in the distant past, when 
these processes were at their peak intensity. 

One route to understanding the relevant baryonic processes is via
simultaneous study of galaxies and the IGM, in
the same cosmic volumes during the epoch when they were arguably exerting the
greatest influence on one another-- near the peak of both universal star formation
and supermassive black hole growth in the redshift range  
$3 \simgt z \simgt  1.5$.
Combining two powerful
lines of investigation provides complementary information on the
state of baryons, both those collapsed into galaxies and those residing outside of
galaxies.  The IGM and ``circumgalactic medium'' (CGM; by which we mean the gas-phase
structures found  within
$\simlt 300$ kpc [physical] of galaxies), together present a laboratory in which the
effects of galaxy formation and AGN accretion (e.g., radiative
and hydrodynamical ``feedback'' and its recent history) can be measured on scales
that are not accessible using direct observations of galaxies. Similarly,
the galaxy distribution relative to the lines of sight to background objects
tells us more about how the physical information garnered from the absorption
line studies should be interpreted.
These ambitious science goals -- to observe both diffuse gas {\it and} galaxies/AGN in the same survey volume at high
fidelity and down to small scales -- require a different approach as compared to most spectroscopic surveys of
the distant universe. Whereas the movement of most galaxy surveys has been toward larger and larger scales, the
most vexing remaining uncertainties are related to phenomena occurring on
$\sim 1-10$ Mpc scales, where the figure of merit is {\it information
density} and not total survey volume.

It has been known for some time that galactic-scale outflows with
velocities of several hundred \kms\ are ubiquitous in star-forming
galaxies at all redshifts for which interstellar (IS) absorption features are accessible
(e.g., \citealt{sgp+96,franx97,pettini00,pss+01,ssp+03,martin05,rupke05,tremonti07,weiner09}.)  
Evidence is found in the offsets
between the redshifts of the nebular emission lines, interstellar
absorption lines, and \lya\ emission \citep{pss+01}, in the relative
velocities of stellar, interstellar and nebular lines in composite UV
spectra \citep{ssp+03}, and in the correlation of 
\ion{C}{4} systems seen in absorption in QSO spectra with the positions of
the galaxies themselves \citep{assp03,ass+05}.  The outflowing gas has
been observed in detail for only a handful of lensed galaxies (\citealt{prs+02,quider09a,quider09b}); 
in the best-observed of these,  MS1512-cB58, the interstellar absorption is
distributed over a very wide range of velocities, $-800 \simlt v \simlt +200$,
with a centroid velocity of $-255$ \kms\ \citep{prs+02}. 
Such outflows are also a
general feature of starburst galaxies in the local universe, where
rest-UV spectroscopic studies reveal similarly complex velocity structure and gas in multiple
phases (e.g.\ \citealt{ham90,lh96,martin99,shc+04,schwartz06,grimes09,chen10}).
However, the outflow phenomenon is so much more widespread at high redshifts that it
is influencing essentially {\it every} galaxy, and potentially every galaxy's local
environment. It must have an influence on both the chemical evolution of the galaxies
and of the IGM, it may well regulate the maximum star formation rate attainable
by a galaxy, and without doubt it is an essential ingredient to basic understanding
of the circulation of baryons as galaxies are forming.

In spite of the substantial observational evidence for galaxy-scale outflows, most of the recent
theoretical work has focused instead on the {\it infall} of cool gas (``Cold Accretion'' or 
``Cold Flows'') via filaments, directly onto the central regions of forming galaxies. According
to much of the recent work, this mode of gas accretion is what feeds (and regulates) high star formation
rates in high redshift galaxies until they attain a particular mass threshold ($M_{tot} \simgt
10^{12}$ M$_{\sun}$) at which point virial shocks develop and accretion of cold material is suppressed 
\citep{keres05,keres09,dekel09a,brooks09,ceverino09}. This transition is believed to be at least 
partially responsible for producing massive ``red and dead'' 
galaxies as early as $z \sim 2.5$.
Similarly, the phenomenon of spatially extended \lya\ emission, including giant \lya\ ``blobs'' \citep{steidel00}
has been ascribed to cooling radiation from the denser portions of the cold streams as 
they are accreting directly onto the galaxy's central regions 
(e.g., \citealt{haiman00,furlanetto05,dijkstra09,goerdt10}).
Although the cold accretion picture may be attractive, the predictions of the observational consequences
for CGM gas are rather model-dependent for both absorption lines and \lya\ emission. The simulations
which predict the cold accretion generally do not account for IS gas that may have been carried to large
galactocentric radii by outflows, nor for \lya\ photons produced by photoionization in a galaxy's
\ion{H}{2} regions which then scatter their way in space through CGM gas before escaping. It may well
be that the observational signatures of infall by cold accretion and outflows via supernova-driven
winds are very subtle, and perhaps indistinguishable. Possibly the most telling differences
would be kinematic-- absorption line signatures of infalling material in galaxy spectra would be expected to be
primarily {\it redshifted}, while outflowing material would be {\it blue-shifted} and could
reach much higher velocities with respect to the galaxy systemic redshift. Thus, accurate determination of
the galaxy systemic redshift is an essential part of understanding the nature of the CGM.

The {\it spatial} distribution of the blue-shifted high-velocity material seen in absorption 
against the galaxy continuum is not yet established. 
We know that virtually every $z>2$ galaxy bright enough to be observed spectroscopically
is driving out material at velocities of at least several hundred \kms, but we do not know how far
this material travels, or even where it is with respect to the galaxy as we observe it. The mass flux
associated with such flows has been measured in only one case at high redshift, MS1512-cB58 (\citealt{pettini00}), 
and even in this case the result depends very sensitively on the assumed physical
location of the absorbing material responsible for the bulk of the observed absorption. 
The best hope for constraining the location of outflowing gas is by observing objects lying in the
background, but at small angular separation, relative to the galaxy of interest. In this case the challenge
is to find background objects bright enough in the rest-frame far-UV but close enough to the foreground
galaxy to provide interesting constraints. A great deal of effort, over a large range of redshifts,
has been invested using QSOs as background sources, where absorption by metallic ions or H{\sc I} in the spectrum
of the QSO is compared with galaxies with known redshifts and projected separations (e.g., \citealt{bergeron91,
steidel94,chen2001,steidel02,lanzetta95,danforth08,bowen95,bouche07,kaczprak10}); most of this work has focused on redshifts $z < 1$
because of the increasing difficulty obtaining spectra of the foreground galaxies at higher
redshifts. Even if
galaxies are identified and have redshifts that correspond closely with observed absorption, 
the association of particular
absorption systems with identified galaxies is almost always ambiguous, since the dynamic range for
identifying faint galaxies is limited, and often the bright background QSOs make it challenging
to observe galaxies within a few arcsec of the QSO sightline. Finally, there is the controversial
issue of whether metals seen near to, but outside of, galaxies are a direct result of recent star formation or
AGN activity, 
or are simply tracing out regions of the universe in which {\it some}
galaxies, perhaps in the distant past, polluted the gas with metals (e.g., \citealt{madau01,scannapieco02,mori02,
ferrara03} ; cf. \citealt{assp03,ass+05}). 


In this paper, our goal is to try to understand the spatial distribution of cool
gas seen in absorption against the stellar continuum
of every galaxy observed at high redshift. 
The objective is to empirically track the kinematics and structure in the CGM from the central parts of
galaxies all the way to large galactocentric radii.  In this work,
we use only galaxy spectra, primarily in the rest-frame far-UV, but we calibrate the velocity
zero-point using a set of nearly 100 H$\alpha$ measurements in the observed-frame near-IR for  
galaxies in the redshift range $1.9 \simlt z \simlt 2.6$. The near-IR measurements are drawn primarily
from the sample of \cite{erb+06b}, after which we use a much more extensive set of rest-UV spectra from a nearly
completed UV-selected galaxy survey targeting the same range of redshifts. 

The paper is organized as follows: in \S\ref{sec:bulkvel} we examine the statistics of the kinematics
of the outflows using a sub-sample of
galaxies for which both near-IR nebular \Ha\ spectroscopy and reasonably high-quality
optical (rest-UV) spectra are available. We also present new empirical formulae
for estimating galaxy systemic redshifts for the typical case in which
only low S/N rest-UV spectra are available. In \S\ref{sec:correlations}, we
seek correlations between the interstellar absorption line kinematics, in
particular the bulk velocities measured from strong low-ionization transitions, 
and other measured galaxy properties. \S\ref{sec:composites} describes 
further inferences on the structure and kinematics of outflowing material 
from high S/N composite far-UV spectra, while \S\ref{sec:lya} 
examines the observed behavior of \lya\ emission and its relationship
to the IS absorption features, and attempts to understand the
observations with simple models. 
We introduce in \S\ref{sec:galgal} the use of close angular pairs
of galaxies at different redshifts for mapping the spatial distribution of 
the circumgalactic gas around the foreground galaxies, while \S\ref{sec:model}
develops a simple geometrical and kinematic model for outflows consistent with both the
line profiles in galaxy spectra and the larger-scale distribution of gas in the CGM. 
\S\ref{sec:model_implications} discusses the observational results and their interpretation
in the context of the models, and \S\ref{sec:discussion} summarizes the conclusions 
and discusses the prospects for improvement in the future. 

We assume a $\Lambda$-CDM cosmology with $\Omega_m=0.3$, $\Omega_{\Lambda}=0.7$, and
$h=0.7$ throughout, unless specified otherwise.

\section{Bulk Outflow Velocities Associated With $z \sim 2-3$ Galaxies}
\label{sec:bulkvel}
\subsection{Spectra in the \Ha\ Galaxy Sample}

The rest-frame far-UV spectra of star forming galaxies include numerous absorption
features which (in principle) provide detailed information on the young OB stars
responsible for the bright continuum, as well as the IS and circum-galactic atomic and ionized gas 
associated with the galaxy (see, e.g., \citealt{pettini00,prs+02}). 
The strongest stellar features are due to stellar winds
from massive stars, producing broad P-Cygni profiles in higher ionization lines   
such as \ion{N}{5} $\lambda\lambda 1238$,1242, \ion{C}{4} $\lambda\lambda1548$,1550, and 
\ion{Si}{4} $\lambda\lambda 1393, 1402$. The absorption is broad ($\Delta v \simgt 2000$ 
\kms) and shallow, with depth dependent on the metallicity of the O-stars.  Photospheric
lines from the same OB stars are also present, but the lines are generally much weaker
than the wind features, and so can be difficult to discern in the spectra of individual
high redshift galaxies. The IS lines are superposed on the integrated stellar spectrum, where 
absorption features of abundant species (e.g., \ion{H}{1}, \ion{O}{1}, \ion{C}{2}, \ion{C}{4}, \ion{Si}{2},
\ion{Si}{3}, \ion{Si}{4}, \ion{Fe}{2}, \ion{Al}{2})  can be extremely strong --- strong enough to be 
useful for redshift identification in low S/N spectra. Unfortunately, the most accessible
absorption features-- the IS and wind features -- are not useful for accurate measurements
of galaxy redshifts because of the non-gravitational origin of their kinematics. Similarly, \lya\ emission,
which is observed in $\simeq 50\%$ of a continuum-selected sample at $z \sim 2-3$ 
(e.g. \citealt{ssp+03,ssp+04,kornei09})
is resonantly scattered, altering the emergent kinematics in a manner that depends on
both the optical depth and velocity distribution of the gas doing the scattering.  

Fortunately, nebular emission lines
lines originating in a galaxy's \ion{H}{2} regions are also relatively accessible observationally,
with the strongest lines (e.g., \Ha\ , \Hb\ , [OIII]) found in the observed-frame near-IR at the
redshifts of interest. The \Ha\ line, which is not resonantly scattered and whose strength
is strongly dependent on both the ionizing UV radiation field and the density, is likely to provide
a reasonable  estimate of the systemic redshift of the stars, as well as a measure of gravitationally-induced
motion within the galaxy \citep{pss+01,erb+06b}. The only disadvantage of the nebular spectra
is that their numbers are currently small in comparison with the available rest-UV spectra.   

The quality of rest-UV survey spectra for typical high redshift galaxies 
is generally not sufficient for detailed spectral analysis, and
so much of what we know about general trends between spectral properties and
other physical parameters of galaxies is based on composites (e.g., \citealt{spa01,ssp+03,erb+06a}).
In order to form these composites for the far-UV spectra, one must assume 
a relationship between the velocities of measured spectral features and the
object's systemic redshift $z_{\rm sys}$; as discussed above, 
this task is made difficult by the fact that the strong
lines of resonance transitions in the far-UV 
are seldom closer than a few hundred \kms\ to the
objects' true redshifts. 

With the current, enlarged sample of \ztwo\ galaxies, including generally higher-quality
UV spectra than available in the past, we revisit the measurement of galaxy
systemic redshifts using far-UV spectral features, building on previous results (\citealt{pss+01,assp03,ass+05}).
In this section, we examine in detail the relationship between observed far-UV IS lines, \lya\ emission,
and the redshifts defined by the \Ha\ line in the rest-frame optical, including careful attention
to possible systematic errors. 

The redshifts (both $\Ha$ and UV-based)
for all but 8 of the galaxies have been
presented in \cite{erb+06b}, where the observations are described in detail. 
The new $\Ha$ spectra 
were obtained in the same manner using NIRSPEC \citep{mbb+98} in 2005 June. 
Using a 0\secpoint76 entrance slit, the typical spectral resolution achieved with NIRSPEC
is $R \simeq 1400$, or FWHM$\simeq 215$ \kms, so that the typical \Ha\ line widths
of $\sigma_v \sim 100$ \kms\ (FWHM$\simeq 240$ \kms) are marginally resolved. 
The rest-UV spectra were obtained with  
the LRIS-B spectrograph on the Keck 1 telescope, primarily using a 400 line/mm grism
blazed at 3400 \AA\ with a dichroic beam splitter sending all wavelengths $\lambda < 6800$ \AA\
into the blue channel, as described in \cite{ssp+04}. 
As discussed in detail below, the typical effective spectral resolution 
achieved in this mode is $\simeq 370$ \kms\ (FWHM), or $\sigma_{res} \simeq 160$ \kms. 

For objects whose UV spectra show detectable interstellar absorption lines, we have included 
only those measurements deemed of sufficient quality to
yield reasonably precise measurements. We have also excluded the handful of objects 
from the \citealt{erb+06b} sample with $z < 1.9$, in order to avoid systematic biases 
caused by differential atmospheric refraction (discussed below), and to focus on the redshift
range for which we have the highest quality ancillary data. 

The resulting \Ha\ sample consists of
a total of 89 galaxies with $\langle z \rangle = 2.27\pm0.16$, of which 48 (54\%) have UV spectra in which only an interstellar absorption
line redshift was measured, 38 (43\%) have measurements of both $\lya$ emission and interstellar absorption, and
3 (3.5\%) have only $\lya$ emission redshifts measured. These fractions are comparable to those
obtained from the full spectroscopic sample of $\simeq 1600$ galaxies in the same range of redshifts,
$1.9 \le z \le 2.6$ (see \citealt{ssp+04}). Most (87\%~) of the galaxies in the parent sample satisfy the ``BX'' 
photometric selection criteria (\citealt{ass+04,ssp+04}), with the remainder satisfying the
``MD'' color criteria defined in \cite{sas+03}. These galaxies selected on the basis of their UV
color have been shown to include all
but the dustiest star-forming galaxies in the redshift range of interest \citep{as2000,reddy08}.  

\subsection{Redshift Uncertainties}

As discussed in detail by \cite{ssp+03}, a large number of interstellar absorption
lines is commonly observed in far--UV spectra of star-forming galaxies. In general, all of the
detected lines are used to verify that the assigned redshift is correct, but the positions of
only 3 lines: \ion{C}{2} $\lambda 1334.53$, \ion{Si}{4} $\lambda 1393.76$, and \ion{Si}{2} $\lambda 1526.72$, have been 
used to measure $\zis$. Of the strong resonance lines in the rest wavelength
range 1000--1600 \AA\ (the range in common for almost all of the spectra discussed in this paper), 
these lines are the
least likely to be blended with other strong lines and most likely to yield a consistent measure of
$\zis$. Clearly, the accuracy of each measurement of $\zis$ depends on the
strength and width of the lines, and on the quality of the spectra. The S/N of the spectra
used for the current study varies considerably. Empirically, from internal agreement of different
absorption lines in the same spectrum, and from repeated measurements of the same galaxy on different slitmasks,
we estimate that the typical measurement uncertainties are $\sim 100$ \kms\ for $\zis$ 
redshifts and $\sim 50$ \kms\ for $\zla$. 

Unfortunately, systematic uncertainties can be more difficult to quantify. For example,  
the measured wavelengths of features depend on the illumination of the spectrograph
slit by the object, whereas the wavelength solutions are determined from calibration lamps
and night sky spectra which illuminate the 1\secpoint2 slits uniformly. This source of error can also be
wavelength-dependent due to the effects of differential atmospheric refraction. The features used 
for redshift measurement generally fall in the observed wavelength range $\sim 3750-5200$ \AA\ for the 
$\Ha$ sample considered here, with $\langle \zha \rangle = 2.27\pm0.16$. Over this wavelength range,
the differential refraction\footnote{All of the spectroscopic observations of the \Ha\ sample
were obtained before the commissioning of the Cassegrain ADC on Keck 1. We are excluding
for the moment galaxies at redshifts where key features fall at wavelengths shorter than
3750 \AA\ due to the rapidly increasing
amplitude of differential refraction. } 
would be $\sim 0.7$ arcsec at an airmass of secz$=1.3$. Slitmasks
were almost always observed within $\sim 10-15$ degrees of the parallactic angle when the airmass
was significantly different from 1.0, but even so, the amplitude of differential refraction
perpendicular to the slit could be as large as $\sim 0.2$ arcsec, which could map into a velocity
shift of up to $\sim \pm 75$ \kms\ depending on the seeing and the spatial profile of the
galaxy. 

The measurements of $\zha$ are subject to a different set of systematic uncertainties.
Differential refraction in the near-IR is negligible for our purposes, 
but the observations were obtained using a 0\secpoint76 slit
after applying a blind offset from a nearby star to the position of the galaxy measured from either
an optical or near-IR continuum image (see \citealt{ess+03,erb+06b}). 
While the observations were taken in a way that should minimize
any systematic offsets in the resulting $\zha$ due to pointing or astrometric inaccuracies,
they of course measure the velocity
of only the flux that entered the slit. Empirically, we found that repeated observations of the
same galaxy (using different slit position angles) suggest an accuracy of $\pm 60$ \kms\ for
$\zha$ (rms), with the largest excursions from consistency applying to objects known
to be spatially extended. Another direct comparison of the redshifts for 14 objects 
from the NIRPSEC sample of \cite{erb+06b} is provided by observations of the same objects with  
the integral-field spectrometer SINFONI at the VLT \citep{fg+06}. The SINFONI redshifts are
based on the velocity centroid of all detected $\Ha$ flux from the object, and are not affected
significantly by pointing errors or by slit losses. This comparison
shows a level of agreement similar to our estimate from multiple observations with NIRSPEC:
$\langle z_{N}-z_{S}\rangle = -34 \pm 59$ \kms (rms), 
where $z_{N}$ is the NIRSPEC redshift and $z_{S}$ is the SINFONI redshift. 
The average offset is marginally significant ($\simeq 2\sigma$), but the scatter is consistent
with our NIRSPEC experience.  Eleven of the galaxies in the present \Ha\ sample
were also observed using OSIRIS ( with Laser Guide Star Adaptive Optics (LGSAO) on the Keck
2 telescope \citep{law09}; the average redshift difference $z_N - z_O = 8\pm41$ \kms, 
indicating no evidence for a systematic difference.   

In summary, adopting an uncertainty in $\zha$ of 60 \kms, 
the typical uncertainties in the measured values of $\dvis= c (\zis - \zha)/(1+\zha)$ 
and $\dvla = c(\zla - \zha)/(1+\zha)$ 
for a given galaxy are $\sim 130$ \kms\ and $\sim 90$ \kms, respectively.

\subsection{Sample Statistics}

\label{sec:sample_stats}

\begin{figure}[htb]
\centerline{\epsfxsize=8cm\epsffile{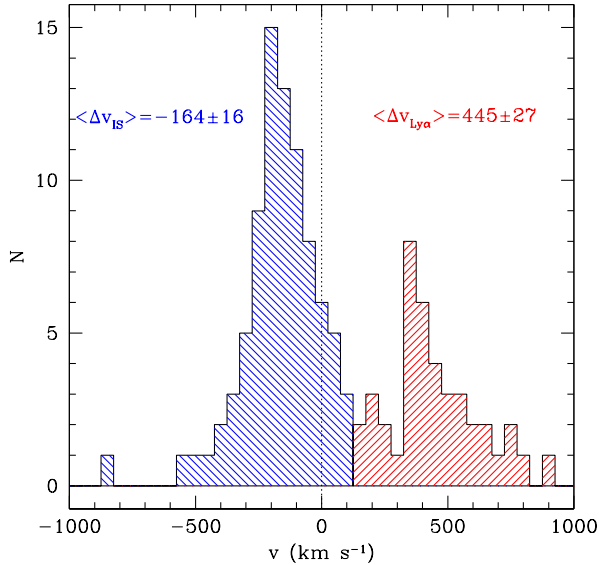}}
\figcaption[fig1.eps]{Histogram of the measured (centroid) velocities of interstellar absorption lines (blue) and
Lyman $\alpha$ emission (red) with respect to the galaxy nebular redshift as defined by the
centroid of the H$\alpha$ emission line, for a sample of 89 galaxies with $\langle z\rangle=2.27\pm 0.16$.
The sample includes only those galaxies having both nebular line redshifts and rest-UV spectra of
adequate quality to measure absorption line centroids.  In this sample, 86 of the 89 galaxies
have measured values of $\zis$, 3 have only $\zla$, and 39 have both.
The mean values of the velocity offsets are indicated.
\label{fig:vhist}
}
\end{figure}

\begin{figure}[htb]
\centerline{\epsfxsize=8cm\epsffile{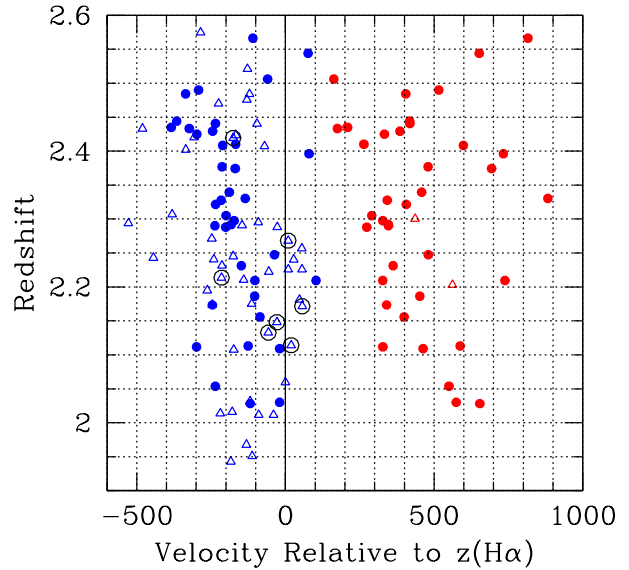}}
\figcaption[fig2.eps]{
Plot showing the interstellar absorption line centroid velocities (blue) and centroid Lyman $\alpha$ emission
velocities relative to the redshift defined by H$\alpha$ for the same sample as in Fig.~\ref{fig:vhist}.
Galaxies for which both interstellar absorption redshifts and Lyman $\alpha$ emission redshifts
are available are indicated with blue (absorption) and red (emission) solid dots; open triangles
show systems for which one or the other measurements is not available. The circled objects
are ones which exhibited measurable velocity shear in the sample of \citealt{erb+06b} (see text for discussion).
\label{fig:v_vs_z}
}
\end{figure}

Fig.~\ref{fig:vhist} shows a histogram of the IS and
\lya\ emission velocities with respect to the systemic redshift defined by $\zha$. The
distributions have 
$\langle \dvis \rangle = -164\pm16$ \kms, and $\langle
\dvla \rangle = 445\pm 27$ \kms, where the quoted uncertainties are errors in the mean. 
Fig.~\ref{fig:v_vs_z} illustrates the same
sample as in Fig.~\ref{fig:vhist}, where different symbols are used depending on the
UV spectral morphology of the galaxies.  

In light of the current \Ha\ sample of $z \sim 2.3$ galaxies, it is worth re--examining
the ``rules'' that one would use to estimate the true systemic redshift of the
galaxies given only information contained in their rest-UV spectra, and assuming
that $\zha$ defines the rest frame. Using a linear regression form similar to that
used by \cite{ass+05}, for galaxies with both $\zla$ and $\zis$ measurements,

\begin{equation}
\zha  = \zis + 0.00289 - 0.0026(2.7-\zis)\; ~~~\sigma_z = 0.00127 
\end{equation}
\begin{equation}
\zha = \zla - 0.0054 + 0.0001(2.7-\zla)\; ~~~\sigma_z=0.00193 ,
\label{eq:zem}
\end{equation}
corresponding to velocity offsets of $\dvis = -170\pm 115$ \kms and $\dvla =+485\pm175$ \kms,
respectively, at the mean redshift of $\zha = 2.27$. 
For objects with a measurement of $\zis$ only, 
\begin{equation}
\zha = \zis+0.00303 -0.0031(2.7-\zis) ; ~~~\sigma_z = 0.00145~,
\end{equation}
or $\dvis = -165 \pm 140$ \kms\ (error is the standard deviation) at the mean redshift of the sample.   

Using all 86 \Ha\ galaxies with measured $\zis$, the best fit single relation
of the form in eqs. 1-3 is 
\begin{equation}
\zha = \zis + 0.00299 -0.00291(2.7-\zis)\; ~~~\sigma_z=0.00138
\label{eq:z_from_zabs}
\end{equation}
or $\dvis = -166 \pm 125$ \kms at $\langle \zha \rangle =2.27$. 
We find that including $\zla$ in the above regression formulae increases the
rms redshift uncertainty over that obtained using only $\zis$, in contrast
to similar estimates at somewhat higher redshift by \cite{ass+05}. One possible
explanation for the difference could be the generally weaker \lya\ lines
in the $z\simeq 2.3$ sample compared to that at $z\simeq 3$ \citep{reddy08}. 
We return in \S\ref{sec:lya} to a discussion of the kinematics of the \lya\ emission line. 
In any case, using {\it only} the
absorption redshift, with a constant offset of $\simeq +165$ \kms, would provide an
estimate of $\zha$ accurate to $\sim 125$ \kms (rms). 

There are too few objects (3 of 89) in the $\Ha$ sample having only $\zla$ to define a significant
relationship for such objects (which are also quite rare in the full $z\simgt 2$
spectroscopic sample), although these 3 objects have $\langle \dvla \rangle = 400\pm183$, consistent
with equation~\ref{eq:zem} above. For this reason, we use equation~\ref{eq:zem} for
subsequent estimates of $\zha$ when only \lya\ emission is available.   

Figs.~\ref{fig:vhist} and \ref{fig:v_vs_z} show that a significant fraction of the
galaxies have IS absorption line centroid velocity shift $\dvis$ consistent with zero.  
Given the uncertainties in $\dvis$,
this is not particularly significant for individual objects, but we discuss the
issue further because of the intriguing behavior of $\dvis$ and $\dvla$ with
respect to one another and because of the greater significance of the result
in higher S/N composite spectra discussed below. 
Three of the 11 galaxies
with measured $\dvis\geq0$ also have \lya\ emission, and have 
$\langle \dvla \rangle =+708\pm50$ \kms, $\sim250$ \kms\ higher than
the average of the full sample; however, the average 
$\langle \dvla-\dvis \rangle = 622\pm40$ \kms\ for
this set of objects is nearly identical to that of the full sample.    
The relative consistency of the difference $\dvla-\dvis$ in the \Ha\ 
sample, as well as in much larger samples without the benefit of $\Ha$ spectroscopy
(e.g., \citealt{ssp+03,ssp+04}), suggests a situation in which $\dvla$ moves
in concert with $\dvis$ irrespective of whether the centroid of the absorption 
line velocities are blue-shifted with respect to systemic. 

Interestingly,
3 other galaxies of the 11 (see Fig.~\ref{fig:v_vs_z})
with $\dvis\ge 0$ are among those
with spatially resolved velocity shear in the \Ha\ emission line, meaning that
the \Ha\ spectrum exhibits a significant velocity offset as a function 
of spatial position; see \cite{erb+06b} for details. Given the overall detection
rate of shear in the sample of \cite{erb+06b}, we would expect to 
find $\sim 1$ such object in a sample of 11.  Furthermore, of the
14 objects with tilted $\Ha$ emission lines in the \citealt{erb+06b} sample, 
eight have high quality
absorption redshifts with $\langle \dvis \rangle =
-47\pm35$ \kms;  6 of 8 have $\dvis >-60$ \kms.  

The apparent connection
between decreased $|\dvis|$ and observation of measurable velocity shear in
$\Ha$ can be interpreted
in at least two ways.  If the velocity shear indicates an unresolved
merger, infalling gas (or an unusual amount of gas near zero velocity)
could significantly reduce the average blue-shift
of the interstellar lines. The interstellar and nebular redshifts
could also arise from different pieces of a merger, or even different galaxies
altogether (see \citealt{quider09b} for a possible example).  Alternatively, if the
tilted emission 
lines are caused by rotation that is most easily detected in nearly edge-on
configurations, the outflows in these objects may be
loosely collimated perpendicular to the disk as for many local starburst galaxies.  
If this were the case, one might expect to observe lower outflow velocities
and perhaps stronger IS absorption near the galaxy systemic redshift 
for objects with smaller inclination angles to the line of sight.  
Both of these effects appear to play a role in the Na D IS line kinematics
observed in a large sample of nearby star-forming galaxies \citep{chen10}. 
While possible inclination effects would not 
naturally account for highly redshifted \lya\ emission, none of the objects
with velocity shear identified from their \Ha\ spectra happen to have \lya\ in emission.  
Inclination effects could be present in
our sample, although the significant bulk outflows 
observed in the vast majority of the sample argue against collimation
and projection effects being a major factor in most cases.  

As we will show in
\S\ref{sec:composites} below, we favor an explanation for many of the observed
trends discussed in this section that hinges on the quantity of gas at or near 
zero velocity (and {\it not} on the overall outflow speed). 

\section{The Relation Between Bulk Outflows and Other Galaxy Properties}
\label{sec:correlations}

One of the advantages of the sample of $z\sim2$ galaxies discussed in this
paper is that a large number of other galaxy properties are available to
look for trends with respect to the bulk outflow properties. 
Most of the measurements and inferred quantities used here are tabulated in
\citealt{erb+06b,erb+06c}. 
Among the parameters available are the $\Ha$ line widths $\sigma_v$, the star formation
rates inferred from the $\Ha$ line fluxes (corrected for extinction according to the method
outlined in \citealt{erb+06c}), the surface density of star formation ${\rm \Sigma_{SFR}}$,
the dynamical mass ${ M_{dyn}}$ measured from a combination of $\sigma_v$ and 
the observed physical size of the $\Ha$ emitting region, and the stellar mass ${ M_*}$
inferred from SED fitting from the rest-UV to the rest optical/IR.
The cold gas mass $ M_{gas}$ is estimated by using the measured \Ha\ 
surface brightness and galaxy size and assuming that the local Schmidt-Kennicutt
\citep{k98schmidt}  
relation between gas surface density and star formation rate applies. Finally, the  total 
baryonic mass $ M_{bar}=M_* + M_{gas}$ 
and the fraction of the inferred baryonic
mass in the form of cold gas, $\mu = {M_{gas}/M_{bar}}$, have been utilized. 

Table~\ref{tab:correlations} summarizes the results of Spearman correlation tests 
between $\dvis$, $\dvla$, $\dvla-\dvis$, and these other physical quantities. The number of galaxies
available in the sample for each test, which is also given in table~\ref{tab:correlations},
varies depending on the quantity being evaluated. The tests have been conducted against
the absolute value of the quantities $\dvis$ and $\dvla$ so that the sense of any
correlations is positive when the bulk velocity differences increase with the other
physical characteristic. 
None of the quantities considered is significantly correlated with
$\dvla$ or $\dvla-\dvis$ at more than the 95\% (2$\sigma$) confidence level, although the measured
\Ha\ velocity dispersion $\sigma_v$ may be marginally correlated, in the sense that
$\dvla$ is larger for objects with larger $\sigma_v$. 

\begin{deluxetable}{l c c c}
\tablewidth{0pt}
\tabletypesize{\footnotesize}
\tablecaption{Correlations of \dvis\ and \dvla\ with Galaxy Properties\tablenotemark{a}\label{tab:correlations}}
\tablehead{
\colhead{Quantity} &
\colhead{${\rm -\Delta v_{IS}}$} &
\colhead{${\rm \Delta v_{\lya}}$}  &
\colhead{${\rm \dvla - \dvis}$}
}
\startdata
$\sigma_v$\tablenotemark{b}     &  $-$2.08  (65)  & +1.81  (29) &  +0.74  (29) \\
SFR\tablenotemark{c}            &  $-$1.52  (87)  & $-$0.04  (42) & $-$0.08  (39)\\
$\Sigma_{\rm SFR}$\tablenotemark{d} & +0.95  (81)  & +0.82  (37)  & +0.49  (35)\\
${\rm M_{dyn}}$\tablenotemark{e} &  $-$2.24  (57)  & +1.02  (24)  & +0.14  (24)\\
${\rm M_{gas}}$\tablenotemark{f} &  $-$1.68  (73)  & +0.71  (36)  & +0.85  (34)\\
${\rm M_{*}}$\tablenotemark{g}   &  $-$1.93  (73)  & $-$0.37  (36) &  $-$0.48  (34)\\
${\rm M_{bar}}$\tablenotemark{h} &  $-$2.66  (73)  & $-$0.10  (36)  &  $-$0.10  (34)\\
$\mu$\tablenotemark{i}           &  +1.72  (73)  & +0.64  (36)  &  +0.93  (34)\\
\enddata
\begin{indent}
\tablenotetext{a}{All values are the number of standard deviations from the null hypothesis that
the quantities are uncorrelated, based on a Spearman rank correlation test. Negative values
indicate anti-correlations between the quantities. The number in
parentheses following each value is the number of galaxies in the sample used to evaluate
the correlation. }
\tablenotetext{b}{Velocity dispersion measured from the \Ha\ emission line.}
\tablenotetext{c}{Star formation rate, in ${\rm M_{\odot}}$ yr$^{-1}$, measured from
the intensity of the $\Ha$ emission line, and corrected for extinction as in \citealt{erb+06c}.}
\tablenotetext{d}{Average star formation surface density, as in \citealt{erb+06c}.}
\tablenotetext{e}{Dynamical mass, as tabulated in \citealt{erb+06a}.}
\tablenotetext{f}{Cold gas mass, estimated from the star formation surface density and
the observed $\Ha$ size, as in \citealt{erb+06a,erb+06c}.}
\tablenotetext{g}{Stellar mass, estimated from SED fitting, from \citealt{erb+06b}}.
\tablenotetext{h}{Total baryonic mass, ${\rm M_*} + {\rm M_{gas}}$.}
\tablenotetext{i}{Gas fraction, ${\rm M_{gas}/M_{bar}}$, as in \citealt{erb+06a,erb+06b}.}
\end{indent}
\label{fig:corr_table}
\end{deluxetable}

\begin{figure}[htb]
\centerline{\epsfxsize=8cm\epsffile{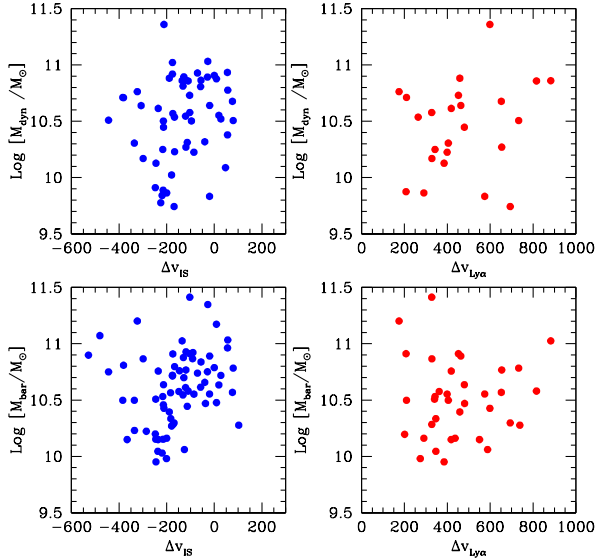}}
\figcaption[fig3.eps]{Plots showing the dependence of bulk outflow velocities on dynamical mass estimated from $\Ha$ line widths and
observed sizes (top panels) and total baryonic mass (i.e., the sum of inferred stellar
and cold gas masses)  (bottom panels).
\label{fig:m_vs_v}
}
\end{figure}

The measured values of $\dvis$, on the other hand, show greater than $2\sigma$ 
deviations from the null hypothesis (no correlation) for the quantities
$\sigma_v$, ${ M_{dyn}}$, and ${ M_{bar}}$. While $\sigma_v$ and ${ M_{dyn}}$
are correlated with one another by virtue of the fact that the former is used to
calculate the latter (see \citealt{erb+06b}), the two mass estimates ${ M_{dyn}}$
and ${M_{bar}}$ are measured using completely independent methods. \citet{erb+06b}
show that these mass estimates track each other very well, inspiring some confidence
that both are providing reasonable estimates of the true masses within the \Ha\ emitting
regions of the galaxies. Within the \Ha\ sample considered here, the inferred 
value of ${M_{bar}}$ ranges from $\simeq 1\times 10^{10}$ to $\simeq 3\times10^{11}$ \msun, and was
shown to track ${ M_{dyn}}$ much more tightly than ${ M_*}$, the stellar mass.
Fig.~\ref{fig:m_vs_v} shows plots of ${ M_{dyn}}$ and ${ M_{bar}}$ versus
$\dvla$ and $\dvis$. The sense of the possible correlations of $\dvis$ with galaxy
mass is that $\dvis$ (the centroid velocity of the interstellar absorption lines) tends to be
closer to $v=0$ for objects with higher mass. The more significant of the
mass correlations, with $M_{bar}$ (significant at the 99\% confidence level), 
appears to arise because none of the galaxies with baryonic masses 
smaller than $\simeq 3\times 10^{10}$ \msun\ 
have measures of $\dvis$ near zero, while at masses ${ M_{bar} > 3\times10^{10}}$
\msun, a significant fraction do. 

There are no significant correlations between $|\dvis|$ and parameters related
to star formation. 
The lack of correlation is interesting in light of recent studies relating outflow 
velocity to galaxy properties in local starbursts; for example, both \citet{martin05} and
\citet{rupke05} find evidence for correlations of outflow speed with SFR and
dynamical mass (represented by the circular velocity), in the sense that
the outflow velocities increase with both quantities (i.e., apparently the opposite
to what is observed in the $z \simeq 2$ sample).  
These relatively local galaxy samples have a 
dynamic range spanning $\sim4$ orders of magnitude in SFR, and
include dwarf starbursts with $v_c\sim30$ \kms\ and ${\rm SFR}<1$
\msunyr, i.e. a much larger range than present in our $z \sim 2.3$ sample. 
The observed correlations in the local sample flatten for galaxies with
${\rm SFR}\gtrsim10$--100 \msunyr\ \citep{rupke05}, the approximate range of SFRs in our
current sample.  In other words, no trends are present in the low-redshift samples when only
galaxies with parameters characteristic of our \ztwo\ galaxies are
considered.  
Also, as we discuss below,
other effects on the kinematics measured from centroid velocities 
would likely mask the presence of a weak correlation if it were present. 

At higher redshift, \citet{weiner09} also found that
the inferred outflow velocity in composite spectra of $z \simeq 1.4$ star-forming
galaxies is a slowly increasing function of SFR, $v_{out} \propto {\rm SFR}^{0.3}$. 
However, because their composite spectra are of relatively high
spectral resolution, \cite{weiner09} measure $\dvis$ as the velocity at which the IS absorption
reaches 90\% of the continuum value, i.e. close to the {\it maximum} blue shifted
velocity rather than the centroid. They decompose the \ion{Mg}{2} IS line into
a ``symmetric'' and ``outflowing'' component, and find that the strength of the
symmetric component is very steeply dependent on stellar mass-- in fact, it is much
{\it steeper} than the variation of the outflowing component. As in our sample,
the lowest-mass galaxies are consistent with having {\it zero} symmetric component 
(see their Fig~13 and Table~1). 

Clearly, the centroid velocity of strongly saturated IS
absorption lines is in many ways a blunt tool for characterizing the velocity
of outflowing gas. First, the lines may include a significant amount
of interstellar gas at or near zero velocity with respect to the galaxy systemic
redshift, which would have the effect of decreasing the measured value of $|\dvis|$
even if outflow velocities were substantial. While 
\citealt{martin05,rupke05} and \citealt{weiner09} removed a component of interstellar absorption  
centered at zero velocity before evaluating the velocity of outflowing material, 
it is generally not possible to do this for individual spectra in our $z \simeq 2$ sample because of
more limited spectral resolution. It is also not clear that subtracting
a symmetric $v=0$ component of IS absorption is the best approach for evaluating the
kinematics of circumgalactic gas. It is possible, or perhaps even likely, that
the IS line profiles would be sensitive to infalling material, which when seen
in absorption against the galaxy's UV continuum would tend to be redshifted by up to a few hundred \kms\ with
respect to a galaxy's systemic redshift; if so, the observed redshifted wing of
IS absorption could be substantially stronger than the blue-shifted portion of
the $v \simeq 0$ profile. As we have mentioned (see also \S\ref{sec:cold_flows} below), 
gas with positive (red-shifted) 
velocities with respect to $z_{sys}$ would be expected in at least a fraction
of galaxies if infalling cool gas is present with a non-negligible covering fraction.  

\begin{figure}[htb]
\centerline{\epsfxsize=8cm\epsffile{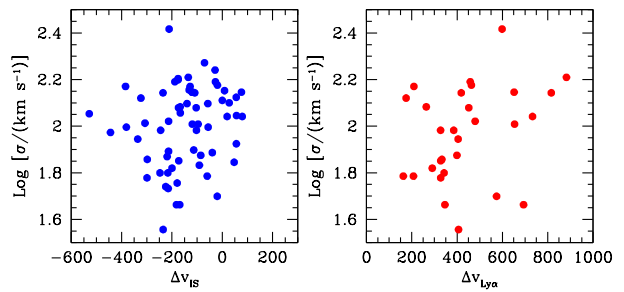}}
\figcaption[fig4.eps]
{Dependence of the \Ha\ velocity dispersion $\sigma_v$ on centroid velocities $\dvla$ and
$\dvis$ measured from the far--UV spectra.
\label{fig:vsig}
}
\end{figure}

Finally, we address whether outflows influence the \Ha\ line
widths used to calculate dynamical masses \citep{erb+06b}.  In Fig.~\ref{fig:vsig}
we show a comparison of $\dvla$ and $\dvis$ with the \Ha\
velocity dispersion $\sigma_v$.  As mentioned above, there is a marginally
significant ($\simeq 2\sigma$) tendency for $|\dvis|$ to be smaller
in galaxies with larger values of $\sigma_v$. The correlation is likely to be
related to the trends with galaxy mass we have already noted. 
If strong outflows were significantly influencing the observed $\Ha$ lines, 
one might expect objects with the largest $|\dvis|$ also to have
larger $\sigma_v$, a trend opposite to what is observed. 
As further evidence of the disassociation of the observable \Ha\ emission and the winds,
composite UV spectra (discussed in \S\ref{sec:composites}) constructed by combining individual spectra
shifted to the nebular redshift show that \Ha\ is at the same redshift
as the stars.  Additionally,
observations of local galaxies  
suggest that \Ha\ emission from outflowing material would fall far below our
detection threshold; for example, \citet{lhw99} studied the
extended \Ha\ emission from the superwind in the starburst galaxy M82,
finding that it has 
a total luminosity of $2.4\times10^{38}$ ergs s$^{-1}$ and comprises only
$\sim 0.3$\% of the total \Ha\ flux.  Since the typical S/N of our $z \sim 2.3$ \Ha\ observations
is $\simlt 10$, it would be extremely difficult to recognize a component of emission
coming from wind material. 
Finally, \citet{cam05} have also found, through integral field spectroscopy of
\Ha\ emission in local ULIRGs, that the central velocity dispersions are
unaffected by outflows.

\section{Inferences from Composite Far-UV Spectra}
\label{sec:composites}

\subsection{The use of \zha\ to measure galaxy systemic redshifts}

\begin{figure*}[ht]
\centerline{\epsfxsize=13cm\epsffile{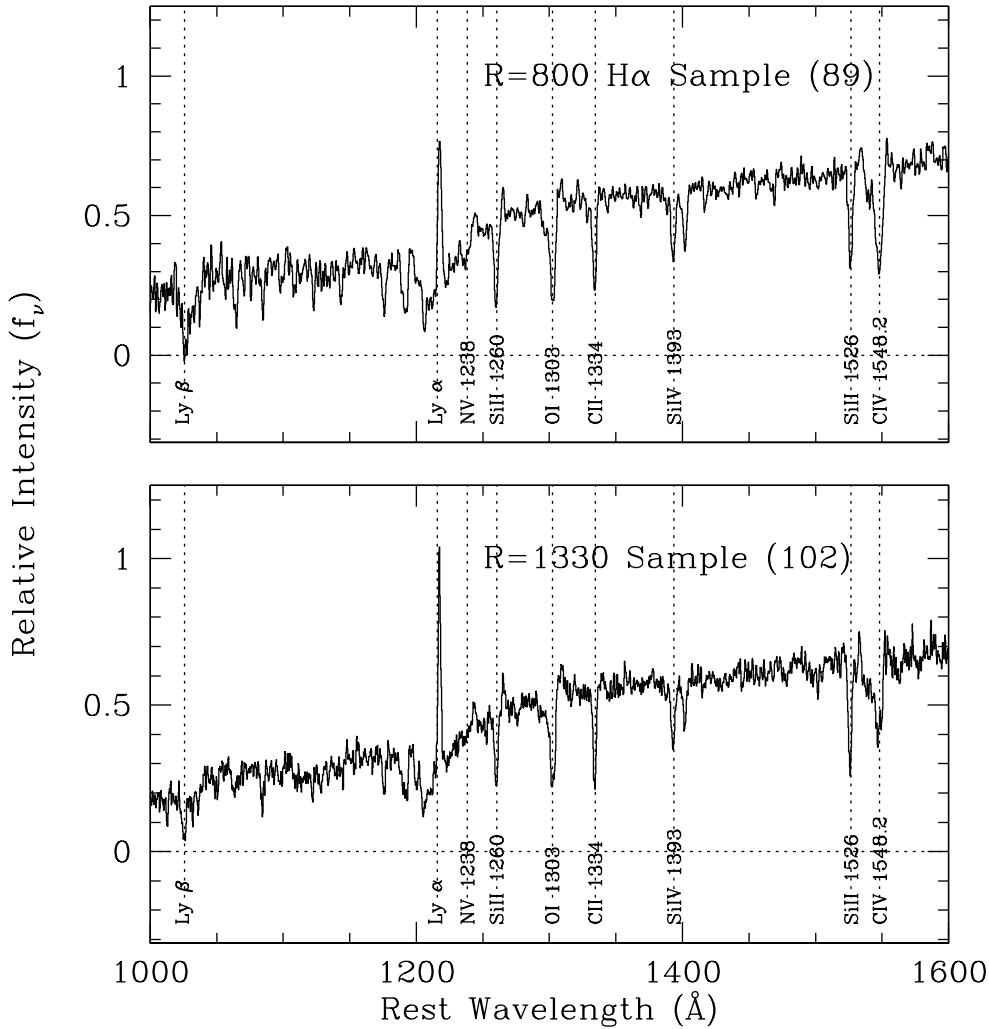}}
\figcaption[fig5.eps]
{Composite rest-frame far-UV spectra for two independent samples of $z \sim 2.3$ galaxies.
The top panel is
an average of the 89 spectra in the \Ha\ sample, with $R=800$, after normalizing each to the same relative intensity in the range
1300--1500 \AA. The bottom panel is a composite of 102 galaxy spectra obtained with higher
spectral resolution ($R=1330$), shifted into the rest frame using equations~\ref{eq:zem} and \ref{eq:z_from_zabs}, and
scaled as the first sample before averaging.
\label{fig:restspec}
}
\end{figure*}

The existence of a large sample of galaxies for which both rest-frame optical (\Ha)
and rest-frame far-UV spectra are available provides an opportunity to evaluate
high S/N composite UV spectra formed from unusually precise knowledge of the systemic
redshifts of the galaxies. A stacked composite UV spectrum was formed by shifting each 
spectrum into the rest-frame using $\zha$,
scaling based on the flux density 
in the range $1300-1500$ \AA, re-sampling onto a common wavelength scale of 0.33 \AA\ per pixel, and averaging 
at each dispersion point, with outlier rejection.  
Thus, the composite is an unweighted average of all 89 galaxies
having both \Ha\ spectra from NIRSPEC and high quality UV spectra from LRIS-B; the result
is shown in Fig.~\ref{fig:restspec}.
A measurement of the centroids of weak stellar photospheric absorption features (\ion{S}{5} 
$\lambda 1501.76$, \ion{O}{4} $\lambda 1343.354$, and \ion{C}{3} $\lambda 1171.76$ were
used for this purpose; see \citealt{ssp+03,pettini00})
in the rest-frame composite spectrum
verifies that the \Ha\ redshifts are very close to the systemic redshifts of the stars
in the galaxies, with a mean velocity of $v_*=13\pm24$ \kms. Composites formed from various
subsets of the data yield a similar level of agreement: for example, a composite formed
from the subset of 28 galaxies with significantly deeper LRIS-B spectra 
(with total integration times ranging from 5-20 hours compared to the typical
1.5 hours) has $v_*=2\pm36$ \kms.  

\subsection{Velocity widths of IS lines and the maximum outflow velocity}

One of the quantities that can be evaluated with increased confidence using the new
sample is the velocity {\it extent} of the interstellar absorption lines and \lya\ emission, in addition
to the centroid velocities discussed in the previous section. Of particular interest are 
${\rm \eta_{IS}}$, the FWHM of the interstellar absorption lines,
and ${\rm \eta_{\lya}}$, the FWHM of the \lya\ emission line, corrected for the instrumental
resolution. Unfortunately, as discussed above, the instrumental resolution is not always known precisely, because
it depends on the seeing-convolved size of the galaxy compared to the 1\secpoint2 slits used for
all of the LRIS-B spectroscopy. For objects illuminating the slit uniformly, the spectral
resolution is measured to be 450 \kms\ (FWHM) for the 400 line/mm grism used for almost all
of the optical spectra of galaxies in the \Ha\ sample; however, for typical seeing of $\simeq 0.6-0.8$\arcs,
galaxies in our sample have FWHM$\simeq$ 1.0\arcs. Given this, we would expect that 
the actual spectral resolution is FWHM$\simeq 370$ \kms ($R=800$). We have verified this for both 
the composite spectrum of all \Ha\ galaxies, and for 
a smaller subset of galaxies with very deep spectroscopic integrations, by measuring the 
spatial size of each object in the slit direction. 
Assuming that
the \Ha\ redshift uncertainties are $\sigma_z\simeq 60$ \kms\ and that individual  
spectra have $\sigma_{\rm res} \simeq 160$ \kms (i.e., FWHM/2.355), the estimated {\it intrinsic} FWHM of
IS absorption features and \lya\ emission in the composite spectrum are $\eta_{IS} \simeq 540-570$ \kms and
${\rm \eta_{\lya}} \simeq 620-650$ \kms, respectively. 

We have recently obtained LRIS-B spectra of galaxies selected in the same way as the \Ha\ sample,
but observed using the 600 line/mm grism instead of the 400 line/mm grism used for the vast majority of the
\Ha\ sample. These observations provide spectral resolution 1.68 times higher for the same slit width and
object size. As a test of our ability to measure line profiles with marginally resolved
data, we assembled a sample of 102 spectra all obtained on the same observing run and with consistent
observing conditions (seeing of $\simeq 0.6-0.7$\arcs\ FWHM evaluated at 4800 \AA) in a field which
remained very close to the zenith at the Keck Observatory, thus minimizing issues of differential
atmospheric refraction. We applied the rules given in equation 4 above to shift the spectra
into the rest frame, and produced a composite spectrum, shown in the bottom panel of Fig.~\ref{fig:restspec}, with
an effective spectral resolution of FWHM$\simeq 225$ \kms ($R\simeq1330$). The stellar photospheric 
absorption features in the stacked spectrum have $v_* = -2 \pm 10$ \kms, illustrating that the
rules for estimating the systemic velocity from the UV spectra work very well on average, and
that the higher resolution spectra are advantageous for producing more accurate wavelengths for
weak (stellar and IS) features.  

\begin{figure}[htb]
\centerline{\epsfxsize=8cm\epsffile{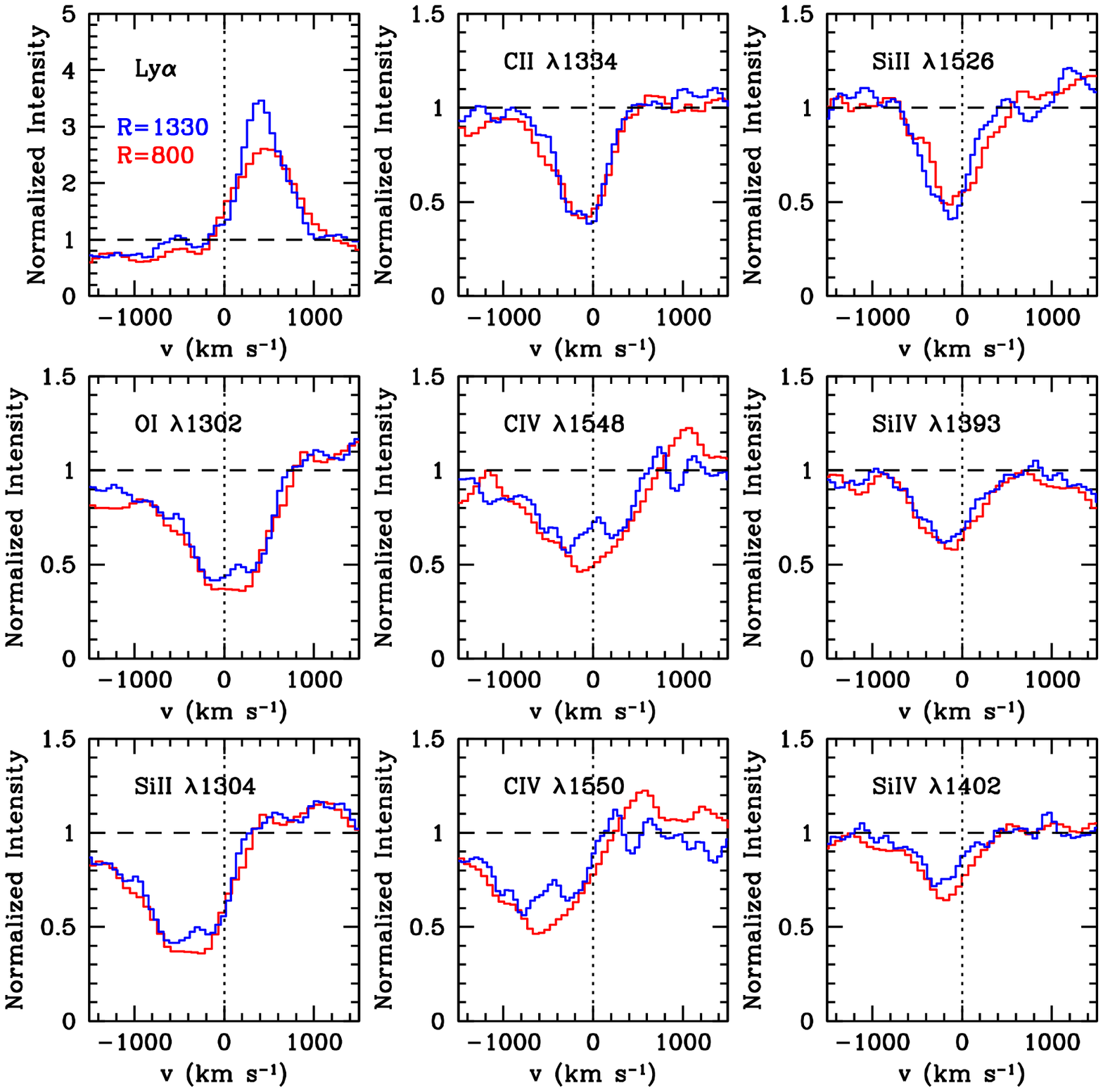}}
\figcaption[fig6.eps]{A comparison of the \lya\ and IS line profiles for the full \Ha\ sample (red)
and the composite of 102 identically-selected objects observed with higher spectral
resolution (blue), where the latter have been shifted into the rest frame using the rules defined in
\S\ref{sec:bulkvel}. For blended features such as \ion{O}{1} $\lambda1302$/\ion{Si}{2} $\lambda1304$ and
\ion{C}{4}  $\lambda\lambda 1548$, 1550, a separate panel is given for each; in both cases,
one should look at the longer of the two components for the profile near $v\sim 0$
and the shorter one to gauge the extent of the most blue-shifted velocities. Note the very
similar velocity profiles for the two independent samples, for both the minimum and
maximum velocities.
\label{fig:velcompare_1}
}
\end{figure}

\begin{figure}[htb]
\centerline{\epsfxsize=8cm\epsffile{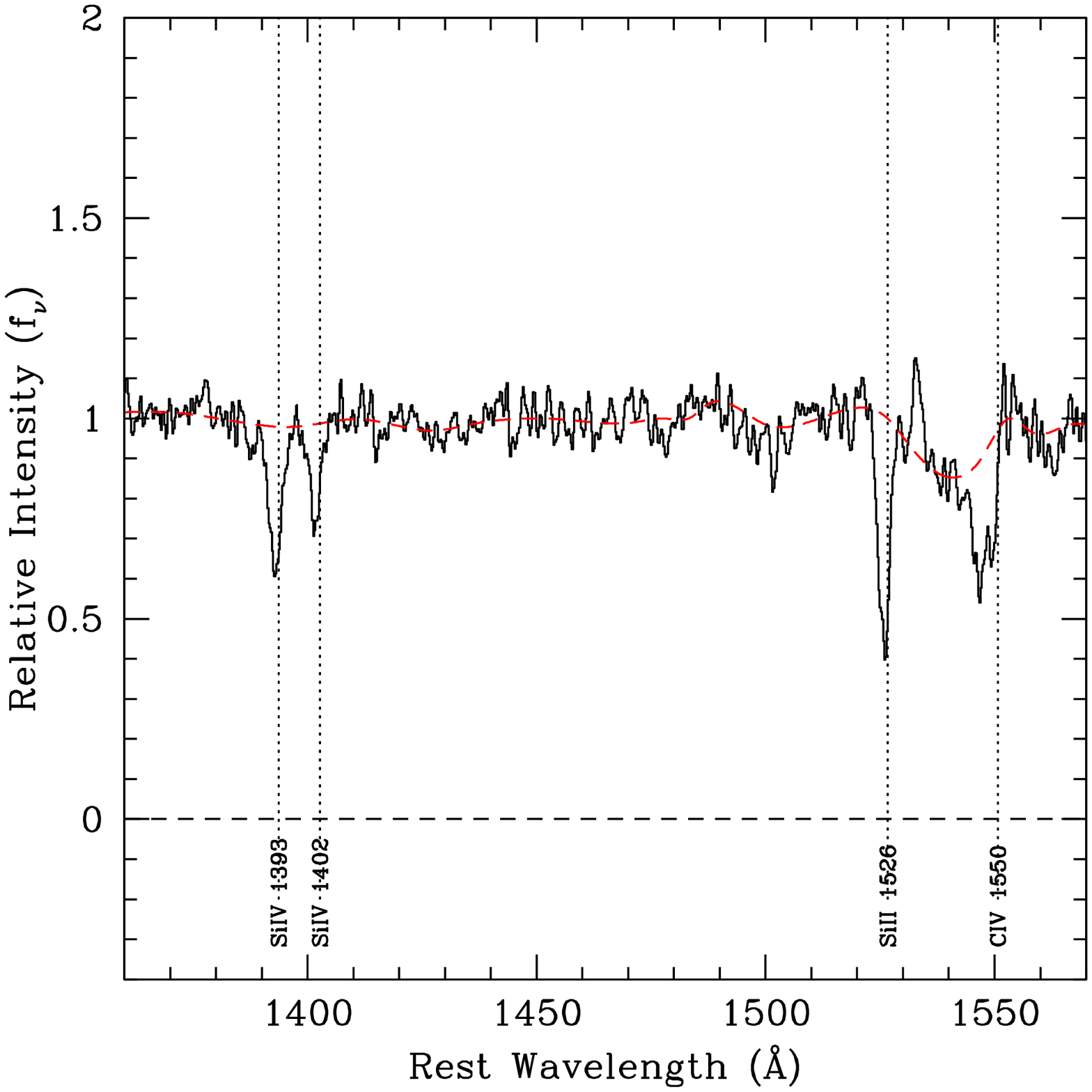}}
\figcaption[fig7.eps]{A portion of the composite spectrum shown in the lower panel of Fig.~\ref{fig:restspec} 
in the vicinity of the \ion{Si}{4} and \ion{C}{4} doublets. The dashed red curve is an example continuum
fit used to normalize the spectrum for measuring the strength of the IS component of these lines. 
Note that the continuum has been adjusted to remove the broad absorption due to the 
stellar wind component of \ion{C}{4}; a similar adjustment was made to remove the stellar component
of \ion{Si}{4}, though it is much weaker than the \ion{C}{4} feature. 
The weak emission line within the broad \ion{C}{4} wind absorption feature is nebular \ion{Si}{2}* $\lambda 1533$, 
one of several weak excited fine structure emission lines observed in the spectra of $z \simeq 2-3$ galaxies
(\citealt{ssp+03}). The weak absorption line near 1501 \AA\ is photospheric \ion{S}{5}. 
}
\label{fig:cont_fit_demo}
\end{figure}

The line profiles of the $R=800$ and $R=1330$ 
composite spectra are remarkably similar, for both \lya\ emission and the strong interstellar absorption features,
as shown in Fig.~\ref{fig:velcompare_1}. The spectra yield the same value of ${\rm \eta_{\lya}}$ and
${\rm \eta_{IS}}$ after accounting for the difference in spectral resolution. One is particularly
interested in the asymmetry of the line profiles and the value of the maximum blue-shifted velocity
${v_{max}}$, which we define as the velocity at which the blue wing of the IS absorption lines
meets the continuum. We find that $v_{max}$, although difficult to measure for typical spectra
of individual galaxies due to limited S/N, is not strongly dependent on spectral resolution;
Figure~\ref{fig:velcompare_1} shows that $|v_{max}| \simeq 700-800$ \kms
for both composite spectra. 
As discussed above, the \ion{C}{4} and (to a lesser extent) \ion{Si}{4} doublets in galaxy spectra
have contributions from both the IS lines and from the P-Cygni stellar wind lines from massive
stars, and thus the IS component must be separated from the stellar feature in the process of
fitting the local continuum. An example of a continuum fit near the \ion{Si}{4} and \ion{C}{4} 
features is shown in Fig.~\ref{fig:cont_fit_demo}. Fortunately, the P Cygni feature is generally
both broader and shallower than the IS components of these lines, so that while the continuum
uncertainties are larger than for unblended features, they do not prevent an accurate measurement
in relatively high S/N composite spectra.    

\begin{figure}[htb]
\centerline{\epsfxsize=8cm\epsffile{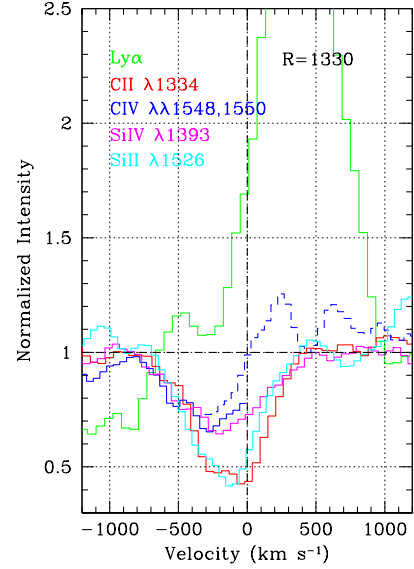}}
\figcaption[fig7.eps]{The velocity profiles of strong IS lines and \lya\ emission features
relative to systemic for the composite UV spectrum of
the $R=1330$ sample. The \ion{C}{4} profile is a solid blue histogram for the $\lambda 1548$ component
and dashed for $\lambda 1550$.
\label{fig:velplots}
}
\end{figure}

The velocity profiles of some of the strongest spectral features in the stacked 
spectrum of the galaxies observed with $R=1330$ are over-plotted in Fig.~\ref{fig:velplots}.
Evidently, $|v_{max|} \simeq 700-800$ \kms is a generic feature of the spectra
of these rapidly star-forming galaxies, in spite of the fact that the average {\it centroid}
of the IS line profiles is more modest, with $\dvis \sim -165$ \kms\ from 
the previous section. The apparent values of $|v_{max}|$ are relatively insensitive
to spectral resolution.   

\begin{figure}[htb]
\centerline{\epsfxsize=8cm\epsffile{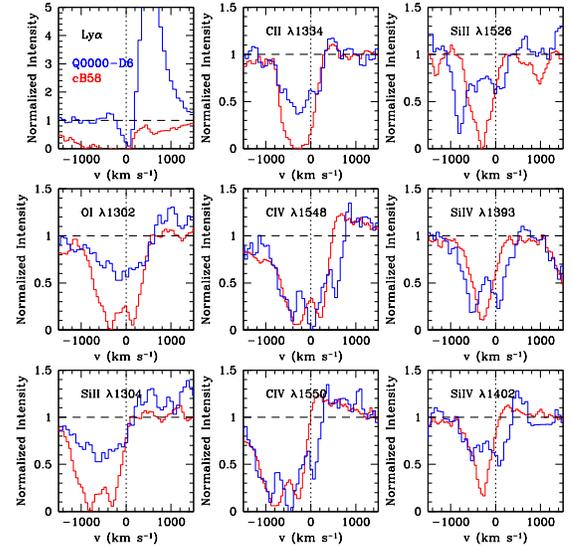}}
\figcaption[fig9.eps] {As for Fig.~\ref{fig:velcompare_1}, comparing the spectra of two individual objects having
particularly high quality UV spectra as well as accurate systemic redshifts from stellar absorption
features. The spectrum of MS1512-cB58 (the $R\simeq 1600$ spectrum from Pettini \et 2000) is in red,
while the spectrum of Q0000-D6, obtained with $R \simeq 1300$, is in blue. Despite significant
differences in spectral ``morphology'' (e.g., D6 has strong Ly $\alpha$ line emission, while cB58
has strong absorption, and the ratio of the strengths of high ionization lines to low-ionization
lines is quite different), the {\it range} of velocities spanned by the outflows are quite
similar in both cases, with the maximum velocities of $v \simeq 800$ \kms.
Note that the Si{\sc II}\ $\lambda 1526$ line of D6 is affected by absorption
from another system in the blue wing of the profile. Both of these galaxies would be
in the ${\rm M_{bar} < 3.7\times 10^{10}}$ \msun\ (i.e, lower baryonic mass) sub-sample.
\label{fig:velcompare_3}
}
\end{figure}

The line profiles in the spectra of individual galaxies can, of course, vary
considerably. Fig.~\ref{fig:velcompare_3} shows the spectra of two
individual galaxies to illustrate the point: one is MS1512-cB58, the $z=2.729$
lensed LBG whose spectrum has been analyzed in detail by \cite{pettini00,prs+02};
the other is Q0000-D6 \citep{ssp+03}, a bright LBG at $z=2.966$ observed
at a comparable spectral resolution of $\simeq 1500$. These spectra
show clear differences in the details of the profiles and with the apparent covering
fraction of the continuum, particularly for the low-ionization species which differ
in apparent optical depth by a factor of $\simeq 2$. Clearly, \lya\ emission is prominent
in D6, but very weak in the spectrum of cB58 (see also 
\citealt{quider09a}).
The spectrum of Q0000-D6 also
has an unusual high ionization component that produces clear
Lyman $\alpha$ absorption in the apparent blue wing of the \lya\ emission line, as
well as in the high ions (but is less prominent in
the lower ions) near $ v = 0$. 
Still, the velocity envelope for the blue-shifted material is remarkably consistent
with that seen in the composites presented above: maximum blue-shifted velocities
of $|v_{max}\simeq 800$ \kms, roughly independent of ionization level.

\subsection{Trends with Baryonic Mass}
\label{sec:barmass}
 
\begin{figure}[htb]
\centerline{\epsfxsize=8cm\epsffile{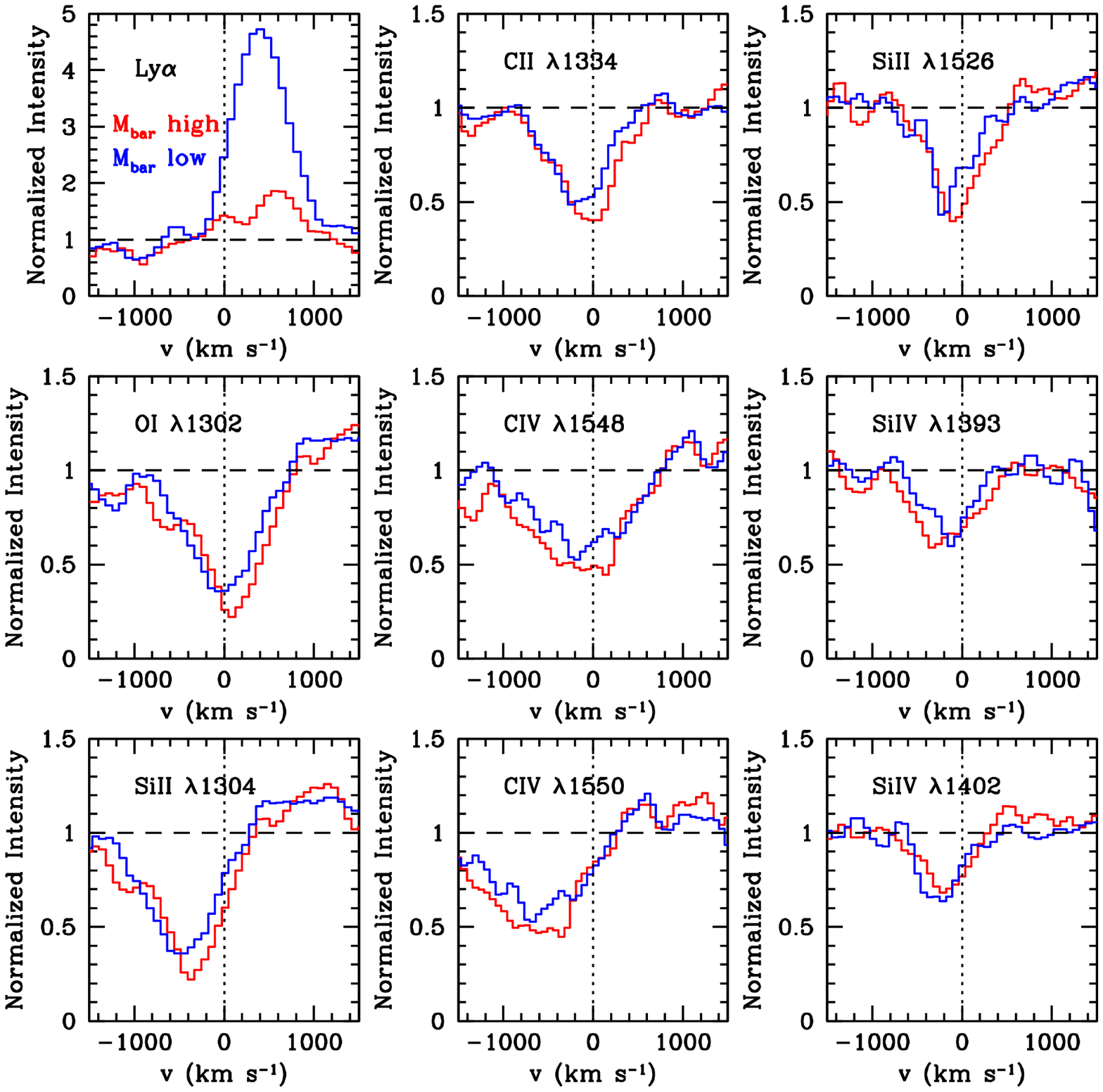}}
\figcaption[fig10.eps]{As for Fig.~\ref{fig:velcompare_1}, comparing the composites of subsets of the \Ha\ sample
depending on their inferred baryonic mass ${\rm M_{bar}}$.  The composite from objects with
${\rm M_{bar} > 3.7\times10^{10}}$
\msun\ is plotted in red, and that from objects with ${\rm M_{bar} < 3.7\times 10^{10}}$ \msun\ in blue.
Note the differences in the profiles of the low-ions (e.g., Si{\sc II}, O{\sc I}, C{\sc II}) for $v \simlt 0$,
where the high baryonic mass objects appear to have stronger low-ion absorption at velocities
closer to systemic. These differences
do not seem to be present for the higher ionization lines (e.g., see the C{\sc IV}  $\lambda 1550$ panel), nor
are there differences in the maximum blue-shifted velocities.
\label{fig:velcompare_2}
}
\end{figure}

Returning to the trends in the centroid velocities of the IS lines noted in the
previous section, it is instructive to examine the mean line profiles of 
composite spectra selected by implied baryonic mass ${ M_{bar}}$, the parameter
most significantly linked to the observed kinematics of the interstellar
absorption features. Figure~\ref{fig:velcompare_2} shows the comparison between
the halves of the \Ha\ sample that are above and below an inferred ${\rm M_{bar}} = 3.7\times
10^{10}$ \msun, the sample median.  There are clear differences in the \lya\ emission
line strength, which is similar to that seen for sub-samples of different metallicity
as in \cite{erb+06a}. The peak of the \lya\ emission line profile is shifted
by $\simeq +200$ \kms (from $\simeq +400$ to $\simeq +600$ \kms) for the higher mass 
sub-sample relative to that
of the lower-mass sub-sample.\footnote{No significant correlation was found between ${\rm M_{bar}}$
and $\dvla$ in \S\ref{sec:correlations}, but many of the galaxies in the
higher mass sub-sample had \lya\ emission that was too weak to measure, resulting in a very
small sample.}
The profiles of the low-ionization interstellar lines (C{\sc ii}\ $\lambda 1334$ is in the
cleanest spectral region and illustrates it best) may be indicating the root cause
of the kinematic trends discussed in \S\ref{sec:correlations}: the higher mass
sample exhibits stronger IS absorption at or near $ v \sim 0$,
while the profiles are nearly identical in their behavior near $v=-|v_{max}|$.
This ``excess'' low velocity material in the higher-mass sub-sample -- a shift of $\simeq 200$ \kms\ in the red
wing of the IS line profiles-- 
systematically shifts the centroid of the IS velocity distribution by $\simeq +100$ \kms relative
to the lower mass sub-sample, while the blue wing of the profile exhibits no clear trend
with ${\rm M_{bar}}$. The excess low-velocity material in the higher
mass sub-sample is not obvious in the
\ion{C}{4} absorption profile, for which the 2 profiles appear to be nearly identical  
for $v \simgt 0$ (note that \ion{C}{4} was not used to measure $\dvis$ 
for any of the galaxies in the sample, because of the dependence of the
rest-wavelength for the blend on the relative strength of the lines of the doublet). 

\begin{figure}[htb]
\centerline{\epsfxsize=8cm\epsffile{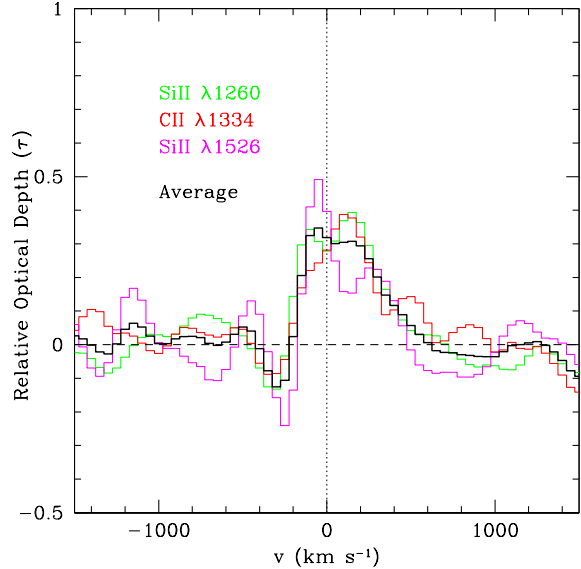}}
\figcaption[fig11.eps]{The relative optical depth $\tau_{\rm high}-\tau_{\rm low}$
for selected low-ionization absorption lines, versus
velocity relative to systemic, where these quantities refer to the composite spectra of the high-${\rm M_{bar}}$ and
low-${\rm M_{bar}}$ sub-samples. The histograms are for
Si{\sc II}\ $\lambda 1260$ (green), C{\sc II}\ $\lambda 1334$
(red), and Si{\sc II}\ $\lambda 1526$ (magenta) interstellar line profiles,  where the heavy black
curve is the average residual optical depth for the 3 transitions.
Note the absence of significant residuals at blue-shifted velocities (with the
possible exception of local ``peak'' near $v=-300$, which represents slightly deeper absorption
in the spectrum of the low-\mbar\ sub-sample).  The excess absorption in the \mbar-high
sub-sample has its centroid at  $v=+154$ \kms and its peak near $v \simeq 0$.
\label{fig:mbar_residuals}
}
\end{figure}

Fig~\ref{fig:mbar_residuals} shows the residual apparent optical depth $\Delta\tau(v)$ for 3 relatively
isolated low-ionization transitions  
for the high-- and low-- ${ M_{bar}}$ sub-samples.  By this we mean the additional  
optical depth as a function of velocity that when added to the line profiles of the
low-$M_{bar}$ sub-sample would produce line profiles identical to those of the high-$M_{bar}$ sub-sample,
i.e., 
\begin{equation}
I_{hm}(v) = I_{lm}(v) {\rm e}^{-\Delta\tau(v)} 
\label{eq:tau}
\end{equation}
where $I_{hm}$ and $I_{lm}$ are the spectral intensity of the high-$M_{bar}$ and low-$M_{bar}$ sub-samples,
respectively. 
The spectrum $\Delta\tau(v)$ has a peak 
at $v \simeq 0$, a centroid at $v = +154$ \kms, and a velocity width of $\sigma_v \simeq 120$  
\kms after correcting for the effective instrumental resolution. The excess apparent optical depth $\Delta\tau$ 
in these transitions accounts for 
$\simeq 25$\% of the equivalent width of the full low-ion profiles of the
high-$M_{bar}$  sub-sample. The additional component of absorption in the high-$M_{bar}$
sub-sample also changes the average interstellar line width from ${\rm \eta_{IS}= 540}$ \kms
for the lower mass sub-sample to ${\rm \eta_{IS} = 660}$ \kms for the higher mass sub-sample.   

We will return to a discussion of the possible implications of the baryonic mass dependence
of the IS line kinematics in \S\ref{sec:model_implications}. 

\section{ Lyman $\alpha$ Emission}
\label{sec:lya}
\subsection{Observed Trends}
We have shown in \S\ref{sec:bulkvel} above that the centroid velocity of \lya\ emission 
$\dvla$ exhibits larger scatter than \dvis\ relative to the systemic redshift
defined by \Ha\ emission, but that the relative consistency of $\dvla - \dvis$ suggests
a causal link between the two kinematic measures. In this section, we attempt to 
understand the nature of this relationship in greater detail. 

In \S\ref{sec:composites} above, we presented evidence that the kinematic profiles of
strong interstellar lines in the spectra of rapidly star forming galaxies exhibit
greater variation near $ v=0$ than at large blue-shifted velocities. Moreover,
the presence of significant low-ion absorption at $v \simgt 0$ strongly
affects both the apparent velocity of \lya\ emission and the centroid velocity
\dvis\ of the IS lines, though $v_{max}$, the maximum blue-shift, remains essentially 
unchanged. 


A commonly-adopted ``toy model'' (see e.g. \citealt{prs+02,assp03}) used to interpret
the kinematics of \lya\ emission and IS absorption lines in starburst galaxy spectra 
involves a roughly spherically-symmetric
outflow resulting in generally blue-shifted absorption as seen from earth. Because \lya\ photons resonantly
scatter, they escape from the nebula only when they acquire a velocity such that the optical
depth to scattering in material which lies (physically) between the observer and the emitted
\lya\ photon becomes small. Perhaps the easiest way for a \lya\ photon to reach earth
is to acquire the velocity of outflowing material on the far side of the galaxy, but
to be emitted in the observer's direction, so that the photon has been redshifted by
several hundred \kms\ relative to the bulk of the material through which it 
must pass to reach us. This picture would explain qualitatively why the dominant component
of \lya\ emission {\it always}
appears redshifted relative to the galaxy systemic velocity. Neglecting radiative transfer 
effects for the moment, if the range of velocities of outflowing gas is similar on the
``far'' side of the galaxy to what we observe (through the IS absorption lines) on the ``near'' side, 
then one would expect the maximum blue-shifted velocity $|v_{max}|$ to be comparable to 
the maximum velocity observed in the red wing of the \lya\ emission line.  One could then explain
different spectral morphology near the \lya\ line by altering the {\it distribution} of \lya\
optical depth as a function of velocity for the material between the observer and the far
side of the outflow.  

\begin{figure}[htb]
\centerline{\epsfxsize=8cm\epsffile{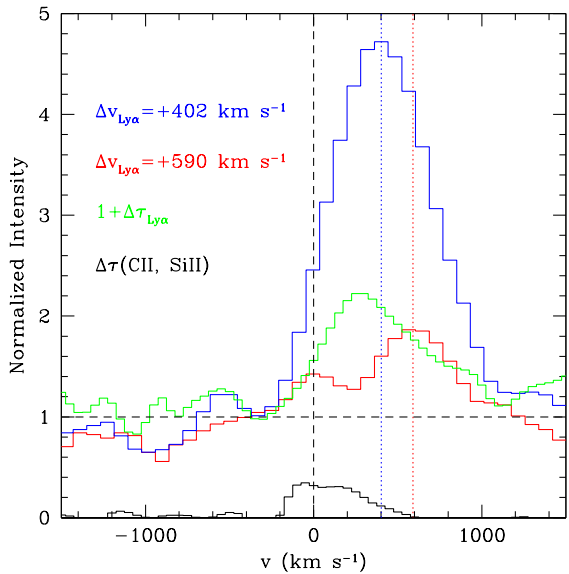}}
\figcaption[fig12.eps]{A comparison of the composite \lya\ emission profiles of the same samples as in Fig.~ \ref{fig:velcompare_2}.
Also shown (green histogram) is the inferred
function $1+\Delta\tau_{\lya}(v)$, where $\Delta\tau_{\lya}(v)$ is the excess apparent optical
depth in \lya\ (as a function of velocity) required to produce the weaker of the two
\lya\ profiles from the stronger. The black histogram reproduces the low-ion residuals
from Fig.~\ref{fig:mbar_residuals}.
\label{fig:lya_mbar}
 }
\end{figure}

For example, in the simplest possible scenario, one could imagine
that the difference in the \lya\ line profiles shown in Fig.\ref{fig:lya_mbar} 
(or any other set of \lya\ profiles) could be explained by altering
the distribution of $\tau_{\lya}(v)$ such that only the most-redshifted photons
have an appreciable chance of making it through the intervening \ion{H}{1}. It is then easy to
see that increasing $\tau_{\lya}(v)$ near $v=0$ will shift
the apparent peak and centroid of \lya\ emission toward the red, since \lya\ photons 
emitted from material with
line-of-sight velocities near $v=0$ will have no chance to reach the
observer-- only those with the most extreme redshifted velocities will find low enough
$\tau_{\lya}$ to make it through in our direction. Figure \ref{fig:lya_mbar} shows  
the \lya\ profiles for the two samples split by ${\rm M_{bar}}$ as in the previous section- 
the green histogram shows $1+\Delta\tau_{\lya}(v)$ with $\Delta\tau_{\lya}$ defined
so as to produce the red profile (high ${\rm M_{bar}}$) from the blue one (low ${\rm M_{bar}}$), i.e.,. 
\begin{equation}
I_{hm}(v) = I_{lm}(v) {\rm e}^{-\Delta\tau_{\lya}(v)} 
\label{eq:taulya}
\end{equation}
Here $\Delta\tau_{\lya}$ is the {\it excess} optical depth, over and above that already present
for the low-${\rm M_{bar}}$ sub-sample; the effective optical depth, including both scattering
and dust opacity, must be very significant even for the latter, since the equivalent width of the observed \lya\
emission line is $\sim 20-40$ times smaller than that expected for Case-B recombination and a normal
population of high mass stars (e.g., \citealt{charlot91}) . 

Fig.~\ref{fig:lya_mbar} also reproduces the excess optical depth $\Delta \tau$(\ion{C}{2},\ion{Si}{2}) 
in the low-ionization metallic
species from Fig.~\ref{fig:mbar_residuals} for comparison to $\Delta\tau$(\lya). While both
have centroids with $v > 0$, $\Delta\tau$(\lya) peaks at $v \sim +270$ \kms, close to the velocity
at which $\Delta \tau$(\ion{C}{2},\ion{Si}{2}) begins to decrease going toward higher positive
velocity. The differences are probably due to a combination of generally much higher optical
depths in the \lya\ transition than for the low-ion metals (\lya\ photons can only
escape from regions having the lowest $\tau$(\lya)), as well as 
geometric effects relating to the possible location of gas seen in absorption (which must
lie in front of the continuum source) versus emission (which may lie either in front of or behind
the continuum source).   

\subsection{Understanding \lya\ Emission Line Kinematics}

The real situation is undoubtedly far more complicated. There have been several recent theoretical
treatments of \lya\ radiative transfer in the context of galactic outflows
(\citealt{verhamme07,verhamme08,schaerer08,hansen06,dijkstra06a,dijkstra06b,zheng02}), and
even in the highly-idealized geometric configurations considered in these papers, a given line
profile does not uniquely specify the combination of velocity field, \ion{H}{1} column density, covering
fraction, and dust opacity that applies to a given observation. For example, both bulk velocity fields and
photon diffusion are capable of accounting for \lya\ photons that acquire large redshifts before escaping
from a model galaxy. The profiles of interstellar absorption lines, particularly in the highest-quality
spectra of individual objects, indicate that material with velocities ranging from $\sim +200$ to
$\sim -800$ \kms exists in most observed ions for most galaxies (as discussed above). Essentially
all of the ions observed in spectra of the quality presented here are strongly saturated, so 
that line profiles are best thought of as maps of covering fraction (hereinafter $f_c$) versus velocity.
The simultaneous presence of neutral and singly ionized species along with higher-ionization
species like \ion{C}{4}, with similar overall velocity envelopes, reinforces the idea that the ISM is a complex
multi-phase medium. To make matters worse, we do not know where, in physical space,  IS absorption at
a given observed velocity actually arises. Since \lya\ photons must traverse this medium, experiencing
typically thousands of scattering events, it
is therefore extremely difficult to predict in detail what the emergent \lya\ profile 
will look like.   

\cite{verhamme07} and \cite{verhamme08} consider a wide range of parameters, examining the effect on emergent
\lya\ line profiles, for models of central (monochromatic) point sources surrounded by an expanding shell of gas
with varying \ion{H}{1} column density and Doppler parameter ($b$).  A generic prediction, as discussed above, is that \lya\
emission (when present) will be very asymmetric, with the details of the line shape depending
on the assumed shell velocity, Doppler parameter, and \ion{H}{1} column density in the shell. For the expanding shell
models, these authors predict that the peak of the \lya\ emission line should appear near
velocity $v \simeq -2\times V_{exp}$, where $V_{exp}$ is the shell velocity, due to the combined
effects of radiative transfer and the bulk velocity of the scattering medium. The red wings of
the \lya\ line are expected to be produced by photons scattering multiple times from the receding
side of the expanding shell (as seen by an observer on earth); most of the migration of \lya\ photons
toward large redshifted velocities arises from absorption in the Lorentzian wings associated
with high values of N(\ion{H}{1}). Larger values of $V_{exp}$ and higher assumed values
of $N(HI)$ also accentuate the red wing of the \lya\ line and suppress regions close to the
systemic redshift, moving both the peak and the centroid of \lya\ emission to higher 
velocities.  A further prediction of the shell models is the presence of additional \lya\ peaks 
correspond to photons escaping in either the blue or the red wing of \lya\ associated
with the approaching side of the shell; photons emitted from the red wing of the \lya\ line
in the approaching side would then lie close to $v=0$, while the blue wing
would tend to form a peak with  $v\simlt -V_{exp}$. 

In a less-idealized situation, where the bulk velocity 
of the outflowing gas is not single-valued, but spans a more-or-less continuous velocity 
range of at least 800 \kms, the dominant mechanism for the migration of \lya\ photons 
in both frequency space and real space becomes simpler, in some respects\footnote{More
general implications of this type of model are discussed in \S\ref{sec:model}.}.
In a scenario with more gradual velocity gradients and a clumpy  medium spread
over a large range in galactocentric distance $r$ (rather than a ``shell'')
associated with an outflow, photons can achieve velocities off-resonance by scattering their way through
gas with a range of $v$; when they escape the nebula and are observed as red-shifted \lya\ photons
by the observer, they would generally exhibit Doppler shifts that directly reflect the velocity
with respect to systemic of the gas from which they were last scattered.  
Since the medium is clumpy, the column density N(\ion{H}{1}) for 
individual ``clumps'' becomes less relevant than their velocity distribution and covering fraction
, since most \lya\ photons will scatter off the ``surfaces'' of clumps, rarely encountering regions where
absorption or emission in the broad wings of a line is important (see \cite{neufeld90} for a discussion
of this type situation). 
Note that the clumpy geometry is qualitatively different from, e.g., the ``Hubble expansion''
model considered by \cite{verhamme07}, which has $v$ increasing smoothly with $r$. In this case,
the radiative transfer is still important because the scattering medium is continuous and has
associated with it a particular N(\ion{H}{1}) through which all \lya\ photons must pass on their
way to larger $r$. The picture we are advocating has photons scattering from the $\tau(\lya) \sim
1$ surfaces of discrete clumps; this causes most \lya\ photons to emerge from scattering
events near line center, and thus to acquire a Doppler shift characteristic of the velocity 
(with respect to systemic) of the most  
recent clump. 
In other words, bulk-motion induced velocity
shifts, rather than radiative transfer effects, may be most responsible for the kinematics of the
observed \lya\ emission line. 

In this context, the IS absorption lines provide a reasonable proxy for the velocity distribution of 
gas that will comprise the medium through which the \lya\ photons must scatter in order
to escape in the observer's direction.\footnote{Note that the velocity field 
information provided by the IS lines has not been exploited in any of the models mentioned above.}. 
If the flows are roughly
spherically symmetric, it should be possible to seek consistency between the kinematics 
of \lya\ emission (which would probe the kinematics of gas on the 
receding side of the flow) and that of the IS absorption, which samples the blue-shifted,
or approaching, side. The \lya\ photons will be most successful in escaping the galaxy when they acquire
velocity shifts well off resonance of whatever material lies between the last scattering surface and
the observer. This effect would tend to produce emission from both the redshifted and blue-shifted
gas. 

\begin{figure}[htb]
\centerline{\epsfxsize=8cm\epsffile{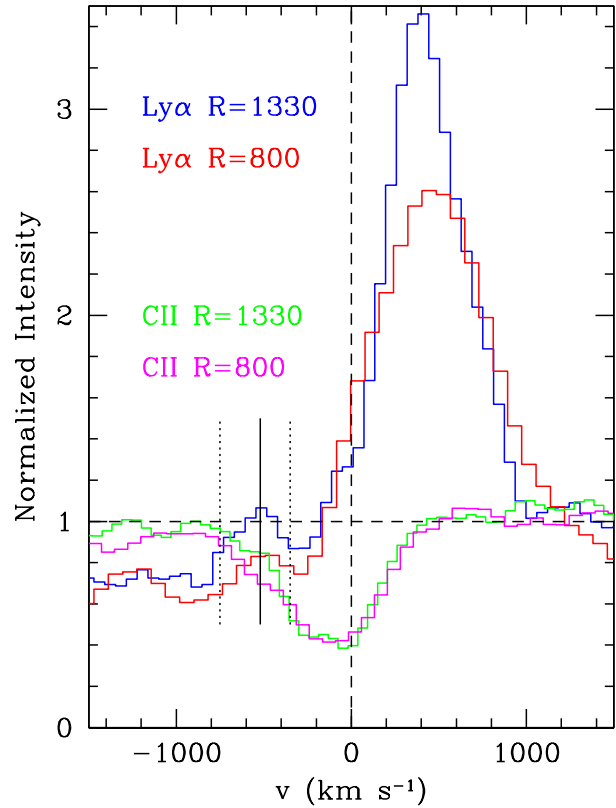}}
\figcaption[fig13.eps]{Comparison of two independent galaxy sample composites illustrating the presence
of a weak blue-shifted component of \lya\ centered at $v=-520$ \kms (indicated with a solid vertical line). The blue
and green histograms are the velocity profiles of \lya\ and C{\sc II} $\lambda 1334$ for the $R=1330$ sample,
while the red and magenta histograms are the same lines for the $R=800$ \Ha\ sample. The position
of the secondary \lya\ peak corresponds to the maximum blue-shifted wing of the IS line
profiles.
\label{fig:lya_bp}
}
\end{figure}

Figure~\ref{fig:lya_bp} compares the velocity profiles of \lya\ emission and \ion{C}{2} $\lambda 1334$
absorption for the same $R=800$ and $R=1330$ composite spectra shown in Fig.~\ref{fig:restspec}. 
First, it does appear that both spectra exhibit
a secondary blue peak in the \lya\ emission line, with centroid velocity
at $v\simeq-520$ \kms (although it is more evident in the spectrum
with higher spectral resolution), close to a ``mirror image'' (in velocity) of
the dominant redshifted emission.  While the strength of this secondary feature in the composites 
is only $\sim 5$\% of the primary redshifted component, it corresponds
closely to the range of velocities seen in the most blue-shifted portion of
the IS \ion{C}{2} profile, with $-350 \simgt v \simgt -750$ \kms. This 
range also corresponds to that over which the apparent optical depth of the IS absorption is
decreasing from its maximum, which extends in both composites over the velocity
range $-350 \simlt v \simlt 0$ \kms.    

\subsection{A Simple Kinematic Model for IS Absorption and \lya\ emission}
\label{sec:lya_model}

In an attempt to produce simultaneously the salient features of the observed IS lines 
{\it and} \lya\ emission lines, we have constructed a very schematic kinematic model
following the line of reasoning outlined above, 
for the purposes of illustration. 
In the models, we assume that optically thick
gas (for either low-ionization metallic species or \lya) 
is present in two kinematic components, each of which is Gaussian in optical depth
$\tau$ for a given transition: one component is centered near $v=0$ and the other is outflowing 
with a velocity distribution $\langle v \rangle = -450\pm150$ \kms. 
We choose 450 \kms\ for the centroid of the outflowing component somewhat arbitrarily, 
but (in addition to producing line profiles that resemble the real ones) it is approximately 
equal to the escape velocity $v_{esc}$ at the virial
radius of a dark matter
halo of total mass $\sim 9\times10^{11}$ M$_{\sun}$ at $z \simeq 2.5$, believed to be typical of the galaxies in our
spectroscopic sample (\citealt{asp+05,conroy08}; see \S\ref{sec:model_implications}). 
For the gas near the galaxy systemic redshift, we assume that the velocity
distribution of the optically thick gas is similar to that of the \ion{H}{2} regions traced by \Ha\ emission, 
i.e., $\sigma_v(\Ha) \sim 100$ \kms (\citealt{erb+06c}). All profiles have been convolved with an
instrumental resolution of $R=1330$.  

\begin{figure}[htb]
\centerline{\epsfxsize=9cm\epsffile{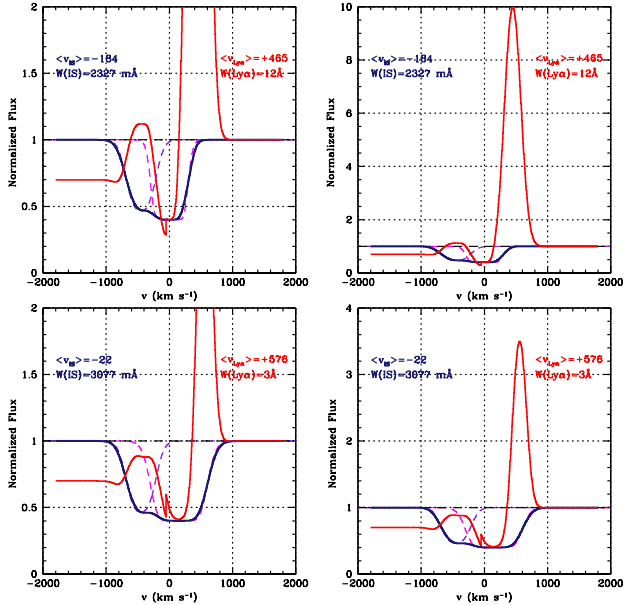}}
\figcaption[fig14.eps]{Examples of model spectra in which the velocity field of optically thick gas in the galaxy is assumed to be
comprised of two Gaussian (in optical depth $\tau$) components, both having a covering
fraction of 0.6 (dashed profiles in each panel). The top 2 panels
show the results for a model with one component outflowing with $v_1=-450\pm150$ \kms, and an additional
component with ${\rm v_2 = 0\pm120}$ \kms ; the
bottom panels show a second example with $v_2=+150\pm200$ \kms, with the velocity $v_1$ and optical depth of the outflowing
component held fixed. Note how the kinematics of the material near $v=0$ influences the measured
centroids of the net absorption lines (heavy blue curve). If the same gas kinematics are assumed for
\ion{H}{1}, the predicted \lya\ emission profiles as seen by an observer would resemble the red curves; the \lya\
centroid is redshifted by an additional $\simgt 100$ \kms in the bottom model as compared to the top.
.
\label{fig:emabs_models}
}
\end{figure}

Fig.~\ref{fig:emabs_models} shows example model line profiles for both IS absorption and \lya\ emission, as seen
by an observer on earth 
(note the righthand panel in each case is just a zoomed-out view). 
One needs to provide the normalization of the $\tau$ distribution (i.e., the maximum optical depth $\tau_0$)
for both components; in the
first example, the $v$=0 component has a central optical depth of $\tau \simeq 20$, and the 
outflowing component has a peak at $\tau \simeq 5$\footnote{The result is relatively insensitive
to these numbers, so long as the lines are saturated}. The dark blue solid curve shows the expected
absorption line profile (assuming that the covering fraction $f_c= 0.6$, for both components) for a saturated
low-ion transition. The combination of the two assumed velocity components produces an absorption
feature with a centroid $\langle \dvis \rangle = -184$ \kms and a rest equivalent width $W_0\simeq 2.2$\AA-- close to the typical 
values measured for the real galaxies. A second example, in the bottom panels of Fig.~\ref{fig:emabs_models}, has an identical
outflowing component, but the $v \simeq 0$ component has been shifted to +150 \kms (to approximately
mimic the ``excess'' optical depth observed in the high-$M_{bar}$ sample discussed above; see
Fig.~\ref{fig:mbar_residuals}), and
broadened to $\sigma_v = 200$ \kms.  The centroid of the resulting absorption feature 
is $\langle \dvis \rangle = -22$ \kms, in spite of the fact that the outflowing component is identical
by construction, with maximum blue-shifts of 
$v_{max} \simeq -800$ \kms.  

The predicted \lya\ emission from the same galaxies is slightly more complex; for the moment
we adopt a simple model in which the $\tau(|v|)$ distribution is assumed to be the same as
for the IS absorption lines. The probability of a scattering event in which a \lya\ photon
is emitted by one atom and then absorbed by another with the same velocity (alternatively, one can think of it
as the relative fraction of time a \lya\ photon spends resonantly trapped in gas with optical depth $\tau(v)$) is roughly 
$\epsilon \propto \tau^{2}(v)$, since $\tau(v) \propto n_{H}(v)$, where $n_{H}(v)$ is the density of neutral H
atoms with velocity between $v$ and $v + dv$.)  
The probability that a resonantly trapped \lya\ photon escapes without
being immediately scattered at or near the same velocity is $\propto {{\rm exp}{[-\tau(v)]}}$. 
Crudely, the probability that a \lya\ photon is scattered by an atom with velocity $v$ {\it and}
is able to escape in the direction of an observer without additional scattering, as a function of $v$, is the  
\lya\ emission profile $I(v)$,  
\begin{eqnarray}
I(v) \simeq C\tau^2(v)~{\rm exp}\left[-\tau(v)\right], 
\label{eq:taudist}
\end{eqnarray}
where $C$ is a scale factor that can be adjusted to roughly account for the destruction of \lya\ photons
due to dust, which we assume is proportional to the total path length traversed
by a photon as it works its way out of the nebula. Finally, we assume that the \lya\ photons reaching an observer who 
sees IS absorption profiles as in Fig.~\ref{fig:emabs_models} have been scattered toward the observer with
probability of ``escape'' according to eq.~\ref{eq:taudist}. To be clear, we do not concern ourselves
with how \lya\ photons made it from the central \ion{H}{2} regions to any part of the outflow-- instead we simply
assume that at any given moment, the relative probability for emission of a \lya\ photon is proportional to $\tau^{2}(v)$. 
We do not know the spatial distribution of the outflowing gas on the redshifted side of the flow as a function of
velocity, but we assume that it is spatially distinct (i.e., farther away from the observer) than the component of ISM near
the galaxy systemic redshift, and that the blue-shifted component is closer. Further, we assume that 
the outflow velocity $|v_{out}|$ is a monotonic function of galactocentric distance $r$ 
on both the red- and blue-shifted sides of the flow. We found
this assumption to be necessary to obtain kinematics and spectral morphology consistent 
with observations; it would also follow from
the assumption that the line depth at a given velocity is directly related to $f_c$, so that the highest velocity
material is farthest away from the continuum source; see also \S\ref{sec:model} and \cite{weiner09,martin09}.

The red curves in the panels of Fig.~\ref{fig:emabs_models} were produced in this way, and 
show the predicted \lya\ emission
profile for a galaxy whose absorption profile (\lya\ or low--ionization metal lines) 
is given by the blue curve, as seen by the observer. 
It is worth noting that (to first order, in the absence of dust) the \lya\ emission profile is not 
strongly dependent on gas covering fraction, since (as seen by an observer) the outflowing material is both the ``source'' and
the ``sink'' for \lya\ photons, both of which would be reduced by the same geometric factor.  
We note a number of interesting features of the predicted
profiles: first, the centroids of \lya\ emission and IS absorption nicely reproduce those typically observed 
in the galaxy sample; second, the extreme velocities of redshifted \lya\ emission and blue-shifted IS absorption
reach 800 \kms, as for the observed galaxies; third, the model in the top panel  
produces a blue-shifted component of \lya\ emission near $v\simeq -500$ \kms -- similar in both
relative intensity and velocity relative to the redshifted component as in the composite 
spectra (see Fig.~\ref{fig:lya_bp}). In the model, the secondary feature arises from photons scattered toward the
observer from gas in the approaching part of the outflow; it is generally weaker
than the redshifted component if there is any dust in the galaxy because photons coming
from the near-side of the outflow do not appear as redshifted relative to the scattering medium
because they are moving in the same direction. As a consequence,  
blue-shifted \lya\ photons have generally experienced a 
larger number of scatterings 
in order to escape in the direction of the observer, thereby increasing the probability of being absorbed by dust. 
Finally, the apparent peak of \lya\ emission is modulated
primarily by the component of gas near $v=0$-- \lya\ emission is more redshifted, and weaker,
when the velocity range spanned by the ISM in the galaxy is broader (see also \citealt{mas-hesse03}). 

In summary, the apparent velocity of \lya\ emission depends primarily on 
gas near the systemic redshift of the galaxy, while the extent of the red wing depends
on the maximum velocity of material with appreciable column density and/or
covering fraction $f_c$. The extent of the red wing of \lya\ appears to be rather consistent over sub-samples
examined.  The relative consistency of $\dvla-\dvis$ reflects this dependency:
as IS absorption lines shift toward more positive velocities, only
the \lya\ photons that scatter off very high velocity material on the far side
of the galaxy can penetrate the \ion{H}{1} gas in the foreground. 
In any case, it appears that the spectral morphology of \lya\ and IS lines in
rapidly star-forming galaxies may be more easily explained by the (observed)bulk velocities of outflowing 
IS gas, as compared to more subtle radiative transfer effects. 

We return to a more physically-motivated outflow model in \S~\ref{sec:model}. 

\section{Where is the Absorbing Gas?}
\label{sec:galgal}
\subsection{Mapping the Circum-Galactic Medium}

Lines of sight to star-forming galaxies can provide a great deal of information on the overall
kinematics, chemical abundances, and (in some cases) estimates of the mass flux of cool material
entrained in an outflow (e.g., \citealt{pettini00,prs+02,quider09a,quider09b}). However, such observations contain
little or no direct information about {\it where} the gas is located relative to the galaxy. A full
appreciation for the physics of the outflows, and their effects on both the ``host'' galaxies and
the local IGM, requires probing them along lines of sight from which the physical location of gas
as a function of ionization level and velocity can be disentangled. In practice, this is
very difficult.

In the absence of spatially-resolved spectroscopy, each galaxy spectrum provides only a
surface-brightness weighted absorption profile in each observed transition, integrated along the 
entire line of sight from the ``center'' of the galaxy. Although ions of differing excitation
appear to have similar velocity profiles when integrated along the line of sight, we do not have
any way of knowing (e.g.) whether the bulk of the C{\sc iv}  ions at a particular $v$ are
distributed in the same general regions as those giving rise to C~II absorption at the same velocity.
The information available from the galaxy spectra themselves is
necessarily crude given the relatively low S/N and spectral resolution of all but a handful
of high redshift galaxy spectra. 
The few high-quality spectra, of which MS1512-cB58 (\citealt{prs+02}) is perhaps the best example,
show that the velocity profiles are complex, and that most of the lines accessible in lower quality
spectra are strongly saturated. If so, then interpretation of crude measures of line strength such
as equivalent width (or apparent optical depth vs. $v$) is ambiguous.  
The absorption arises in gas which may not be uniform across the entire
continuum source. The absorbing gas may be close to the galaxy, which is typically
several kpc across, and some lines of sight through the (outflowing) material may be completely absorbed, while others
may be altogether free of absorption.  
As discussed by \citet{ssp+03}, the equivalent width of a strong line
is modulated primarily by the fraction of the continuum source covered by gas giving rise
to the particular transition (which controls the depth of the lines relative to the continuum) and 
the velocity range in the foreground gas.  

Since the absorbing gas is almost certainly clumpy (e.g., the range of ionization level
seen in absorption cannot coexist in the same gas, and yet they share a comparable overall
velocity ``envelope''), 
the apparent optical depth for saturated lines 
(as seen by an observer on Earth) arising from 
material at a given galactocentric
radius $r$ will depend on the characteristic physical scale of the ``clumps''  
$\sigma_c(r)$, and their number density $n_c(r)$. The observed line strength (in the present case,
the rest equivalent width $W_0$) will also depend on the distribution of line-of-sight
velocity-- large values of $W_0$ {\it require} a large line of sight velocity dispersion, the
width of which depends on the sampled distribution of covering fraction. 
Clearly, any model for the outflows
is under-constrained by the spectrum of the galaxy itself, even for high quality
spectra (e.g., \citealt{prs+02,quider09a}). 

However, high-resolution spectra of very closely-spaced sightlines toward gravitationally-lensed QSOs
have provided constraints on the
characteristic physical scales over which column densities of various transitions such as \lya, C{\sc IV}, 
and lower-ionization
species vary significantly (\citealt{rauch99,rauch01,ellison04}). 
These scales appear to be $\simeq0.5-1$ kpc for \ion{C}{4} (thus, comparable to the ``beam footprint''
of the galaxy) 
and $\simeq 50-100$ pc for
low ionization metallic species (much smaller than the typical beam footprint). 
If it is assumed that the QSO spectra are sampling metal-enriched
gas similar to that found near galaxies like the ones in the current sample, these scales,    
together with measures of covering fraction,  provide some useful information on the structure of
the gas in the flows. Information on the distribution of gas transverse to the line of sight
breaks some of the remaining degeneracies hindering a full phase-space (velocity, position) 
understanding of the circumgalactic gas.  

\subsection{Galaxy-Galaxy Pair Samples}
\label{sec:pairs}

One method for probing the geometry of outflows is to observe foreground galaxies found near the lines of sight to
background QSOs, and to compare the kinematics of absorbing gas in the galaxy spectrum with that seen in
the QSO spectrum. A picture of the structure of the gaseous envelope surrounding galaxies can then be built
up statistically, by observing a sample of galaxies with a range of projected galactocentric distances from
the background QSOs (e.g., \citealt{ass+05,assp03}). This method has the advantage that one can choose the
background QSOs to be very bright, so that very high S/N and spectral resolution is possible, but the
disadvantage is that such QSOs are rare, and one is ``stuck'' with whatever galaxies happen to lie within
$\simeq 1$\arcm\ of the QSO. Of course, the smallest separations (e.g., $\theta \le 15$\arcs, or
a projected physical galactocentric radius of $\simlt 120$ kpc ) will be
especially rare. Ultimately, one is also limited by the quality of the (faint) galaxy spectrum 
if one is interested in a kinematic comparison using two different lines of sight through the galaxy.   

An alternative and complementary method is to use close angular pairs of {\it galaxies} at discrepant redshifts
to vastly increase the sample sizes, at 
the expense of information quality (because both the foreground and the background objects are
faint).  An initial attempt to use this method was presented by \citet{ass+05}; here we investigate it
using a significantly larger, higher quality, sample of galaxy pairs drawn from an ongoing 
densely-sampled spectroscopic survey 
(see \citealt{ssp+04} for early results; the full survey will be described 
in detail elsewhere).  

The survey has been conducted over a total of 18 independent
fields, 15 of which are centered on bright QSOs with $z_Q \simeq 2.6-2.8$ (high resolution spectra
of the bright QSOs are used in \S\ref{sec:hires_lya}). The spectroscopic sample 
includes $\simgt 2500$ galaxy spectra in the redshift range $1.5 \le z \le 3.6$, with $\langle z \rangle = 2.28\pm 0.42$. 
We used this galaxy survey catalog to identify pairs of spectroscopically-confirmed galaxies in each of 3 bins
in angular separation: $\theta < 5$\arcs, 5\arcs\ $< \theta \le$10\arcs, and 10\arcs\ $< \theta 
\le 15$\arcs, which will be identified as samples P1, P2, and P3, respectively 
(see Table~\ref{table:sample_table}). At the mean redshift of the foreground galaxies of $\langle z \rangle = 2.2$, 
the angular bins correspond to projected physical distances of $b \le 41$ kpc, $41 < b \le 83$ kpc, and $83 < b \le 124$ kpc
for samples P1, P2, and P3, respectively (see Table~\ref{table:sample_table}).  
The conversion from angular separation to physical impact parameter 
changes by only $\pm 2$\% over the full redshift range of the foreground
galaxy samples, so that we will use angular separation and impact parameter interchangeably.

\begin{deluxetable*}{l c c c c c c c}
\tablewidth{0pt}
\tabletypesize{\footnotesize}
\tablecaption{Foreground/Background Galaxy Pair Statistics\tablenotemark{a}}
\tablehead{
\colhead{Sample} &
\colhead{$\theta$ Range\tablenotemark{b}} &
\colhead{$b$ Range\tablenotemark{c}} &
\colhead{Number} &
\colhead{$\langle \theta \rangle$} &
\colhead{$\langle  b \rangle$}  &
\colhead{$\langle {\rm z_{fg}} \rangle $\tablenotemark{d}} &
\colhead{$\langle {\rm z_{bg}} \rangle $\tablenotemark{e}}
}
\startdata
P1 & $<5$\arcs\   &  $10-41$ kpc  & ~42 & $3\secpoint8 \pm 0\secpoint8$ & $31\pm~7$ kpc & $2.21\pm0.32$ & $2.64\pm0.36$  \\
P2 & 5\arcs $-$10\arcs\ & $41-82$ kpc & 164 & $7\secpoint6\pm1\secpoint5$ & $63\pm12$ kpc & $2.18\pm0.36$ & $2.64\pm0.41$ \\
P3 & 10\arcs $-$15\arcs\ & $82-125$ kpc & 306 & $12\secpoint5\pm1\secpoint5$ & $103\pm12$ kpc & $2.14\pm0.33$ & $2.65\pm0.39$ \\
\enddata
\tablenotetext{a}{Galaxy pairs in the spectroscopic sample for which both the foreground and
background galaxies have unambiguously measured redshifts which differ by $>3000$ \kms.}
\tablenotetext{b}{Range of angular separation of the galaxies, in arc seconds.}
\tablenotetext{c}{Range of impact parameter, in physical kpc.}
\tablenotetext{d}{Mean redshift for the foreground galaxies in pairs with the specified range of $\theta$.}
\tablenotetext{e}{Mean redshift for the background galaxies in pairs with the specified range of $\theta$.}
\label{table:sample_table}
\end{deluxetable*}

Within each pair sample, a number of additional criteria
were imposed: 1) both galaxy spectra within the pair must have accurately-determined redshifts; the redshift-space separation
of the two galaxies in each pair must be $> 3000$ \kms to ensure that their geometry is unambiguous (i.e., so that
we know which galaxy is ``behind'' the other); 2) the redshift separation $z_{bg} - z_{fg} \le 1.0$ to ensure that the
spectra contain a significant rest wavelength interval in common after both are shifted to the rest frame of the
foreground object. Redshifts were assigned to each galaxy using their \Ha\ redshift if available, and
otherwise using equations 2 or 4, depending on spectral morphology. As discussed in 
\S\ref{sec:sample_stats}, the uncertainty in the redshifts for the bulk of the galaxies in the
samples are expected to be $\simeq 125$ \kms. 
After culling, the final pair samples include 42, 164, and 306 pairs for samples
P1, P2, and P3, respectively. 
The properties of these three angular pair samples are summarized in
Table~\ref{table:sample_table}. 
 
\subsection{A Case Study: GWS-BX201 and GWS-BM115}

\begin{figure}[htb]
\centerline{\epsfxsize=7cm\epsffile{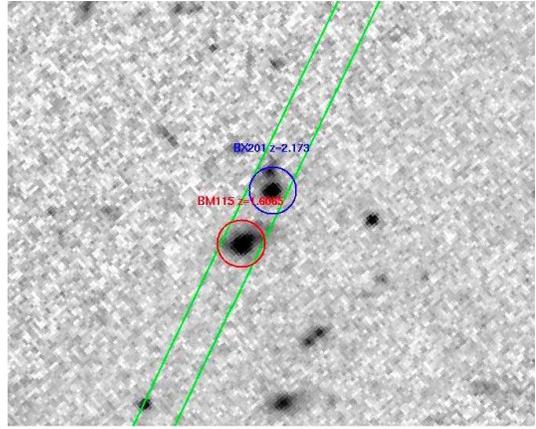}}
\figcaption[fig15.eps]{HST/ACS image of the GWS-BM115/BX201 galaxy pair, in the F814W filter. BM115 has
$z_{\rm sys} = 1.6065$, based on the H$\alpha$ line observed with Keck/NIRSPEC. BX201 lies
1\secpoint9 away on the plane of the sky, corresponding to an impact parameter at $z=1.6065$
of $\simeq 16.1$ kpc (physical). Fig.~\ref{fig:vel_gwspair} and Table~\ref{table:gws_table}
compare the interstellar lines observed in the spectrum of the foreground galaxy BM115 and
those observed in the spectrum of BX201 at the redshift of BM115.
\label{fig:pair_img}
}
\end{figure}

\begin{figure}[htb]
\centerline{\epsfxsize=9cm\epsffile{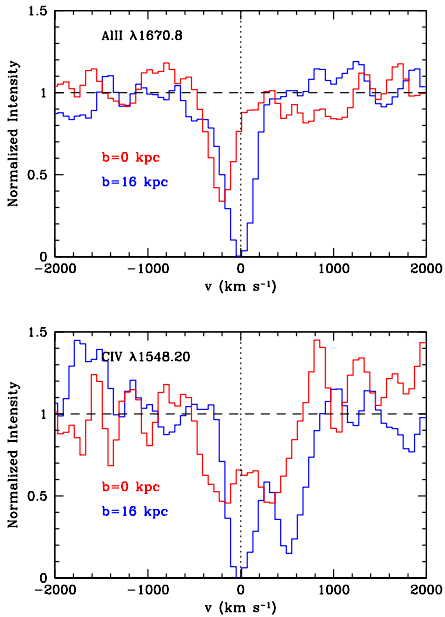}}
\figcaption[fig16.eps]{Comparison of the kinematics of representative interstellar absorption lines in the spectrum of the
$z=1.6065$ galaxy BM115 (red) to the absorption in the spectrum of the background galaxy
BX201 ($z=2.173$, 1\secpoint9 on the plane of the sky, or a projected transverse separation
of $\simeq 16$ kpc at $z=1.607$), shifted into the rest frame of
BM115 using the same systemic redshift (blue). The rest wavelengths of the transitions shown are indicated (the
\ion{C}{4} profiles are plotted with respect to $v=0$ for the $\lambda 1548.20$ line).  Note
that in both cases the absorption in the background galaxy spectrum is {\it stronger} and shifted by $\simeq +190$ \kms compared to the features in the foreground galaxy spectrum, placing the centroid
very close to the foreground galaxy systemic redshift.
\label{fig:vel_gwspair}
}
\end{figure}

Among these pair samples, there is a handful for which the spectra were of sufficient quality 
to be analyzed individually. One example, illustrated in Figs.~\ref{fig:pair_img} and
Fig.~\ref{fig:vel_gwspair} (see also Fig.~4 of 
\citealt{ass+05}), allows us to compare the interstellar absorption lines in the spectrum of 
the $z=1.6065$ galaxy GWS-BM115 with the same features observed in the spectrum of a background
galaxy projected only 1\secpoint9 away on the plane of the sky, GWS-BX201 ($z=2.173$). In this case,
the systemic redshift of GWS-BM115 is known from \Ha\ spectroscopy \citep{erb+06c}, and 
the optical spectra of both
objects were obtained with spectral resolution $R\simeq 1500$ (i.e., $\simeq 2$ times higher than
most of the other spectra in the sample). These galaxies have apparent magnitudes $g'=23.5$ (BX201) and
$g'=23.7$ (BM115), $\simeq 1$ magnitude brighter than typical galaxies in our sample. 
Using the systemic redshift of the foreground galaxy (GWS-BM115) measured from its \Ha\ line,
both spectra were shifted into the foreground galaxy's rest-frame, and continuum normalized,
to produce the comparison in Fig.~\ref{fig:vel_gwspair} 
which shows the \ion{Al}{2} $\lambda 1670$ and \ion{C}{4} $\lambda\lambda 1548$,1550 transitions\footnote{Because
of the low redshift of BM115, there are fewer lines in common for these two galaxies than for most of the
pairs in the 3 samples; the transitions shown are representative of low-ionization and high-ionization
lines, respectively.}. A summary
of relevant measurements is given in 
Table~\ref{table:gws_table}. 

\begin{deluxetable*}{c  c c c c c }
\tablewidth{0pt}
\tabletypesize{\footnotesize}
\tablecaption{Absorption Lines at ${\rm z_{fg}}$ in GWS-BX201/BM115 Pair\tablenotemark{a}}
\tablehead{
\colhead{Spectrum} &
\colhead{$\langle  b \rangle$} &
\colhead{$W_0$(\ion{C}{4}~1549)\tablenotemark{b}}  &
\colhead{$W_0$(\ion{Al}{2}~1670)} &
\colhead{\dvis\ \tablenotemark{c}} &
\colhead{FWHM\tablenotemark{d}} \\
\colhead{} & \colhead{(kpc)} & \colhead{(\AA )} & \colhead{(\AA )} &
\colhead{\kms} & \colhead{\kms}
}
\startdata
GWS-BM115 ($z=1.6065$)  &  ~0.0 &  2.78 & 1.11 & $-192\pm 15$ & 400  \\
GWS-BX201 (@$z=1.6065$) &  16.1 &  3.53 & 2.20 & ~~$0\pm 18$  & 270 \\
\enddata
\begin{indent}
\tablenotetext{a}{For this galaxy pair, the spectra were both shifted to the rest frame of the foreground
galaxy (BM115) and then evaluated with respect to $z=1.6065$.}
\tablenotetext{b}{Values include both components of the C{\sc iv}  doublet; the doublet ratios are 1.20 and
1.05, respectively.}
\tablenotetext{c}{Velocity of IS line centroids relative to $z=1.6065$.}
\tablenotetext{d}{Velocity widths of IS lines, corrected for instrumental resolution}
\end{indent}
\label{table:gws_table}
\end{deluxetable*}

There are several points worth making for this particular case: first,
the lines seen in the spectrum of the foreground galaxy have blue-shifted centroids, with
$\dvis \simeq -192$ \kms, which we have shown is typical of the objects in the \Ha\ sample.
Although the \ion{C}{4} lines are very broad (FWHM$\simeq 400$ \kms\ after correcting for
the instrumental resolution), one can still discern a separation of
the two components of the doublet. The line of sight to the background galaxy (GWS-BX201) passes
the foreground galaxy at a projected physical distance (hereinafter ``impact parameter'', $b$) of
$b=16.1$ kpc. Along this line of sight, the centroids of \ion{Al}{2} and \ion{C}{4} are consistent 
with $\dvis = 0$ 
relative to the foreground galaxy GWS-BM115-- fitting the two lines of the \ion{C}{4} doublet
and the \ion{Al}{2} feature yields a velocity offset $\dvis=0\pm 15$ \kms and velocity width
FWHM$\simeq 300$ \kms after accounting for the instrumental resolution. For both transitions, 
the ``offset'' line of sight produces
lines which are significantly {\it stronger} (in terms of equivalent width) than 
those seen directly toward  BM115 itself (see Table ~\ref{table:gws_table})! 
Of course, the path length through the galaxy is different in the two cases;
one line of sight passes through only the ``front half'' of whatever
gas is associated with the foreground galaxy, while the sightline to the background object, although spatially
offset, samples material both ``behind'' and ``in front of'' BM115 as seen by an observer on earth. 
The fact that the centroid of 
the absorption has $\dvis \simeq 0$ is consistent with any picture in which material is distributed
reasonably symmetrically around BM115; e.g., this would be expected for axisymmetric radial outflow
from (or infall onto) BM115, reaching a galactocentric distance of at least $ r_{g} \ge 16.1$ kpc.  

As mentioned above, strong lines such as \ion{Al}{2} $\lambda 1670$ and 
\ion{C}{4} are very likely to be saturated; they clearly exhibit saturation along the $b=16$ kpc line of sight, but 
the lines in the foreground galaxy spectrum itself do not reach zero intensity. 
\ion{Al}{2} $\lambda 1670$ and both components
of the \ion{C}{4} doublet
have approximately the same residual intensity of 
$\simeq 0.5$ (Fig.~\ref{fig:vel_gwspair}). 
The inference of $f_c < 1$ for cool outflowing material is not unusual, and in fact
a covering fraction of $\sim 50$\% is quite typical of $z \sim 2-3$
galaxies (see, e.g., \citealt{ssp+03,quider09a}). 

Based on the HST/ACS image (I band) shown in Figure~\ref{fig:pair_img}, 
although the 2 galaxies have similar apparent magnitudes they have quite different surface
brightness distributions: 
BX201 has FWHM $\simeq 0.22$\arcs\ ($\simeq 1.8$ kpc, physical), 
while the UV light of  BM115 has FWHM$\simeq  0.40$\arcs\ ($\sim 3.4$ kpc, physical). 
Thus, the spectrum of BM115 itself indicates that, averaged over an area of $\simeq 10$ kpc$^{2}$
centered at $b=0$ 
(assuming that the intensity-weighted diameter of the light distribution is
given approximately by the FWHM), 
but through only ``half'' of the outflow, only $\simeq$ 50\% of
the ``beam'' area intercepts appreciable outflowing low and intermediate ionization gas. 
The ``offset'' line of sight ($b=16.1$ kpc) produces absorption reaching zero intensity in both relatively
low (\ion{Al}{2}) and relatively high (\ion{C}{4}) ionization gas, indicating $f_c \simeq 1$ when 
averaged on a scale of $\simeq 2.5$ kpc$^2$, at a galactocentric radius of $\simeq 16.1$ kpc. 

It is difficult to know from observations of a single galaxy pair whether differences in the kinematics
and/or depth of absorption lines arise from averaging over a smaller spatial region (e.g., a more compact
background source might subtend a region within which the gas is optically thick everywhere)
or from a changing gas-phase covering fraction with impact parameter $b$ (such that the 
coherence scale of absorbing gas may be physically larger or smaller). This degeneracy
is one of the disadvantages of resolved background sources. 
However, as we now demonstrate, it is possible to use a large {\it ensemble} of pairs to construct a physical 
picture of the size and structure
of an average galaxy's circumgalactic gas distribution, again using stacked composite spectra. 

\subsection{Composite Spectra of Galaxy-Galaxy Pairs}
\label{sec:composites_galgal}

\begin{figure*}[htb]
\centerline{\epsfxsize=13cm\epsffile{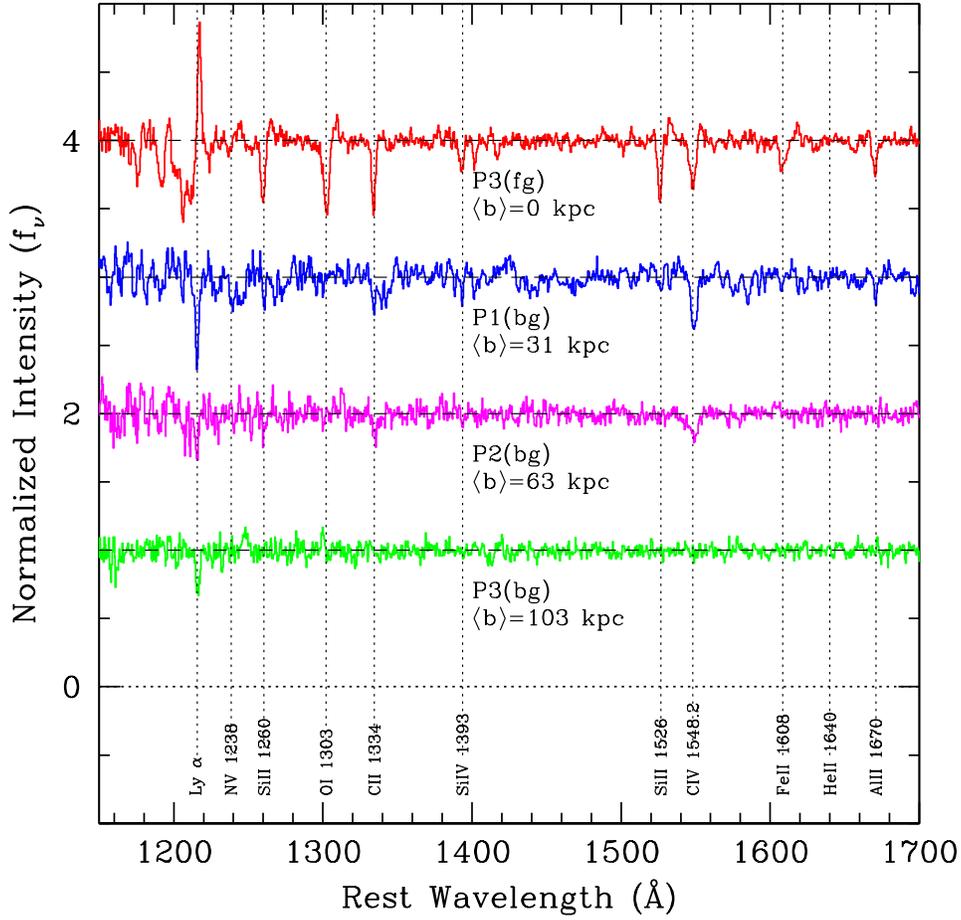}}
\figcaption[fig17.eps]{Comparison of the absorption lines in composite spectra formed from galaxy-galaxy pairs.
The top spectrum is the composite of all foreground galaxies in pairs from sample P3 with redshift
differences $>3000$ \kms.
The second spectrum is the composite of all {\it background} galaxies
in pairs with angular separation $\Delta \theta \le 5$\arcs\ [sample P1(bg)], formed by shifting
each to the redshift of the foreground galaxy and averaging, as described
in the text. The third and fourth spectra were formed the same way, for the background galaxies in pairs with
5-10\arcs\ and 10-15\arcs\ angular separations [samples P2(bg) and P3(bg)], respectively.
\label{fig:plot_galgal}
}
\end{figure*}

Unfortunately, most of the galaxy pairs in the current spectroscopic sample are considerably fainter than
the GWS-BX201/BM115 pair, and were observed with somewhat lower spectral resolution. Nevertheless, the large
size of the pair samples, coupled with the improved accuracy of systemic 
redshifts estimated from far-UV spectral 
features (most of the galaxies in pairs do not have \Ha\ measurements), allow for interesting 
measurements from composite spectra. For each angular pair of galaxies, the spectrum of the
foreground galaxy in each pair was shifted to its own 
rest frame using the measured values of $\dvis$ and the rules given in eqs. 2 or 4, and continuum-normalized. 
The resulting foreground spectra were then averaged within each pair sample, 
producing stacked spectra which we refer to as  
P1(fg), P2(fg), and P3(fg), respectively. A similar approach was used
to produce stacked spectra of the background galaxies:  each individual background galaxy 
spectrum was shifted into the
{\it foreground} galaxy's rest frame using the same 
systemic redshift applied to the foreground galaxy spectrum, 
and continuum normalized. The strong IS absorption features at the redshift of each background galaxy
constitute a source of noise that can affect the composite spectrum of the background
galaxies (after shifting the spectra to $z_{fg}$, the lines will appear at different wavelengths for
each spectrum). To minimize this effect, the strongest interstellar lines in each background
galaxy spectrum (at the galaxy redshift, $z_{\rm bg}$) were masked when producing the average spectrum at 
$z_{fg}$. The resulting composites P1(bg), P2(bg), and P3(bg) are shown in Figure~\ref{fig:plot_galgal},
together with (for display purposes only; in the analysis below we use a distinct stack of
the foreground spectra within each pair sample) the spectrum of the average of all 
foreground galaxy spectra in sample P3.

\begin{figure}[htb]
\centerline{\epsfxsize=8cm\epsffile{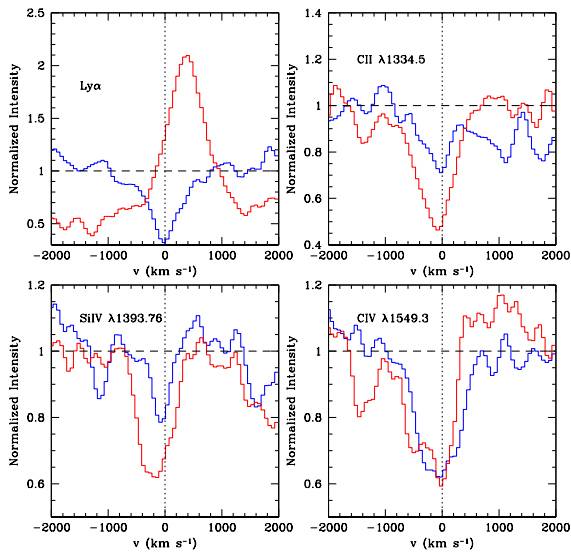}}
\figcaption[fig18.eps]{A comparison of the absorption profiles in the composite spectra of sample P1(fg) [red] and
P1(bg) [blue]. For each of the 42 pairs, the same foreground galaxy systemic redshift was used to shift
both the foreground and background galaxy spectra to ${\rm z_{fg}}$. Thus, the red curve represents
the average galaxy absorption line spectrum (i.e., ${\rm b=0}$)
while the blue curve is the average spectrum of the same
ensemble of galaxies at mean impact parameter of ${\rm \langle b \rangle =31}$ kpc (see
Table~\ref{table:sample_table} and Table~\ref{table:abs_table} for sample descriptions and statistics.
Note that the rest wavelength of the \ion{C}{4} doublet blend has assumed $W_0(1548)/W_0(1550) =1.4$.)
\label{fig:vel_plt5}
}
\end{figure}

\begin{figure}[htb]
\centerline{\epsfxsize=8cm\epsffile{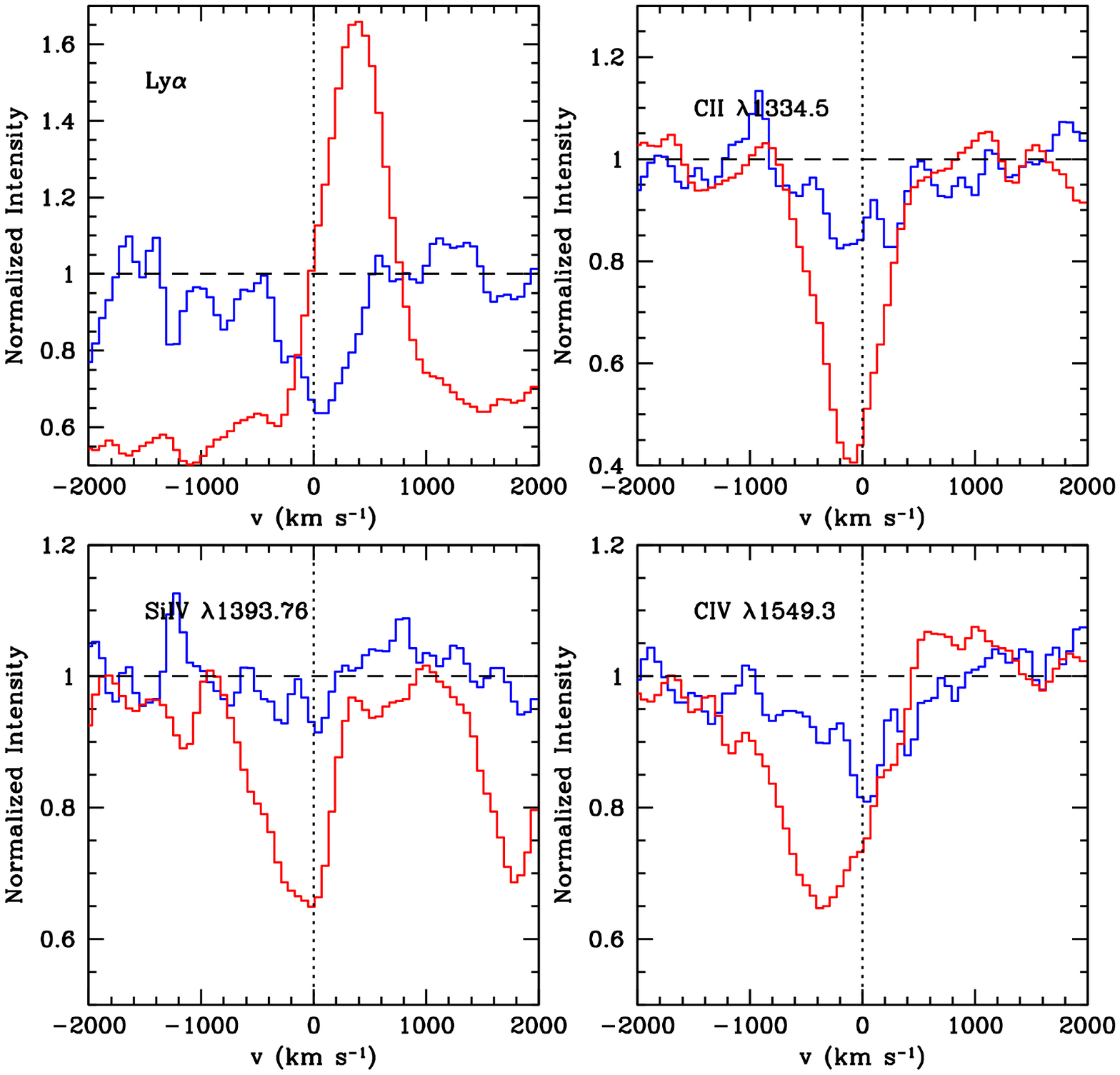}}
\figcaption[fig19.eps]{Same as for Figure~\ref{fig:vel_plt5}, for sample P2(fg) (red) and P2(bg) ([blue). In this case,
the blue spectrum represents the average absorption profile at impact parameters ${\rm b=41-82}$ kpc,
for the same 162 galaxies which comprise the mean galaxy spectrum in red. (see
Table~\ref{table:sample_table} and table~\ref{table:abs_table} for sample descriptions and statistics.)
\label{fig:vel_p5to10}
}
\end{figure}

\begin{figure}[htb]
\centerline{\epsfxsize=8cm\epsffile{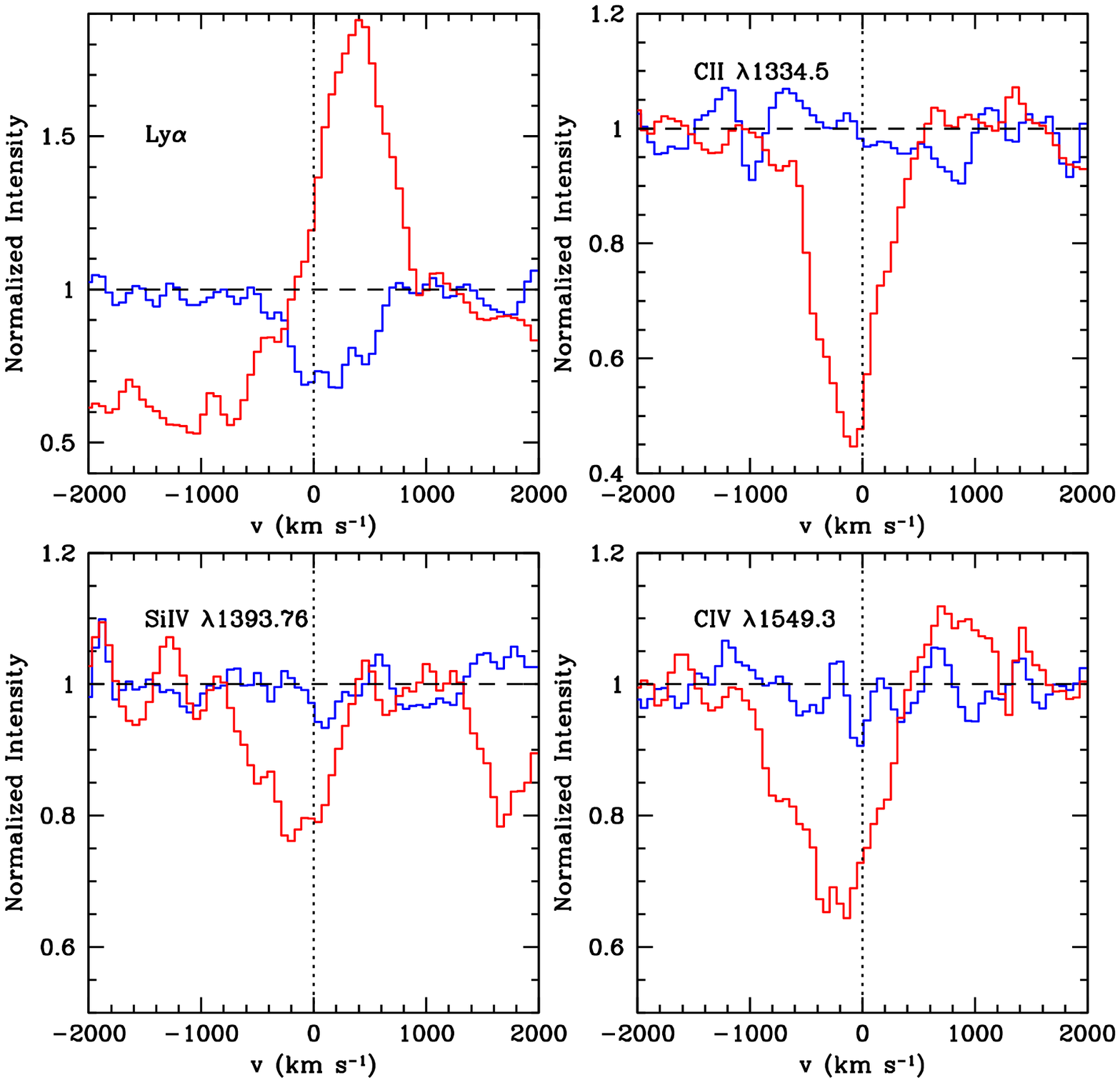}}
\figcaption[fig20.eps]{Same as Figure~\ref{fig:vel_plt5}, for sample P3(fg) (red) and P3(bg) (blue). In this case,
the blue spectrum represents the average absorption profile at impact parameters ${\rm b=82-125}$ kpc,
for the same 306 galaxies which comprise the mean galaxy spectrum in red. (see
Table~\ref{table:sample_table} and table~\ref{table:abs_table} for sample descriptions and statistics.)
\label{fig:vel_p10to15}
}
\end{figure}

Figures ~\ref{fig:vel_plt5}, ~\ref{fig:vel_p5to10}, and ~\ref{fig:vel_p10to15} show line profile comparisons of 
P1(bg) and P1(fg), P2(bg) and P2(fg), and P3(bg) and P3(fg), respectively, for selected transitions. 
Table~\ref{table:abs_table} summarizes the measured rest-frame line equivalent widths for
the same stacked spectra. The quoted errors on the equivalent widths in Table~\ref{table:abs_table}
account for sample variance, continuum uncertainties, and measurement errors
based on repeated measurements with spectra formed from subsets of the data.  

\begin{deluxetable*}{c c c c c c c c c}
\tablewidth{0pt}
\tabletypesize{\footnotesize}
\tablecaption{Absorption Line Strengths at ${\rm z_{fg}}$ in Galaxy Pairs\tablenotemark{a}}
\tablehead{
\colhead{Sample} &
\colhead{$\langle  b \rangle$} &
\colhead{Ly$\alpha$}  &
\colhead{\ion{Si}{2}~1260}  &
\colhead{\ion{C}{2}~1334}  &
\colhead{\ion{Si}{4}~1393}  &
\colhead{\ion{Si}{2}~1526}  &
\colhead{\ion{C}{4}~1549\tablenotemark{b}}  &
\colhead{\ion{Al}{2}~1670}
}
\startdata
P1(fg)  &  ~~0   & \nodata & $1.38\pm0.12$ & $1.79\pm0.15$  &  $1.20\pm 0.12$  & $1.34\pm0.12$ & $1.95\pm0.15$  & $0.97\pm0.10$ \\
        & ~~0\tablenotemark{c} & ($4.7\pm1.0$)\tablenotemark{d}    & ($2.01\pm0.18$) & ($2.61\pm0.22$) & ($2.04\pm0.20$) & ($1.96\pm0.18$) & ($3.90\pm0.30$) & ($1.42\pm0.15$) \\
P1(bg)  &  ~31   & $2.01\pm 0.15$ & $0.42\pm 0.06$  & $0.90\pm0.08$  &  $0.39\pm0.08$  & $0.37\pm0.06$ & $2.13\pm0.15$  & $0.40\pm 0.08$ \\
        &          &  & &       &        &       &   \\
P2(fg)  &  ~~0   & \nodata  & $1.42\pm0.10$ & $1.74\pm0.15$  &  $1.12\pm0.12$  & $1.52\pm0.12$ & $1.90\pm 0.10$  & $1.10\pm 0.12$ \\
        &  ~~0\tablenotemark{c} & ($4.8\pm1.0$)\tablenotemark{d}  & ($2.07\pm0.15$) & ($2.54\pm0.22$) & ($1.90\pm0.20$) & ($2.22 \pm 0.18$) & ($3.80\pm0.20$)  & ($1.61\pm0.18$) \\
P2(bg)  &  ~63   &  $1.23\pm0.20$   & $0.41\pm0.09$ & $0.67\pm0.12$  &  $0.19\pm0.08$  & $<0.15$  & $1.18\pm0.15$  & $<0.20$ \\
        &          &      & &   &        &       &   \\
P3(fg)  &  ~~0   & \nodata\ & $1.40\pm0.10$ & $1.62\pm0.10$  & $1.10\pm0.12$   & $1.41\pm0.10$ & $1.90\pm0.10$   & $1.01\pm0.11$ \\
        &  ~~0\tablenotemark{c} & ($4.9\pm1.0$)\tablenotemark{d} & ($2.04\pm0.15$) & ($2.37\pm0.15$) & ($1.87\pm0.20$) & ($2.06 \pm 0.18$) & ($3.80\pm0.20$)  & ($1.47\pm0.22$) \\
P3(bg)  &  103   &  $0.92\pm0.12$   & $<0.05$ & $<0.12$     &  $0.12\pm0.06$  & $<0.04$  & $0.13\pm0.05$  & $<0.10$ \\
\enddata
\tablenotetext{a}{For each pair sample, the rest-frame equivalent width (in \AA\ ) for the stack of the foreground
galaxy spectra are given in the first row; the results from the composite of the background spectra,
{\it shifted to rest frame of the foreground galaxy}, are in the second row.}
\tablenotetext{b}{Values include both components of the \ion{C}{4} doublet.}
\tablenotetext{c}{ Rest equivalent widths after applying corrections (see text) to represent sightlines through the entire
galaxy at $b=0$}
\tablenotetext{d}{Values for \lya\ at $b=0$ are estimated from the observed strength of Ly$\beta$ absorption.}
\label{table:abs_table}
\end{deluxetable*}

As might be expected, the composites P1(fg), P2(fg), and P3(fg) are very
similar, since they are averages of galaxy spectra for 3 sets of galaxies selected
using identical criteria, with very similar redshift ranges (see Table~\ref{table:sample_table}). 
Nevertheless, they are completely independent, having been  
formed from the spectra of distinct samples of galaxies. The differences in measured line
strength for the foreground galaxy composites indicate 
the approximate level of sample variance ($<10$\% for the strongest absorption
lines).  
It is of interest to predict what the absorption line profiles of the spectra of the foreground
galaxies would look like if the line of sight were to probe the full galaxy rather than only
the part in the foreground. If one assumes that the strong lines are saturated, and that the kinematics
of outflowing gas are similar on the far side of the galaxy as observed on the near side, the
full line profile should include a reflection of the portion of the line profile with $v <0$
about $v=0$. Generally, accounting for this effect increases the equivalent widths 
by a factor of 1.45 for low ionization species (e.g., C{\sc II}, Si{\sc II}), 1.70 for \ion{Si}{4}, 
and $\simeq 2$ for \ion{C}{4}. These factors vary because of the differing strength of the absorption 
near $v=0$ -- typically low-ionization species have stronger $v \simeq 0$ absorption, while for
\ion{C}{4} there is very little. The corrected values of the line strengths for the foreground 
galaxy spectra are also indicated in Table~\ref{table:abs_table}; in the discussion below, we
adopt the corrected values for the $b \simeq 0$ sightlines.  

The comparison in Fig.~\ref{fig:vel_plt5} thus shows the difference in absorption line
profiles for lines of sight at $b=0$, and those offset by 1\secpoint5$-5$\arcs\ ($\langle \theta
\rangle = 3\secpoint8$, or $\langle b \rangle \simeq 30$ kpc), for precisely
the same set of 42 galaxies. As in the example of GWS-BM115/BX201 above, the absorption line centroids in
the $\langle b \rangle \simeq 30$ kpc line of sight are close to the foreground galaxy systemic redshift. 
For instance, the observed \lya\ absorption line in the P1(bg) spectrum has a measured wavelength of 1215.62 \AA, only
12 \kms\ from $v=0$; the estimated error in the line centroid
from propagation of the uncertainties in $z_{sys}$ for individual objects is $\sigma(\Delta v)
\approx 20-25$ \kms.  The other lines in the P1(bg) spectrum show similar behavior, in the sense
that their centroids are much closer to $v=0$ than the same lines in the P1(fg) spectrum,
which have $ v\simeq -150$ to $-200$ \kms. It is also clear from Table~\ref{table:abs_table} and
Figure~\ref{fig:vel_plt5} that the strength of low and intermediate ionization absorption lines is
significantly weaker for lines of sight with an average impact parameter of $\langle b \rangle =31$ 
kpc (physical). Specifically,
the strength of \ion{C}{2} $\lambda 1334$ is reduced by a factor of $\sim 3$, and Si{\sc iv} $\lambda 1393$
by more than a factor of 5, for sample P1(bg) ($\langle b\rangle = 31$ kpc) relative to P1(fg). Interestingly,
the strength of the \ion{C}{4} absorption falls much less steeply in P1(bg) versus P1(fg), declining
by only a factor of $\simeq 1.8$.  

The same general trend continues in the comparison of P2(fg) and P2(bg) (Figure ~\ref{fig:vel_p5to10}
and Table~\ref{table:abs_table}), where P2(bg) represents an average spectrum for impact parameters
$b=41-82$ kpc with $\langle b \rangle = 63$ kpc.
P3(bg), sensitive to absorption from gas with $b=82-125$ kpc ($\langle b \rangle = 103$ kpc), has
reached the point where the composite spectra formed from the low-resolution, low S/N galaxy
spectra are just barely
sensitive enough to detect absorption lines other than \lya\ . 
\ion{C}{4} absorption is detected with only marginal significance, and the low-ionization metal lines fall below the
threshold of detectability. Interestingly, there is no significant change in equivalent width for
the \lya\ absorption line between sample P2 and P3: both have ${\rm W_0(\lya)}$ reduced by only
a factor of $\simeq 2$ relative to sample P1(bg). We will discuss possible explanations for this 
behavior in \S\ref{sec:model} below.  

\begin{figure*}[htb]
\centerline{\epsfxsize=13cm\epsffile{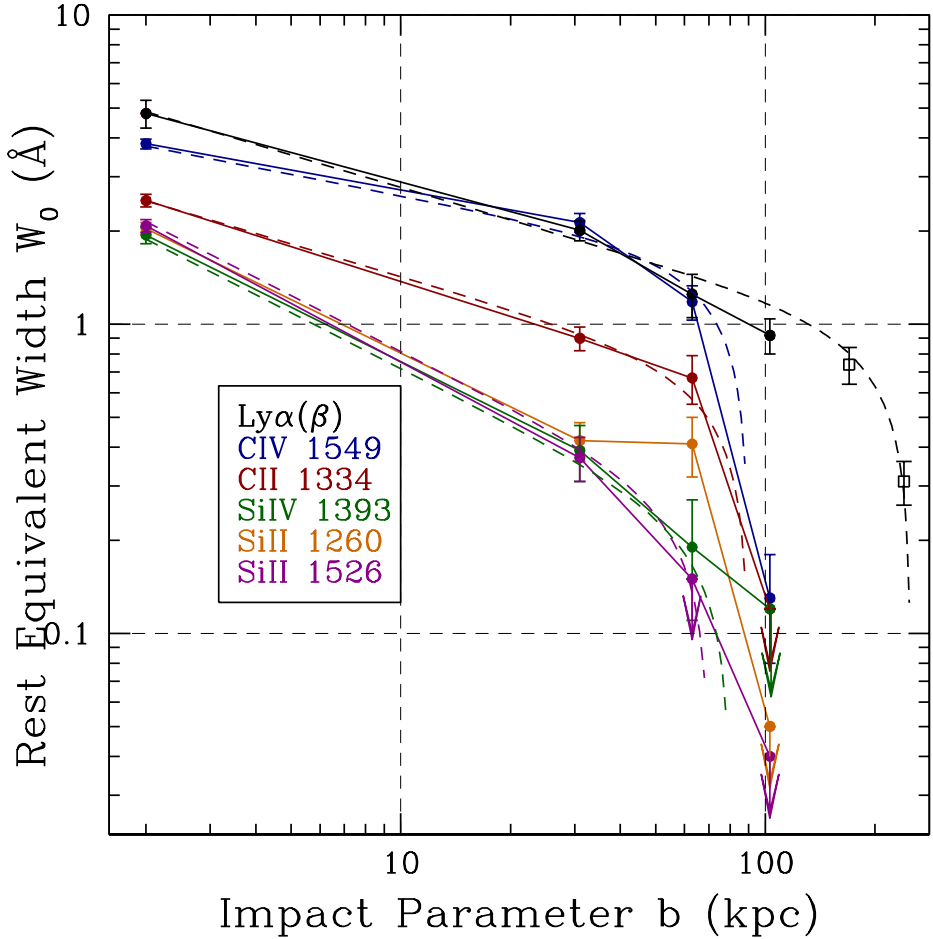}}
\figcaption[fig21.eps]{The dependence of IS absorption line strengths on the galactocentric impact parameter in
physical kpc, for the galaxy-galaxy pair samples outlined in Tables~\ref{table:sample_table}
and \ref{table:abs_table}. The values near $b\simeq 0$ are an average of the points for the 3 distinct
foreground galaxy samples P1(fg), P2(fg), and P3(fg). These values have each
been corrected upward to account for the estimated contribution to the line strength from
the ``far side'' of the gas distribution, assuming symmetric kinematics of outflowing material.
Each color (for points and connecting line segments) represents a different ISM transition as
indicated in the box legend; downward arrows on points
indicate upper limits. The dashed curves using the same color coding are predictions for $W_0(b)$ using the model
described in the text (summarized in Table~\ref{table:model_table}). The two points for \lya\ represented
by open squares were measured from HIRES spectra, as described in the text. The corresponding spectra
are plotted in Fig.~\ref{fig:plot_gal_hires}.
\label{fig:logw_vs_logb}
 }
\end{figure*}

\begin{figure}[htb]
\centerline{\epsfxsize=8cm\epsffile{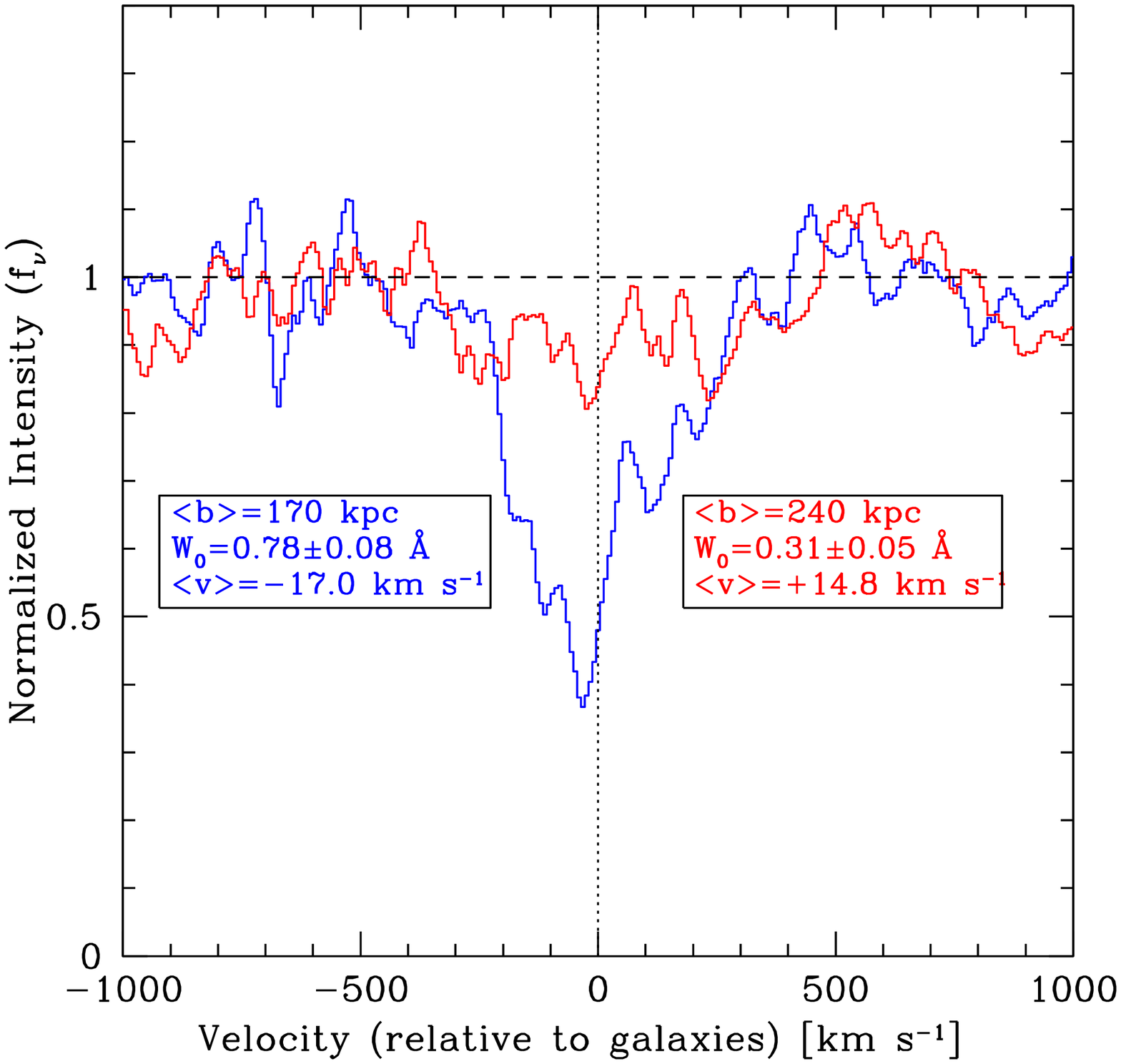}}
\figcaption[fig22.eps]{Average spectrum near \lya\ in regions of HIRES spectra centered on the systemic redshifts
of galaxies with QSO line of sight impact parameters of
$120 \le b \le 200$ kpc (Blue; $\langle b \rangle = 170$ kpc) and $200 < b \le 280$ kpc
(Red; $\langle b \rangle = 240$ kpc). Each spectrum represents an average of 21 galaxy-QSO sightline
angular pairs. The centroid velocities are very close to to $v=0$, indicating that our
estimates of the galaxy systemic redshifts are accurate. The equivalent widths fall on an
extrapolation to larger $b$ of the relationship obtained from galaxy pairs.
\label{fig:plot_gal_hires}
}
\end{figure}

A graphical summary of the contents of Table~\ref{table:abs_table} is shown in Fig.~\ref{fig:logw_vs_logb}. 
Points representing lines of sight with $b\simeq 0$  [i.e.,  P1(fg), P2(fg),
and P3(fg)] have been averaged, the result of which is plotted near $b=0$. The small error bars show
that the scatter among the {\it independent} sub-samples of foreground
galaxy spectra is small. Ly$\beta$ has been used in lieu of \lya\ for a measure of the strength of \ion{H}{1}
absorption in the foreground 
galaxy spectra, because of the contamination of \lya\ absorption by \lya\ emission. Because both lines probably
fall on the flat part of the curve of growth, the strength of \lya\ absorption is likely to be comparable
to that of \lyb.   

In spite of the limited sensitivity to weak absorption lines in the composite spectra, there are several 
trends worth noting.
The strength of \ion{Si}{4} and both \ion{Si}{2} lines track each other very closely out to $b \sim 40$ kpc. This is noteworthy
because it is to be expected when all 3 lines are strongly saturated and sample the same
range of velocities; in particular, the two measured lines
of Si{\sc ii}\ ($\lambda 1260$, $\lambda 1526$) have oscillator strengths $f=1.18$ and $f=0.133$, respectively. For a given
column density N(Si{\sc ii}) the ratio of $\lambda^{2}f$ for the two lines would differ by a factor of $\simeq 6.1$; the fact
that the equivalent widths of these two lines are equal out to $\sim 40$ kpc indeed supports strong saturation. 
This suggests the trend toward weaker absorption with increasing $b$ is 
dominated by either a decreasing velocity spread or
a decreasing $f_c$;  the galaxy spectra are not of high enough resolution to distinguish between the two. 
By $\langle b \rangle  \simeq 63$ kpc, however,
the measured Si{\sc ii}\ $\lambda 1526$ line strength has decreased 
significantly with respect to Si{\sc ii}\ $\lambda 1260$ (it is
at least 3 times weaker) indicating that the $\lambda 1526$ line is becoming unsaturated.  
The equivalent width ratio $W_0(\lambda 1526)/W_0(\lambda 1260)$ implies an average total \ion{Si}{2}\ column density at $\langle b \rangle \simeq 63$ kpc of 
log~N(Si{\sc ii})$\simeq 13.3$ cm$^{-2}$. 
In principle the two lines of the Si{\sc iv} doublet should allow an assessment of the degree of saturation, but 
the Si{\sc iv} $\lambda 1402$ absorption line is sufficiently weak in the composite spectra that it cannot be
measured with useful precision. 

The other lines in Fig.~\ref{fig:logw_vs_logb} exhibit trends similar to that of the Si lines, albeit with
a shallower decline in line strength 
with increasing impact parameter.  The equivalent widths for 
all of the ions in Fig.~\ref{fig:logw_vs_logb} fall precipitously somewhere  
between $b \simeq 60$ and $b \simeq 90$ kpc, with the exception of \lya.  
All of the species decline with galactocentric  
radius (or impact parameter) as $W_0 \propto b^{-\beta}$ with $0.2 \simlt \beta \simlt 0.6$ out to
$b \sim 50$ kpc, whereas $\beta \simgt 1$ at larger $b$ (again, \lya\ excepted). 
Because of line saturation, and the fact that only relatively strong lines can be detected using
low-resolution spectra of faint galaxies, we know that the trends seen in Fig.~\ref{fig:logw_vs_logb} are dominated
by a declining value of $f_c$ with galactocentric radius, along with a smaller 
range of line-of-sight velocities with increasing $b$. Schematic models which reproduce the
curves in Fig.~\ref{fig:logw_vs_logb} are discussed in \S\ref{sec:geo_model}.  

\subsection{\lya\ Absorption at Larger Impact Parameter}
\label{sec:hires_lya}

In order to trace the average \lya\ absorption line strength to larger galactocentric
radii ($b > 125$ kpc), we made use of the ${\rm S/N \simeq 40-100}$, high resolution ($R \simeq 30000$) 
spectra of bright
QSOs lying in 15 of the galaxy survey fields introduced in \S\ref{sec:pairs}. 
The QSO spectra were obtained using the HIRES spectrograph on the Keck 1 telescope, with
the UV cross-disperser providing simultaneous wavelength coverage from 3100-6000 \AA\ with
very high throughput. 
For the time being, we degrade these QSO spectra (using only their high S/N) in order
to obtain absorption line measurements 
analogous to those using the galaxy pairs above. 
As for the galaxy pairs, foreground galaxy redshifts were assigned based on
equations 2 or 4, and the spectral regions of the QSO spectra corresponding to the \lya\ transition
at the redshift of the foreground galaxy were extracted. The foreground galaxies were chosen to
lie in two impact parameter bins:  $125 \le b \le 200$ kpc and $200 \le b \le 280$ kpc.
Within each bin, 21 QSO spectral regions were shifted into the rest frame of 21 distinct
foreground galaxies, i.e. so that $v=0$ lies at the galaxy redshift. The spectra within each bin
were then averaged to obtain the mean \lya\ absorption profile. The resulting composites were then 
continuum normalized\footnote{The strength of the absorption associated with the galaxies is
measured relative to a new effective continuum level reduced by $\simeq 20$\%, the average \lya\ forest
decrement at $z \sim 2.2$.} and averaged to obtain the mean \lya\ absorption profile. 
Although the background QSOs are point sources, and therefore not suitable for measuring gas
covering fraction in a single line of sight, the ensemble of QSO sightlines should provide
an average absorption profile equivalent to what would be obtained using resolved background
galaxies.  The two new points have been added to Fig.~\ref{fig:logw_vs_logb}
to extend the measurement to $b=280$ kpc, or an angular separation of $\simeq 34$\arcs\ at $\langle z \rangle = 2.2$.  
The averaged spectra in the galaxy rest frame for the two bins are shown in Fig.~\ref{fig:plot_gal_hires}; 
the centroids of the resulting \lya\ absorption features are remarkably close to $v=0$, another indication
that the estimated galaxy redshifts are relatively free of systematics \citep{assp03}. 

Returning to Fig.~\ref{fig:logw_vs_logb}, the high-quality HIRES spectra have allowed us to extend the
$W_0$ vs. $b$ relation for \lya\ to see the point where the line strength decreases rapidly, evidently $b
\simeq 250$ kpc.   
It seems likely that the rapid fall-off
in $W_0$ indicates that \lya\ is becoming optically thin. Indeed, under this assumption,
the line strength in the bin with the largest $b$ is $W_0 = 0.31\pm0.05$ \AA\, which
would correspond roughly to 
to log N(\ion{H}{1}) $\simeq 13.25$.  Thus, as for the metal line transitions, a combination 
of decreasing covering fraction
and linear curve of growth effects steepens the $W_0$ vs $b$ relation.  

We now explore the possibility that
a simple schematic outflow model might reasonably 
reproduce the observations of both the $b=0$ constraints (line strength and line shape
in galaxy spectra) {\it and} the observed variation of absorption line strength with impact parameter.  

\section{A Simple Model for Outflows and Circum-galactic Gas} 
\label{sec:model}

\subsection{Line Strength and the Spatial Extent of Circum-galactic Gas}
\label{sec:geo_model}

Assuming a spherically symmetric gas flow with (galactocentric) radial velocity outward,
$v_{out}(r)$ [note that this is {\it not} a ``shell''], 
the strength of absorption lines produced by material along an observers line of sight at 
impact parameter $b$ will depend
on the line of sight component of $v_{\rm out}$ and on the radial and velocity 
dependence of the covering fraction,
$f_c(r,v)$. 
For the moment, we characterize the rapid fall-off at some galactocentric
radius $R_{eff}$ (which may differ for each ionic species) as an ``edge''
beyond which no absorption would be detected.  

\begin{figure}[htb]
\centerline{\epsfxsize=8cm\epsffile{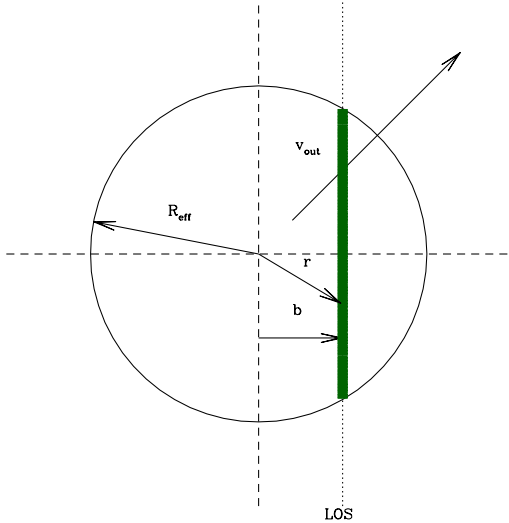}}
\figcaption[fig23.eps]{Diagram illustrating the simple model for estimating the relative interstellar absorption line
strength $W_0$ as a function of impact parameter $b$ used to produce the model curves in Fig.~\ref{fig:logw_vs_logb}.
${\rm R_{\rm eff}}$ is the characteristic size of the gas distribution producing saturated lines of a particular
transition. Any systematic dependence of ${\rm v_{out}}$ on radius has not been included in this model, but
the covering fraction $f_c$ of gas giving rise to the transition of interest is assumed to have a radial
dependence $f_c(r) \propto r^{-\gamma}$ (see text for additional details).
\label{fig:coords}
}
\end{figure}

Modeling the effective value of the covering fraction (and the resulting line equivalent width) 
for gas along an observed line of sight must include an assumption about
the radial dependence of the outflow velocity of absorbing material $v_{\rm out}(r)$ and $f_c(r,v)$, 
where the radial coordinate $r$ takes on values between $b$ and ${\rm R_{\rm eff}}$,
as illustrated in Fig~\ref{fig:coords}. For a saturated transition, the observed line profile 
(normalized intensity $I$ vs. velocity) is given by  
\begin{eqnarray}
I(v) =  1-\int{f_c(r,v)dl} ~~,
\end{eqnarray}    
where $v$ is the line-of-sight component of $v_{out}(r)$ relative to the galaxy systemic redshift, 
and the integral is evaluated along the line of sight (shaded region in Fig.~\ref{fig:coords}.)
If the form of $v_{out}(r)$ is assumed, then one can associate a line of sight component of 
velocity ($v$) with each value of $r$ along the sightline, so that
$f_c(r,v)$ becomes a function of only one variable (either $r$ or $v$).  
The line strength is then given by the integral of $I(v)$ evaluated between the minimum
and maximum line-of-sight velocity expected. For a velocity field which has $v_{out}$ 
increasing with $r$, 
the extreme velocities expected would be 
\begin{eqnarray}
| \Delta v_{max(b)}| = v_{out}(R_{\rm eff}) [1-b^2/R_{\rm eff}^2]^{1/2} 
\end{eqnarray}
and the associated equivalent width (with continuum normalized to unity) is just
\begin{eqnarray}
W_0(v) = 2 \int_0^{\Delta v_{\rm max}} f_c(v) dv
\end{eqnarray}
which can be integrated over wavelength to give the expected absorption line equivalent width. 
Assuming for the moment that $f_c(r) =f_{c,max}r^{-\gamma}$, and that $v_{out}(r)$ is constant, 
one can adjust the values of $v_{out}$, $\gamma$, and $R_{\rm eff}$ in order to reproduce
the observed $W_0(b)$ for each species. The maximum covering fraction $f_{c,max}$ can be
measured from the stacked spectra of the foreground galaxies-- it is just the maximum depth
of the absorption lines relative to the continuum for each species. These have been measured
from an average of all 3 foreground galaxy samples, P1(fg), P2(fg), and P3(fg), 
and collected in Table~\ref{table:model_table}. Table~\ref{table:model_table} also contains 
the parameter values used to produce the model curves shown in Fig.~\ref{fig:logw_vs_logb} for
each species. 

\begin{deluxetable}{l c c c c }
\tablewidth{0pt}
\tabletypesize{\footnotesize}
\tablecaption{$W_0$ vs. $b$ Model Parameters\tablenotemark{a}}
\tablehead{
\colhead{Line} &
\colhead{$\gamma$\tablenotemark{b}} &
\colhead{$R_{\rm eff}$ (kpc)}  &
\colhead{$v_{\rm out}$}  &
\colhead{$f_{c,max}$\tablenotemark{c}}
}
\startdata
Ly$\alpha$(1216) & 0.37 & 250 & 820 & 0.80 \\
\ion{C}{4}(1549) & 0.23 & ~80 & 800 & 0.35/0.25\tablenotemark{d} \\
\ion{C}{2}(1334) & 0.35 & ~90 & 650 & 0.52 \\
\ion{Si}{2}(1526) &  0.60 & ~70 & 750 & 0.40 \\
\ion{Si}{4}(1393) & 0.60 & ~80 & 820 & 0.33 \\
\enddata
\begin{indent}
\tablenotetext{a}{Parameters used to produce the model curves shown in Fig.~\ref{fig:logw_vs_logb}}
\tablenotetext{b}{Power law exponent in the expression $f_c(r) =f_{c,max}r^{-\gamma}$}
\tablenotetext{c}{Maximum value of the covering fraction for each transition, measured from the composite
spectrum (see Fig.~\ref{fig:velplots})}
\tablenotetext{d}{Includes contributions from \ion{C}{4} $\lambda 1548$ and \ion{C}{4} $\lambda 1550$ of
0.35 and 0.25, respectively. }
\end{indent}
\label{table:model_table}
\end{deluxetable}

The examples
plotted using dashed lines in Fig~\ref{fig:logw_vs_logb} show that the model does a reasonable job
reproducing the relatively shallow dependence of $W_0$ on $b$
, and the steep decline at $b\simeq 70-250$ kpc (depending on the transition). 
The best values of $\gamma$ are in the range $0.2 \le \gamma \le 0.6$, depending on the transition 
(Table~\ref{table:model_table}); 
note that the assumption of $f_c(r) \propto r^{-2}$, which
would apply if the absorbing clouds retained the same characteristic physical size as they move to larger
$r$, appears strongly ruled out by the relatively shallow observed dependence of $W_0$ on $b$
\footnote{It turns out (perhaps counter-intuitively) that the power law 
slopes relating $W_0$ and $b$ in the model curves in Fig.~\ref{fig:logw_vs_logb} are very close to the
power law index $\gamma$ assumed in the expression for $f_c(r)$ because
of the relationship between impact parameter $b$ and the line-of-sight dependence of the sampled velocity range.}.  
Obviously, a distinct ``edge'' to the gas distribution at some radius ${\rm R_{\rm eff}}$ is not physical; however,
this radius could (e.g.) mark the point at which a transition is becoming optically thin, so that both
declining covering fraction {\it and} decreasing optical depth lead to a rapid diminution of ${\rm W_0}$ with $b$.  
This explanation could also account for the fact that \lya\ does not exhibit a break in the
near-power-law dependence of $W_0$ on $b$ until $b \sim 250$ kpc: \lya\ 
remains strongly saturated as long as log N(\ion{H}{1})$\simgt 14.5$, so that geometric
dilution dominates over changes in optical depth to radii beyond the limit of sensitivity for
the the galaxy-galaxy pair samples;
even with unity covering fraction a rest equivalent width of $\simeq 0.9$ \AA\ would correspond to
a saturated \lya\ line.  

For this particular model in which $v_{out}$ is independent of galactocentric
radius $r$, the combination of the measured $f_{c,max}$ and $v_{out}$ control the overall normalization, 
while the slope of the $W_0$ vs. $b$ function is dictated entirely by the choice of 
$\gamma$. As shown in Table~\ref{table:model_table}, 
\lya, \ion{C}{4}, and \ion{Si}{4} are best matched if $v_{out}=800$ \kms, while 
\ion{Si}{2} and \ion{C}{2} suggest $v_{out}=750$ \kms and $v_{out} = 650$ \kms, respectively.
The fact that the latter two values are somewhat lower may be related to the influence 
of (non-outflowing) absorption near $v=0$ on the measurement
of $f_{c,max}$ (see discussion in \S~\ref{sec:composites}); 
for example, changing the value of $f_{c,max}$ for \ion{C}{2} from 0.52 to 0.42 (i.e.,
by less than 20\%) favors $v_{out}= 800$ \kms\ rather than $v_{out}=650$ \kms. 

So far, the model has made a number of idealized assumptions: spherical symmetry, 
uniform outflowing velocity field (we will assess the validity of this assumption in \S~\ref{sec:vel_model}), radial 
dependence of the covering fraction $f_c(r) \propto r^{-\gamma}$, and saturated lines
whose equivalent widths are  controlled by the covering fraction coupled with the range of velocities expected to
be sampled along the line of sight. 
It is worth a cautionary reminder at this point that $W_0(b)$ for a particular transition 
cannot be converted unambiguously into a map of total column density vs. $b$ so long
as the lines remain optically thick. Instead, one measures the average covering fraction
of gas hosting that particular ion; conversely, when the optical depths become of order unity
it becomes feasible to measure total column density, but at that point the constraints on covering fraction
are diminished.  

As mentioned above, \ion{C}{4}, with $\gamma \simeq 0.20$, has a much more gradual decline with $b$
than \ion{Si}{2} and \ion{Si}{4}.
This behavior might be attributed to a changing ionization state 
of more diffuse gas entrained in the outward flow, so that ``new'' pockets of gas become more likely to absorb in C{\sc iv}  
as the flow becomes more diffuse at larger radii (i.e., it is not just the same ``clouds'' with larger
cross-section). The strength of \ion{C}{4} absorption 
often appears ``de-coupled'' from the strength and depth of the lower ionization species in galaxy spectra 
(see \citealt{ssp+03}), which may be related to the very different radial distribution of absorbing
gas inferred above, as well as to the much larger physical coherence scale of \ion{C}{4}-absorbing gas
compared to gas giving rise to lower ionization species observed in high resolution studies of very
small-separation lines of sight. Nevertheless, C{\sc iv} absorption usually spans a very 
similar range of blue-shifted velocities in the galaxy spectra, 
strongly suggesting it is being carried in the same outflow. 
The quality of the present data  does not justify a more detailed 
analysis. 

\subsection{Total Absorption Cross-Section}

A very simple (and model-independent) calculation that can be made using the 
information collected in Fig.~\ref{fig:logw_vs_logb} 
is the expected incidence of absorption lines exceeding a particular threshold in $W_0$ 
associated with extended gas around galaxies
similar to those in the current sample. This exercise is the converse of one undertaken many
times in the past for various types of QSO absorbers; typically, one takes an observed
value of $dN/dz$, the number of systems observed per unit redshift, and calculates the
galaxy cross-section and number density required to account for the rate of incidence,
assuming the cross-section is all contributed by galaxies. In the present
case, we have a well-defined population of galaxies, uniformly selected, with a
well-established far-UV luminosity function.  
\cite{reddy09} have produced the most up-to-date LF, determined from the same data set and
using the same selection criteria as the galaxies used in the galaxy-galaxy pair analysis. 
They found that a Schechter function
with $\phi*=2.75\pm0.54\times10^{-3}$ Mpc$^{-3}$, $M^* = -20.7$, and $\alpha  = -1.73\pm 0.07$ is a good
description of the rest-frame 1700 \AA\ luminosity function of UV-selected galaxies in the redshift range 
$1.9 \le z \le 2.7$. Integrating
this function only over the luminosity range of the actual spectroscopic sample (i.e., to ${\cal R} \le 25.5$,
which corresponds to $L \sim 0.3 L^*$ at $\langle z \rangle = 2.3$,)
the number density is $n_{\rm gal} \simeq 3.7\times10^{-3}$ Mpc$^{-3}$. Taking $b\simeq 90$ kpc 
as the point where ${\rm W_0}$(\ion{C}{4}~$\lambda 1548) \simeq 0.15$ \AA\ (see Fig.~\ref{fig:logw_vs_logb}), 
at the mean redshift of the foreground galaxies, the co-moving cross-section for absorption per galaxy
is $\sigma_{\rm gal} \simeq 0.26$ Mpc$^{2}$. 
The relevant co-moving path length per unit redshift at $\langle z \rangle = 2.20$ 
is $l \simeq 1320$ Mpc, so that the expected number of $W_0$(\ion{C}{4})$ > 0.15$ \AA\ absorption systems\footnote{
This equivalent width threshold is roughly equivalent to systems having log~N(C{\sc iv})$\simgt 
13.5$ cm$^{-2}$} per unit redshift is 
\begin{eqnarray}
{dN \over dz} ({\rm CIV,~pred})  \simeq (3.7\times 10^{-3})~l \sigma_{\rm gal} \simeq 1.1 ~.  
\end{eqnarray}
The observed incidence of \ion{C}{4} systems at the same equivalent width threshold
at $z \simeq 2.2$ is ${\rm dN/dz({\rm obs}) = 2.44\pm0.29}$ (\citealt{sbs88,steidel90}), 
so that galaxies {\it in the current spectroscopic sample alone} account for $\sim 45$\% of
strong C{\sc iv} systems at $z \sim 2$. Although the statistics for the incidence of other lines are not
as well-established, \cite{ss92} found that, for the \ion{C}{2} $\lambda 1334$ absorption with $W_0 \ge 0.15$ \AA\ 
and $\langle z \rangle = 2.35$,
\begin{eqnarray}
\label{eq:dn_dz_c2}
{dN \over dz}({\rm CII,~obs})  \simeq 0.94\pm0.33. 
\end{eqnarray}
At face value, since the extent of \ion{C}{2} is approximately the same as \ion{C}{4} 
(Table~\ref{table:model_table} and Fig.~\ref{fig:logw_vs_logb}), comparison with eq.~\ref{eq:dn_dz_c2} 
suggests that essentially {\it all} \ion{C}{2} 
$\lambda 1334$ absorption with $W_0 > 0.15$ \AA\ is within 
$\simeq 90$ kpc (physical) of a galaxy similar to those in the sample used here.       
The presence of strong \ion{C}{2} absorption generally depends on the self-shielding provided by
\ion{H}{1} with log N$\simgt 17.2$ (e.g., \citealt{steidel90b}), so that although we cannot measure
N(\ion{H}{1}) from the galaxy pair spectra, one might expect a similar cross-section for Lyman Limit System (LLS)
absorption. According to \cite{steidel92}, 
\begin{eqnarray}
{dN \over dz}({\rm LLS,~obs}) \simeq 1.4 
\end{eqnarray}
at $z\simeq 2$, suggesting that
the total absorption cross-section of these galaxies accounts for as much as $\sim 70$\% of all LLS absorption
at similar redshifts.   

It is important to emphasize that the absorption line statistics obtained using galaxies
as background sources should be identical to what would be obtained using QSOs, in the limit
of a suitably large number of QSO sightlines. In other words, a given average rest-frame
equivalent width would be obtained at the same characteristic impact parameter from a
galaxy, independent of the morphology of the background probes. It follows from the above 
discussion that most of the cross-section for strong low-ionization metal-line absorption (and
by extension, LLSs) is provided by rapidly star-forming galaxies, at least at $z \simeq 2.2$. 
A very similar connection between strong QSO absorbers and rapid star formation has
been observed recently by \cite{menard09}, who used a very large sample of $W_0 > 0.7$\AA\ MgII absorbers
(in SDSS QSO spectra) over the redshift range $0.4 \simlt z \simlt 1.2$, and compared with the stacked [OII] 
$\lambda 3727$ emission signature detected within the SDSS fibers. These authors conclude that 
the strong absorption is tightly correlated with the presence of a rapidly star forming galaxy
within $\simeq 50$ kpc of the QSO sightline. By way of comparison, the statistical impact parameter
for producing \ion{C}{2} absorption with $W_0 \ge 0.7$ \AA\ at $\langle z \rangle = 2.2$ is $b \le 60$ kpc. 

We now turn to additional consistency checks by combining the spatial information from the preceding
discussion with information extracted from the profiles of the IS absorption lines in the
$b=0$ spectra of typical LBGs.

\subsection{Constraints from IS Absorption Line Profiles}
\label{sec:vel_model}

From the previous section, the galaxy pair data indicate a covering fraction
of absorbing gas with a radial dependence $f_c(r) \propto r^{-\gamma}$, with $0.2 \simlt \gamma \simlt 
0.6$. All measured lines except Ly$\alpha$ have
dropped below the current detection limit for $W_0$ by $b \simeq 100$ kpc; for larger
impact parameters, spectra of higher resolution and S/N are required-- these measurements are
being made using the bright QSO sightlines in each of the survey fields (e.g., Rakic et al 2010; Rudie et al 2010, in prep.).
If we focus for the moment on scales $r \le 125$ kpc, the dominant importance of the covering fraction
in modulating absorption line strength
means that there are only weak constraints on changes in total column
density with radius. 
However, the kinematics of the absorption lines observed along sightlines at $b=0$-- the spectra
of the galaxies themselves-- provide a means
of checking the models for consistency. Ideally, one should be able to reproduce the
absorption line shape (i.e., covering fraction as a function of velocity) using a model 
for $v_{\rm out}(r)$ and $f_c(r)$ consistent with the observations of both direct and offset lines of sight.  

The boundary conditions provided by the direct line of sight ($b=0$) observations are that the maximum observed $|\dvis|$ 
should be
$\simeq 800$ \kms, the maximum covering fraction is typically smaller than unity, and the line shapes
generally imply that larger blue-shifted velocities are associated with gas having smaller covering fraction. As
we have shown, the
simplest (although probably not unique) physical explanation would
be that the higher velocity gas is located predominantly at larger radii\footnote{Note that the simple
models presented in \S\ref{sec:lya} above assumed a monotonic relationship between physical location
and velocity to simultaneously explain the kinematics of observed \lya\ emission and IS absorption} (see also
\citealt{weiner09,martin09}). Since the outflows
are believed to be driven by energy and/or momentum originating in the inner regions of the
galaxy, a plausible model might have the acceleration experienced by a gas cloud
depend on galactocentric distance $r$. The details of this dependence are likely
to be complex, since the force experienced by cool gas clouds would depend on the energy and/or momentum
acquired by the gas, the ambient pressure gradient, 
the radial dependence of the clouds' physical sizes\footnote{Recall that the 
radial dependence of the covering fraction is inconsistent with clouds maintaining the same
size; we show below that the cloud dimensions $R_c$ must increase with $r$. }, and the entrainment of ambient gas
in the flow (sometimes referred to as ``mass-loading'') as it moves outward. 

In an idealized picture of a multi-phase outflow, gas clouds, perhaps pressure confined
by a hotter, more diffuse gaseous medium, are accelerated by thermal and/or radiation
pressure to larger radii and higher velocities. 
The fact that the inferred radial dependence of the absorbing gas covering fraction remains
at $f_c(r) \propto r^{-\gamma}$ with $\gamma \le 0.6$ (at least for radii $r < R_{\rm eff} $)  
implies that the clouds must expand as they move out (otherwise $f_c(r) \propto r^{-2}$). 
Even the most steeply declining species, with $\gamma \simeq 0.6$, requires that the total cross-section
for absorption per cloud increases as $\sigma_c \propto r^{4/3}$, or that the 
characteristic cloud size $R_c(r) \propto r^{2/3}$.  
For clouds of constant mass $M_c=4/3\pi R_c^3 \rho_c$ and assumed pressure gradient 
$p(r) \propto r^{-2}$,
maintaining local pressure equilibrium leads to the same dependence, 
$R_c \propto r^{2/3}$; thus the behavior of $f_c(r)$ for Si{\sc ii}\ and Si{\sc iv}  
is very close to what one might expect for cooler, denser ``cloudlets'' confined by a hotter medium
whose pressure decreases with galactocentric radius.   
One might think of the clouds giving rise to low-ionization absorption 
as ``expanding bullets'', surviving as long as there
is sufficient ambient pressure to confine them, or until they expand to the point that they no longer
produce significant low-ionization absorption (see \citealt{schaye07} for additional observational
evidence of such phenomena). 

Strickland (2009) has suggested
that the force experienced by cool clouds in a wind model obeying
local pressure equilibrium would have radial dependence $F(r) \propto r^{-4/3}$; this is
precisely what would be expected under the circumstances described above (i.e., $R_c \propto r^{2/3}$ and
$p(r) \propto r^{-2}$). 
For clouds of constant mass, the radial dependence of cloud acceleration $a(r)$ would have 
the same form as $F(r)$; 
if additional mass (e.g. ambient ISM) is entrained in the wind, then $a(r)$ would decrease more rapidly with
increasing radius. 
Once the radial dependence of cloud acceleration is
specified, one obtains $v_{\rm out}(r)$, and thus 
$f_c(v)$, the expected velocity dependence of the covering fraction. 

Rather than develop a more detailed model for the physics of the multiphase
outflowing material (which is beyond the scope of the current work), 
for simplicity we parametrize the cloud acceleration $a(r)$ as
\begin{eqnarray}
a(r) =A r^{-\alpha}~~
\end{eqnarray}
where $A$ is a constant, and seek the value of $\alpha$ that can reproduce the
IS line profiles given $f_c(r)$ inferred from the galaxy-galaxy pair data
(Fig.~\ref{fig:logw_vs_logb} and Table~\ref{table:model_table}). Using this
parametrization of $a(r)$, 
\begin{eqnarray}
a(r) = Ar^{-\alpha} = {dv(r) \over dt} = \left[{dv(r) \over dr}\right] \left({dr \over dt}\right) = v(r)\left({dv \over dr}\right) .
\label{eqn:ar} 
\end{eqnarray}
Assuming the clouds are ``launched'' from $r=r_{\rm min}$, equation~\ref{eqn:ar}
 can
be rearranged and integrated to provide an expression for $v_{out}(r)$:
\begin{eqnarray} 
v_{out}(r) = \left({2A \over \alpha -1 }\right)^{0.5} \left(r_{\rm min}^{1-\alpha} - r^{1-\alpha}\right)^{0.5}.
\end{eqnarray}
The constant $A$ can be obtained by using the boundary condition provided by $|v_{max}|$, which is the
velocity at which $f_c(r)$ approaches zero\footnote{In other words, this is the maximum velocity beyond which
the covering fraction is too small to produce absorption that can be distinguished from the
continuum.} and the value of $R_{eff}$ estimated for the transition of interest (Table~\ref{table:model_table}). 

We are now in a position to attempt to match observed IS absorption line profiles while
simultaneously adhering to the observationally-inferred form of $f_c(r)$. For simplicity, we assume
that the spectra of galaxies sample their own outflow kinematics at $b=0$ (i.e., we ignore the
spatial extent of the galaxy continuum and treat observed spectra as a spatial average
over a region typically 2-3 kpc in diameter). 
 
\begin{figure*}[htb]
\centerline{\epsfxsize=13cm\epsffile{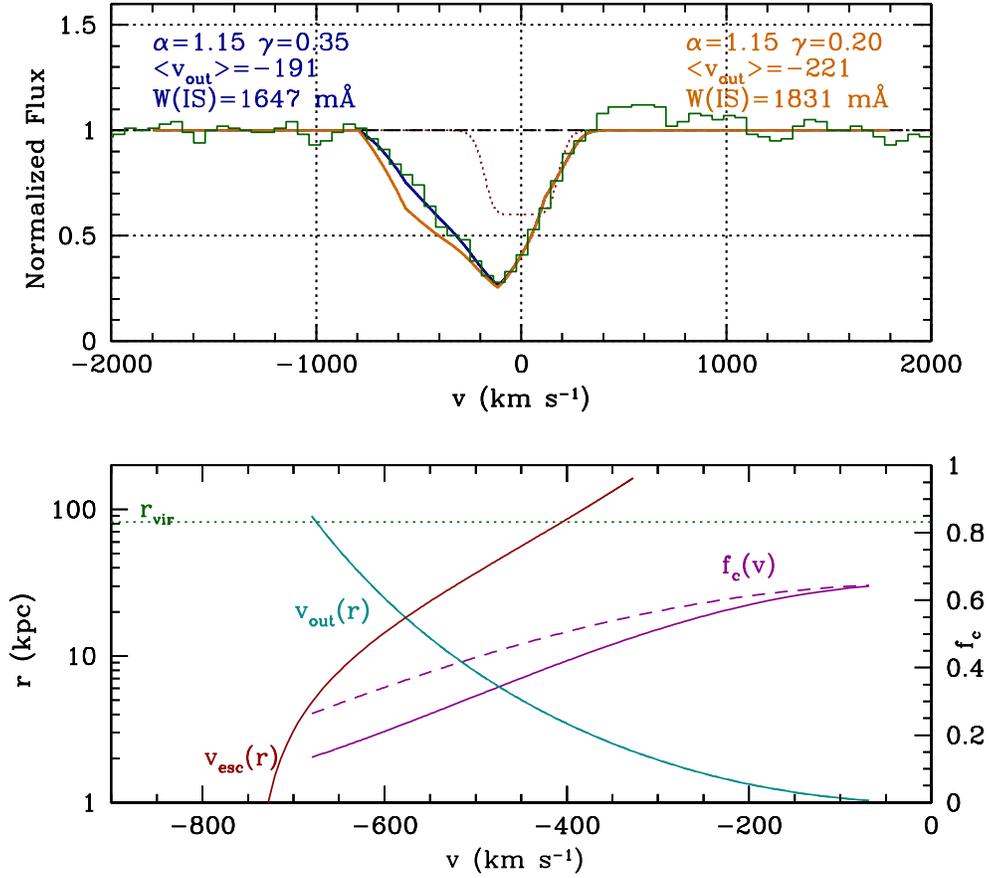}}
\figcaption[fig24.eps]
{{\it (Top):} Model line profiles (convolved with the appropriate instrumental resolution)
for an IS absorption line with radial covering fraction
$f_c(r) \propto r^{-\gamma}$, maximum covering fraction for outflowing material $f_{c,max} = 0.59$, and radial
cloud acceleration $a(r) \propto r^{-\alpha}$, compared to
the \ion{C}{2}\ $\lambda 1334.53$ profile from a composite of $R=1330$ galaxy spectra.
Two models are plotted, both of which include an absorption
component with $v=0$, $\sigma_v=80$ \kms, and $f_{c,max}$ = 0.4 :
the dark blue profile assumes $\gamma = 0.35$ as inferred from the galaxy pair measurements
for the \ion{C}{2} line, and $\alpha = 1.15$, which provides a good match to the observed
line shape. The orange curve represents the model line profile if $\gamma=0.2$ and all
other parameters are held fixed. Note that
the relatively small change in $\gamma$ produces a line profile inconsistent with the
data, with a high velocity wing that is too strong.
Both are constrained to
have $v_{max}=-700$ \kms at the point where the line profile becomes indistinguishable from
the continuum. The covering fraction of outflowing material is assumed to go to zero at $r=R_{eff}=90$ kpc
{\it Bottom:} The dependence of the covering fraction (purple,
righthand axis)
$f_c(v)$ and the relation between $v_{out}$ and galactocentric radius $r$ (cyan, left-hand axis).
The solid curves correspond to the dark blue profile, with the dashed curves indicating values for the orange profile.
Note that the velocity profiles are identical for the two models, since they use the same value of $\alpha$.
The red curve shows the estimated escape velocity as a function of galactocentric
radius $r$ for an NFW halo of mass $9 \times 10^{11}$ M$_{\sun}$; the green dotted line is
the virial radius $r_{vir}$ for the same halo. Note that at $r=r_{vir}$ (in the context of this model) the outflow
velocity exceeds $v_{esc}(r_{vir})$.
\label{fig:vvI}
    }
\end{figure*}

For demonstration purposes, to the outflow profile is added an absorption component at $v=0$ with
a maximum covering fraction varied to reproduce the $v>0$ portion of the line profile. The
width assigned to this component is taken to be the same as the nebular line width associated
with the \ion{H}{2} regions in a typical galaxy ($\sigma \simeq 100$ \kms.) 
Fig.~\ref{fig:vvI} shows
a comparison of the predicted velocity profiles for an IS line given the above model (assuming
that $r_{\rm min} =1$ kpc),
compared to the observed velocity profile of the \ion{C}{2} $1334.53$ line 
in the composite spectrum of $z \sim 2-3$ LBGs with $R\simeq 1330$. Also shown in Fig.~\ref{fig:vvI} are the
resulting relationships between $r$ and $v$ and between $f_c$ and $v$. The values
of $\gamma$ were chosen based on the models (Table~\ref{table:model_table}) that
adequately describe how $W_0$ varies with $b$ for the relevant ion:
a self-consistent kinematic model may be obtained for 
$\gamma=0.35$ (the value inferred for \ion{C}{2} $\lambda 1334$; see Table~\ref{table:model_table}) if $\alpha= 1.15$. 
To provide some intuition on how the line profile would be altered with different parameter values,
the predicted profile with $\alpha$ held fixed and $\gamma=0.20$ instead of $\gamma = 0.35$ is also
plotted. It shows that assuming an incorrect value for the radial dependence of $f_c$ leads (in this case)
to an absorption feature that has too much apparent optical depth at high values of $|v|$ and therefore
significantly over-predicts the values of $|\langle v_{out} \rangle|$ and $W_0$.   

\begin{figure*}[htb]
\centerline{\epsfxsize=13cm\epsffile{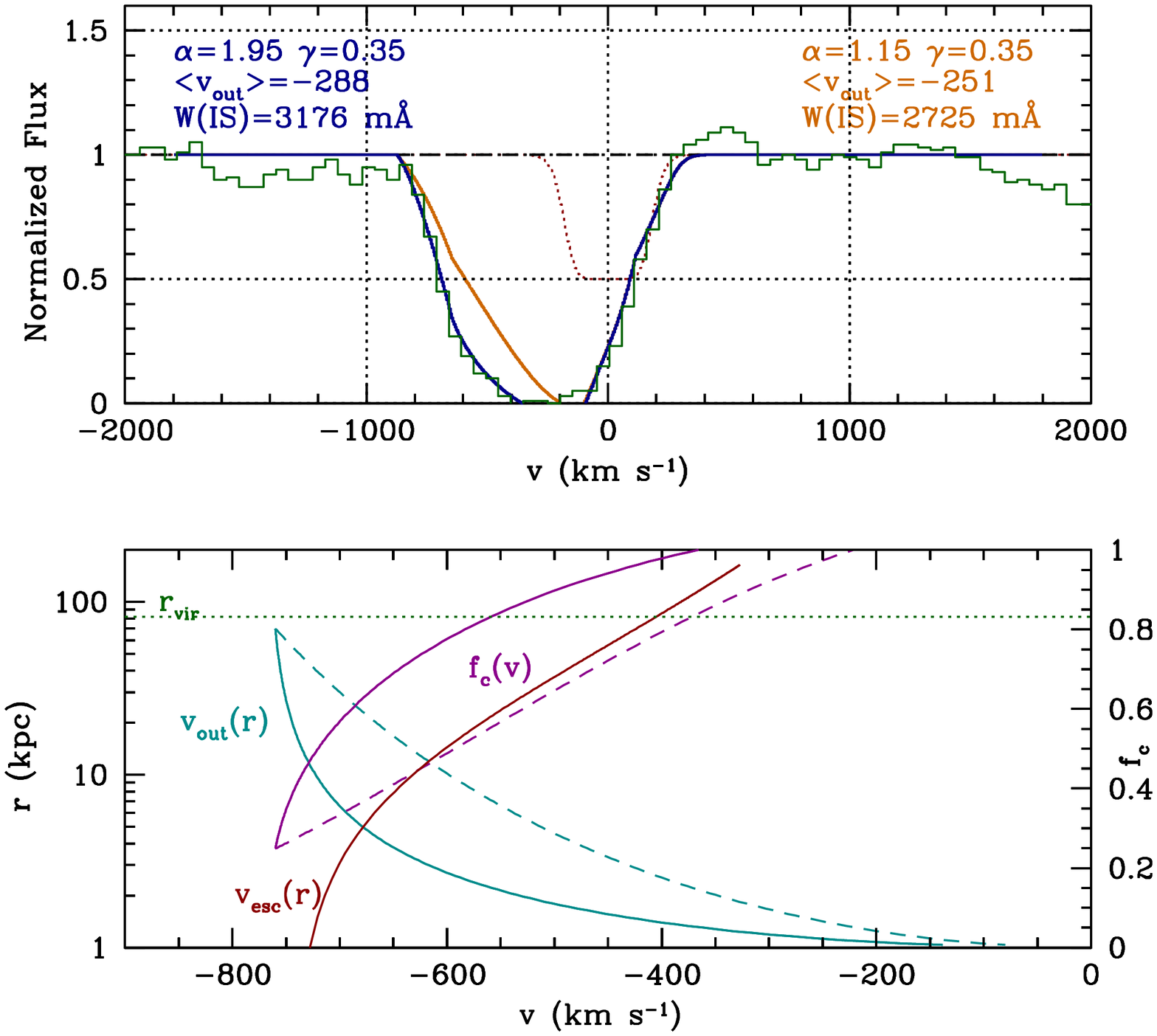}}
\figcaption[fig25.eps]{Same as Fig.~\ref{fig:vvI}, but with increased maximum covering fraction of the outflowing
gas $f_{c,max} = 1$, and a slightly larger covering fraction $f_{c,max} = 0.5$ for the component at $v =0$. The dark blue model profile
shows that for $\gamma = 0.35$, a good match to the observed \ion{C}{2} line in the spectrum
of the lensed LBG MS1512-cB58 ($R \simeq 1500$, to match Fig.~\ref{fig:vvI}; see \citealt{pettini00}) is obtained if
$\alpha = 1.95$. The orange curve illustrates the change in the model line profile
if $\alpha = 1.15$ is assumed instead, with all other parameters held fixed.
Clearly the latter model fails to produce the depth of the
absorption at high values of $|v|$.  As for the example in Fig.~\ref{fig:vvI}, the outflow velocity
is significantly higher than $v_{esc}$ at the virial radius of the assumed halo.
\label{fig:vvI2}
}
\end{figure*}

By altering the maximum covering fraction of the outflowing gas and the
optical depth and covering fraction of the $v = 0$ component, it is possible to reproduce
a wide range of observed line profiles. Fig.~\ref{fig:vvI2} (dark blue model profile) shows
another example with $\alpha$ chosen 
to produce a very strong line with a large blue-shifted centroid 
such as those observed in the spectrum of the lensed LBG MS1512-cB58 \citep{prs+02}. In this case,
it was necessary to change  $f_{c,max}$
to unity and to adjust the function $a(r)$ to be steeper than for the example 
in Fig.~\ref{fig:vvI}. Since the line is again \ion{C}{2} $\lambda 1334.53$, once again  
$\gamma = 0.35$. The orange model profile is that obtained when all parameters
are held fixed but $\alpha$ is reduced to the value that provided a good fit to
the profile in Fig.~\ref{fig:vvI}.   
Qualitatively, as the function $a(r)$ 
becomes steeper, the covering fraction at high velocity increases along with the line equivalent
width (given the imposed boundary conditions); the physical meaning of a steeper $a(r)$ 
is that material is accelerated more rapidly, such that it reaches high velocity before it
becomes geometrically diluted. 

The bottom panels of Figs.~\ref{fig:vvI} and \ref{fig:vvI2} are also interesting to consider. 
They show that, at least in the context of the current models, the outflowing gas attains
close to its maximum velocity within 
$\sim 3-10$ kpc (depending on the model) of the center of the galaxy, where
whatever mechanism is at work in accelerating the clouds is most effective. At larger
radii, 
the outflow velocities change slowly with increasing radius. Recall that our simple model for inferring
$f_c(r)$ from $W_0$ vs. $b$ assumed that $v_{\rm out}$ is independent of $r$. Since the current
model requires $a(r) = dv_{out}(r)/dt \propto r^{-\alpha}$ with $1.15 \simlt \alpha \simlt 1.95$,
most of the acceleration up to 
the maximum observed velocity occurs when $r << 10$ kpc. Thus, we conclude that having assumed 
a fixed $v_{out}$ as in \S\ref{sec:geo_model} was justified, given other sources
of uncertainty.    

We caution that in the particular family of models described above, 
the parameters $\gamma$ and $\alpha$ are covariant 
and combinations of the two can usually be found that will adequately reproduce observed IS line profiles. It is
only because we have information on the radial dependence of $f(r)$ from the galaxy-galaxy pairs
(constraining $\gamma$) that anything like a unique solution results. Since the value of $\gamma$
is statistical (and so may not apply for particular individual galaxies)  
it would be unwise to read too much physical significance into the inferred values of $\alpha$ 
needed to fit a particular line profile.  For example, the \ion{C}{2} line profile in Fig.~\ref{fig:vvI}
can be well-fit a model with the parameters $(\alpha,\gamma) = (1.33,0.60)$, and that
shown in Fig.~\ref{fig:vvI2} 
can be equally well-fit by the combination $(\alpha,\gamma) = (1.25,0.20)$.

\section{Discussion}
\label{sec:model_implications} 

\subsection{Connection Between $b=0$ and $b>>0$ Sightlines}

The models discussed above are schematic only, and probably not
unique in adequately describing the relatively crude observations. 
Nevertheless, we have seen that a reasonably successful model for the kinematics and overall geometry
of the outflows includes a characteristic asymptotic velocity of $\simeq 700-800$ \kms. 
Such high velocities are required to explain the shapes of IS absorption line
profiles ``down the barrel'' for $b=0$ sightlines, as well as the strength
of the IS absorption and its relatively shallow dependence on impact parameter
for sightlines with $b \ge 10$ kpc. This suggests that the absorption
at large $b$ is {\it causally related} to the outflowing material observed at $b=0$,
in which case the gas at large $r$ would most naturally be identified with cool gas carried
out in an earlier phase of the current episode of star formation. If this were the case,
one might expect the geometry of the metal-enriched gaseous envelope of a galaxy
to be time-dependent. 

For example,  
galaxies with spectral morphology similar to that of MS1512-cB58 (and represented by the model
shown in Fig.~\ref{fig:vvI2}) 
tend to be those with estimated stellar population ages (star formation episodes) of $t_{sf}< 100$Myr 
(\citealt{shapley01,kornei09}).
Even for gas moving at 800 \kms, reaching $r\sim 100$ kpc requires $\sim 120$ Myr, i.e. longer
than the inferred duration of the current star formation episode and $\simeq 20\%$ of $t_{sf}$ typical
of UV-selected galaxies at $z \simeq 2-3$ (\citealt{shapley05}). Since the inferences
about the CGM geometry from the galaxy-galaxy pairs are statistical, and the
subset of the pair sample having ancillary information on stellar populations is small, we are not in a position 
at present for a definitive test. Significant variations in star formation rate or of the 
mass flux of outflowing material might also be expected; if star formation were to shut down completely,
the corresponding diminution of outflowing gas might eventually be recognized as ``gaps'' in the
CGM near to the galaxy, but lingering high-velocity material at larger radii. Such remnant
wind material may have been observed in the spectra of massive post-starburst galaxies at
intermediate redshift (\citealt{tremonti07}). At $z \simeq 2-3$, our sample does not currently
include galaxies without significant star formation, though it does include some
galaxies with small inferred gas fractions and low specific star formation rate SFR$/M*$ (\citealt{erb+06c}).  
Unfortunately, these too do not have high enough surface density to have contributed significantly
to our galaxy-galaxy pair samples. 

One of the expectations of the simple kinematic model outlined above is 
that most of the absorption seen in the spectrum of a LBG is due to gas located within
$\simeq 10$ kpc, while most of the absorption observed in offset lines of sight is due
to gas that makes a relatively small contribution to the line profile in the $b=0$ 
spectra. At large $r$,  because gas giving rise
to a particular absorption line transition has lower $f_c$, the constraints on 
its velocity from the $b=0$ line profiles become progressively weaker. This means
that there could be lower {\it or} higher velocity material at large radii and small covering
fraction that might go unnoticed compared
to material with the same velocity but much smaller galactocentric radius (because it would have
a less-diluted $f_c$.) 

In this context, it is interesting to consider what is responsible for the ``extra'' IS absorption
in the higher baryonic mass sub-sample discussed in \S\ref{sec:composites} above. 
It is intriguing that the strength of the $v \sim 0$ components of IS absorption
is such a strong function of galaxy mass in both the current sample  
and in the $z \sim 1.4$ sample of \cite{weiner09}. A comparison of the highest and lowest
stellar mass bins in Table 1 of \cite{weiner09} shows that the line strength 
of the symmetric component increases with stellar mass by a factor of at least
16.6. \cite{weiner09} argue that the absorption near $v=0$ is  due
primarily to stellar photospheric \ion{Mg}{2} absorption. However, the set of far-UV transitions used
for our $z \sim 2.3$ sample are not contaminated by significant stellar photospheric
absorption, and yet the line profiles exhibit a very similar trend with galaxy mass. 
We therefore conclude that the ``excess'' absorption near $v=0$ arises from IS gas, at
least in our $z\sim 2-3$ sample..

There is generally little evidence for infalling (redshifted) gas in the far-UV spectra of the galaxies, and
even less evidence that $v(r)$ could be decreasing with increasing $r$ (the line {\it shapes} would
be very different if this were the case). It would also be difficult to explain the spectral
morphology of \lya\ emission (see \S~\ref{sec:lya}) if the relationship between $v_{out}$ and $r$
were inverted, as might be expected if stalling winds and/or infall dominated the gas flows. 
The expected signature of infall as seen in the galaxy spectra is IS absorption velocities
{\it redshifted} with respect to the galaxy systemic velocity.  Such absorption would
be expected if gas were prevented from escaping the galaxy potential and eventually 
began to fall back onto the central regions; a similar signature could also result
from ``fresh'' (rather than recycled) infalling cool gas that is expected to have
physical conditions ($T \simeq 10^4-10^5$K; e.g., \citealt{goerdt10}) very similar to the cool 
outflowing gas. The absorption
line strength for the metallic species we have been able to measure would be relatively
insensitive to the metal abundance of infalling material, once again due to the
high degree of saturation in the strong IS lines. 
In other words, the presence of infalling gas should be easily detected in the galaxy spectra (both at $b=0$ and
at large galactocentric radii). Redshifted absorption should be observed in the velocity range
$ 0 \simlt v \simlt +300$ \kms\ as long as the covering fraction is within a factor of a few
of that inferred for outflowing gas.  
Under the current model, infalling gas should have increasingly redshifted velocities
as its covering fraction {\it increases} (i.e., as $r$ becomes smaller); the
expected line profiles would then be inverted with respect to the observations, in
the sense that line profiles would be deepest at the high velocity end of
the distribution.

We have discussed evidence from the composite spectra in \S\ref{sec:composites} that
galaxies with higher baryonic mass are more likely to have stronger absorption
at $v \simgt 0$, which could possibly be a signature of stalled outflowing
material and/or infalling gas. This possibility is discussed further in \S\ref{sec:cold_flows} below. 

\subsection{The CGM and Dark Matter Halos}

\label{sec:dark_halos}

Fig.~\ref{fig:logw_vs_logb} shows that among the strong lines we can measure from the composite
galaxy spectra, all except \lya\ appear to become much weaker at impact parameters of $b\simgt 70-100$
physical kpc even though the rate of decline of ${\rm W_0}(b)$ (at smaller values of $b$) varies among
the observed lines.  It is interesting to ask whether this particular scale has physical significance 
(as opposed to being an artifact of the limited sensitivity of the data) given
that we know something about the properties of the galaxies associated with the extended gas. 
The particular objects in the present sample, drawn from a spectroscopic survey
of ``BX'' UV-color-selected galaxies, have been well-characterized in terms of their
spatial clustering (\citealt{asp+05}) and stellar population parameters (\citealt{erb+06b,shapley05,res+05}).
\cite{conroy08} have used the observational results of \cite{asp+05} to match these $z \simeq 1.9-2.6$
galaxies to dark matter halos in the Millennium simulation \citep{springel05}. They find a good match to both
the clustering strength and the space density of the ``BX'' galaxies with dark matter halos 
having ${\rm M_{halo} > 4.2\times 10^{11}}$ M$_{\sun}$ and ${\rm \langle M_{halo} \rangle} = 9.0\times10^{11}$
M$_{\sun}$. If we assume that the dark matter
halos ``formed'' several hundred million years prior to the epoch of their observation (e.g.,
a galaxy observed at $z \sim 2.4$ might have formed at $z \sim 2.8$ for a typical inferred
stellar population age of $\simeq 500$ Myr), one can estimate the corresponding virial radius\footnote{Here we
are assuming that the virial radius is the radius within which the average halo matter density
is $\simeq 178$ times higher than the mean density of the universe at the time of initial
collapse.}, which is $\sim 64$ kpc for the minimum halo mass of  $4.2\times10^{11}$ M$_{\sun}$ and
$\simeq 82$ kpc for a halo with the expected average mass of $9\times10^{11}$. These values of $r_{vir}$
would increase by $\simeq 10-15$\% if the ``formation'' redshift were assumed to be $z=2.3$ instead. 
It is possible that 
the similarity of $r_{vir}$ and the radius at which the $W_0$ vs. $b$ curve begins to steepen for
some of the low ionization absorption lines is not a coincidence. If the clouds are confined by
the pressure of a hotter, more diffuse medium, then $r_{vir}$ may be approximately
the radius at which the clouds become too diffuse and highly-ionized to produce significant
columns of low-ionization metals.   

By assuming a radial density profile for the characteristic dark matter halo, we can 
estimate the ``escape velocity'' ${\rm v_{esc}}(r)$ from a point at   
radius $r$ relative to the center of mass. Assuming a \cite{navarro97} (NFW) dark matter density profile,
\begin{eqnarray}
\rho_{DM}(r) \propto {1 \over \left( 1+cr/r_{vir}\right)^2cr/r_{vir}} ,
\label{eqn:nfw_dens}
\end{eqnarray}
the corresponding escape velocity from radius $r$ is given by 
\begin{eqnarray}
v^2_{esc}(r) = {2GM_{vir} \over r} {{\rm ln}(1+cr/r_{vir} ) \over {\rm ln}(1+c)-c/(1+c)}
\end{eqnarray} 
where $c$ is the NFW concentration parameter.  Assuming a ``formation'' redshift of
$z=2.8$, $M_{vir} = 4.2\times 10^{11}$ M$_{\sun}$, and $r=1$ kpc, $v_{esc} \simeq 561$ \kms, while 
for the expected average halo mass from above, $v_{esc} \simeq 730$ \kms, 
both assuming $c=7$.~~\footnote{ $v_{esc}$ is relatively
insensitive to $c$, changing by $<10$\% as $c$ goes from 5 to 9, but is 
somewhat more sensitive to the value assumed for $z_f$ (e.g., assuming $z_f=4$ changes the velocities
for the same halos above to $v_{esc}(r=1) \simeq 640$ \kms and $830$ \kms\, respectively.)}  
Perhaps more relevant, 
the escape velocity for material at $r=r_{vir}$ is significantly
lower, $v_{esc} \simeq 310$ \kms for the minimum mass halo ($r_{vir}=64$ kpc) and $v_{esc} \simeq 
400$ \kms for the average mass halo ($r_{vir}=82$ kpc). The lower panels of Figs.~\ref{fig:vvI} and \ref{fig:vvI2}
show the estimated $v_{esc}$ as a function of $r$; note that in
the context of the model, both cases have  
$v_{out}(r_{vir}) > v_{esc}(r_{vir})$.   

\subsection{Comparison to Observations at Lower Redshifts}
\label{sec:lower_z}

The observed value $|v_{max}|\sim 800$ \kms so common among the galaxies in the UV
selected samples discussed above suggests that much of the outflowing gas is destined to become unbound from 
the parent galaxy. One might naturally ask {\it why} $v_{max}$ seems to be so consistent within the
current $z \sim 2-3$ spectroscopic samples.   
For halos forming at a given redshift, the expected dependence of $v_{esc}$ on mass is fairly
shallow, $v_{esc}(r_{vir}) \propto M_{vir}^{0.3}$. The inferred
range of baryonic mass $M_{bar}$ (stars + cold gas) among the 
``BX'' sample at $z \simeq 2-2.6$ is reasonably described by a log-normal
distribution with  ${\rm log}~M_{bar} =10.61\pm0.34$ (\citealt{erb+06b}); assuming the cosmic baryon-to-dark matter
ratio for all of the galaxies, the expected 
range in $v_{esc}$ among the sample would then be $\pm 25$\%, which is marginally compatible
with the observed consistency of $v_{max}$ among the BX sample of galaxies. Similarly, the range of  
star formation rates among the NIRSPEC \Ha\ sample is $SFR \simeq 30\pm15$ M$_{\sun}$ yr$^{-1}$ (\citealt{erb+06c})
meaning that if the correlation of $v_{max}\propto$ SFR$^{0.25}$ (found by \cite{weiner09} for galaxies at 
$z \sim 1.4$) applied to the
$z \sim 2-3$ galaxies, the expected variation of $v_{max}$ would be $\pm 6$\%, which is indistinguishable 
from ``constant'' given the uncertainties in the measurement.  

\cite{strick09} have emphasized that outflow
velocities observed in cool gas entrained in super-wind-driven outflows will be significantly
smaller than that of the hot ``wind fluid'', which for M82 they infer to have characteristic
velocity of $1400-2200$ \kms based on hard-X-ray observations. 
The highest velocity observed in cool material (via \Ha\ filaments) in M82 is $v_{out} \simeq 500-600$ \kms. 
The typical galaxy in our sample has a bolometric luminosity of $\simeq 3\times 10^{11}$ 
L$_{\sun}$, roughly 5 times higher than that of M82 (cf. \citealt{reddy06,sanders03}). 
Applying the scaling expected for momentum-driven winds\footnote{\cite{murray05}
point out that cool clouds entrained in a hot flow behave like momentum-driven winds whether 
the source of the flow is primarily momentum-driven or energy-driven.} (\citealt{murray05}), $v_{max} \propto
L^{0.25}$, by analogy with M82 one would expect the cool component of the outflows for a typical $z \sim 2-3$
galaxy in our sample to reach 750-900 \kms. This range is clearly consistent with the observations. 

\subsection{Evidence for ``Cold Accretion''?}

\label{sec:cold_flows}

It is clearly of interest to understand the expected observational signature of 
the {\it accretion} of gas in the context of the proposed schematic model
of the CGM. The current models of cold accretion predict characteristic gas temperatures
of $\sim 1-5 \times 10^4$ K (e.g. \citealt{goerdt10}), which is the same range 
of temperatures expected for the cool outflowing material seen in absorption against
the galaxy far-UV continuum. The bulk of the cool accreting gas is predicted to lie within
$\simeq 100$ kpc for halos of mass $10^{12}$ M$_{\sun}$. According to \cite{dekel09a}, 
absorption due to cold stream accretion should have a covering fraction as seen by
background sources of
$\simeq 25$\% for $20 \le r \le 100$ kpc with N(\ion{H}{1})$>10^{20}$ cm$^{-2}$, with 
infall velocities $\sim 200$ \kms. Along lines of sight with $b>>0$, one might
expect absorption centered near 
$v_{los} \simeq 0$ relative to the galaxy systemic redshift (because on average the sightlines should intersect
both blue-shifted and red-shifted IS material), with velocity range of
perhaps $\pm 150$ \kms. Sightlines 
with $b \simeq 0$ (i.e., in the spectra of the galaxies themselves) 
should intersect inflowing gas with $0 \simlt v_{los} \simlt +200$ \kms.  

As mentioned in \S\ref{sec:barmass} above (and illustrated in Fig.~\ref{fig:mbar_residuals}), 
the ``excess'' absorption in the composite spectrum of 
galaxies with $M_{bar} \simgt 4 \times 10^{10}$ M$_{\sun}$ 
may be consistent with both the expected
kinematics and optical depth (or covering fraction) for infalling gas. It is notable
that the $v_{los} \ge 0$  component of absorption is seen in low-ionization species, but not in more highly ionized
species such as \ion{C}{4} that are otherwise present over all velocities $v_{los} < 0$ \kms.  
This may imply that the gas in the $v_{los} > 0$  kinematic component is more self-shielded than the IS material
that contributes most of the observed line equivalent width, suggesting it lies primarily
at small $r$. 
In any case, even when it is present, a putative infalling IS component cannot account for 
the observation of strong absorption in both \lya\ and metal lines whose strength requires
high velocity dispersion and covering fraction. 
Nor can accretion explain the observed kinematics of \lya\ emission discussed
extensively in \S\ref{sec:lya}. 
Nevertheless, we {\it know} that galaxies with halo masses of $\simeq 10^{12}$ M$_{\sun}$ 
must accrete material in some form over the typical star formation timescale of $\sim 500$ Myr
in any reasonable hierarchical model; infall of low-metallicity gas may also be necessary
for producing  the observed mass-metallicity relation at $z \simeq 2$ (e.g. \citealt{erb08}) . 

Possibly more intriguing than the dominance of outflowing IS material is the 
observation (both in the present $\langle z \rangle = 2.3$ sample and
in the $z \sim 1.4$ sample of \cite{weiner09}) that galaxies below a threshold in baryonic mass
appear to lack the $v_{los} > 0$ IS absorption component. In the $z \sim 2.3$ sample the mass threshold appears to be
near $M_{bar} \simeq 4\times 10^{10}$ M$_{\sun}$, below which essentially all IS absorption is blue-shifted. We have
seen that the probable range in dark matter halo mass (assuming correspondence between baryonic mass
and halo mass)  for the lower-M$_{bar}$  half of the \Ha\ sample 
is $4\times 10^{11}$ M$_{\sun} \simlt M_{halo} \simlt 9\times 10^{11}$ M$_{\sun}$-- a range in mass
over which cold accretion is believed to be near its peak at $z \simeq 2$. Paradoxically, 
the sub-sample showing possible evidence for an accreting component of cool gas is the
half with  $M_{halo} \simgt 10^{12}$ M$_{\sun}$, i.e. close to the critical mass
at which virial shocks begin to prevent cold gas from streaming to the central regions of
the galaxy (e.g., \citealt{dekel06,ceverino09}, but cf. \citealt{keres09}).  
Lower baryonic mass galaxies with similar star formation
rates show no evidence for an infalling component with significant covering fraction, at least
as compared to that of out-flowing material.  The extended nature of the galaxies in whose spectra
the IS lines are measured provides a spatially-averaged IS absorption profile to detect both
the coolest regions of inflowing material as well as more highly ionized gas (i.e., having 
N(\ion{H}{1}) as low as 10$^{14}$ cm$^{-2}$) that would present a larger covering fraction. It would
be very hard to miss infalling cool gas, which would appear  
as redshifted absorption for $b=0$ sightlines. Moreover, sightlines at $b>>0$ would produce
lines too weak to be consistent with the $W_0$ vs. $b$ results from \S\ref{sec:galgal}
due to the quieter velocity fields expected for infalling material; see \cite{dekel09a}. 

With the possible exception of the $v >0$ absorption in the more massive half of the
current sample (which could have alternative explanations; see discussion in \S\ref{sec:bulkvel}),
the observations reveal an absence of evidence supporting 
cold flow accretion as currently envisioned. It is possible that the streams cover a 
much smaller fraction of 4$\pi$ steradians than the $\sim 20-25$\% estimated from simulations 
\citep{goerdt10,dekel08} in order to remain undetected by the absorption probes; 
however, since the absorption line probes are sensitive to \ion{H}{1} column densities as low
as N(\ion{H}{1})$ \simlt 10^{15}$ cm$^{-2}$, they would be sensitive to much more than the
coolest, most highly collimated regions of the accretion flow.
In any case, there seems to be no way to reconcile 
the observed CGM absorption line strength and kinematics with the results of simulations which
seem consistently to predict that accretion of cool gas should be  dominant over outflow for
galaxies with $M_{tot} \sim 10^{12}$ M$_\sun$. Taken at face value, it seems that
the importance of cold accretion has been significantly over-estimated -- or at least 
that its observational signature must be more subtle than suggested by the early predictions.  
Of equal concern is that 
the influence of outflows, affecting large regions of the CGM of relatively
massive galaxies, seems to have been seriously under-estimated; 
in many cases, outflowing material has been completely
ignored or deemed negligible for galaxies in the same mass range as in our sample. 
There is strong empirical evidence for high
velocity outflows whose influence extends to galactocentric radii of at least 125 kpc around an
average galaxy in our sample-- roughly the same range of $r$ over which cold accretion is
supposed to be most observable. It would clearly be interesting to understand how high velocity outflows
would interact with cool accreting gas in the case that both are occurring simultaneously in the same
galaxies and where the relevant physics of both processes have been realistically modeled.
At present, the bulk of the observational results do not support the direction in which
the theory has been moving-- a situation which clearly needs to be resolved. 

\subsection{\lya\ Emission Re-Visited}

In \S\ref{sec:lya} we presented a simple 1-D model in an effort to understand the kinematics
of \lya\ emission in the galaxy spectra. The model assumed that the observed blue-shifted
IS absorption is providing information on the relevant gas-phase kinematics on the ``far side''
of the galaxy. We showed that the scattering of \lya\ photons in circumgalactic gas having
large bulk velocities and steep velocity gradients would produce redshifted \lya\ emission
similar to that observed. A somewhat more physically-motivated CGM model was presented in
\S\ref{sec:model}. The outflowing material is multi-phase, where the covering 
fraction of gas giving rise to a particular transition (rather than the optical depth) 
versus velocity and galactocentric
radius modulates the line profile shapes and the line strength as a function of impact
parameter.   A more sophisticated model for \lya\ emission, including photon diffusion in
3-D physical space as well as velocity space (e.g., \citealt{laursen07,verhamme08}), would be needed for a detailed understanding
of the \lya\ morphology. Such modeling is beyond the scope of the present work; however, the 
multi-phase gas involved in galaxy-scale outflows is likely to lead, qualitatively, to \lya\
kinematics similar to the observations. 

As discussed in \S\ref{sec:lya}, the high velocity and (more importantly) the large
velocity {\it range} evident from the observed IS absorption lines allow \lya\ photons initially
produced by recombination in the galaxy \ion{H}{2} regions to work their way outward
in both real space and velocity space via multiple scattering events. Rather than a shell
or a continuous medium of high \ion{H}{1} optical depth, the velocity gradient due to the
acceleration of the clumpy outflowing gas allows \lya\ photons to migrate to larger radii
and higher velocities until they are sufficiently redshifted that photons emitted
toward an observer can escape the galaxy and its circumgalactic gas. 
Since we know that the clouds giving rise to low-ionization absorption lines
typically extend to at least $r \simeq 70$ kpc, the model we have proposed implies
that there should be a diffuse \lya\ halo on the same physical scales. The total luminosity
in the diffuse halo may, in some cases, represent a
significant fraction of the galaxy's production of \lya, albeit distributed over a projected
surface area as much as $\simgt 1000$ times larger than the galaxy continuum light. 
Of course, most of this \lya\ emission would not be included in the narrow slit typically used
for galaxy spectroscopy. It is possible that the diffuse emission has already been observed in
very deep \lya\ images in the form of \lya\ ``Blobs'' (\citealt{steidel00,matsuda04})-- 
the most extreme examples--  or as
diffuse emission evident only after stacking spatial regions surrounding normal LBGs 
(\citealt{hayashino04}, Steidel et al, in preparation).   

The main point is that the structure and kinematics of the CGM ultimately control
the morphology of escaping \lya\ emission, and the \lya\ ``photosphere'' of a galaxy
is expected to be roughly coincident with the distribution of gas responsible for the
observed IS absorption features. This ``prediction'' of 
very extended \lya\ halos on scales of $\sim 50-100$ kpc is remarkably similar
to recent predictions of cooling \lya\ emission associated with cold accretion onto
galaxies of very similar mass scale to those in the current sample \citep{goerdt10,dijkstra09}. 
\lya\ emission is expected to be both redshifted and relatively axisymmetric 
for \lya\ originating in the galaxy and scattering through the CGM, whereas \lya\ emission from accreting gas should favor
blue-shifts and will be concentrated in a few dense filaments. However, it remains unclear 
how {\it observationally} distinct the extended \lya\ emission arising from these two
very different processes will be. 

We intend to address these issues in greater detail in a separate paper.   

\subsection{Outflow Mass Flux and the Gas Content of the CGM}

In principle, the type of outflow model described above 
can be used to estimate the outward mass flux $\dot{M}_{out}$ as well as the total (cool) gas content of the CGM,
$M_{CGM}$.  
Mass outflow rates of cool gas have been estimated for nearby starbursts (e.g., \citealt{martin05}), for
rare individual examples of high redshift galaxies (e.g., \citealt{pettini00}), and for composite
spectra of star-forming galaxies at intermediate redshifts \citep{weiner09}. In all cases, the mass
flux is estimated to be $\dot{M}_{out} \simgt \dot{M}_*$ where $\dot{M}_*$ is the galaxy
star formation rate. The calculations generally assume that the outflowing material is in a thin
shell located at a galactocentric radius $r$ just beyond the observed stars, traveling with outward
velocity given by the centroid of the outflowing component of the IS absorption lines, where 
the total gas column densities have been estimated from the strength of low-ionization metal absorption
lines (for $z < 1.6$) or from the observed N(\ion{H}{1}) measured from the \lya\ absorption,
in the case of MS1512-cB58.  The uncertainties in these calculations are very large even in the
context of the simple shell model, since they depend on the assumed shell radius as $r^2$ and
the total gas columns are poorly constrained even when N(\ion{H}{1}) has been directly
measured, due to unknown ionization corrections. 

Unfortunately, the less-idealized model
introduced above does not necessarily improve the situation; the IS absorption is produced by 
gas with a large range of both radius and velocity, rather than a shell at a single radius. This means
that a galaxy spectrum contains information on the integral absorption along its line of sight, and
therefore includes material deposited by both current and past winds. Since the depth of absorption
lines is significant out to $R_{eff} \sim 100$ kpc in our outflow model, gas that was launched up to
$\simgt 100$ Myr ago may still contribute to the line profiles, albeit diluted by a decreasing
covering fraction with increasing $r$.  
 
In the context of our model, cool gas deposited by outflows to $r \sim 100$ kpc 
represents an appreciable fraction of the expected $M_{bar,tot} \simeq 1.5 \times 10^{11}$ M$_{\sun}$ 
associated with a typical dark matter halo mass of $\simeq 9 \times 10^{11}$ M$_{\sun}$. 
If we assume for definiteness that $R_{eff} \simeq 80$ kpc and that the asymptotic outflow
velocity is $v_{out} \simeq 750$ \kms,  a packet of cool gas
launched from $r_{min} \sim 1-2$ kpc will reach $r\simeq 80$ kpc after $t_{80} \sim 1.3 \times 10^{8}$ years. 
If there is a steady-state flow with mass flux $\dot{M}_{out}$, the total cool gas content within $R_{eff} = 80$
kpc deposited by the outflow would be given approximately by
\begin{eqnarray}
\langle M_{CGM}\rangle  =  \langle \dot{M}_{out} \rangle  t_{80} ~.
\end{eqnarray}

The model that best reproduces the radial behavior of $f_c$ for low-ionization species 
in the CGM of typical galaxies implies cloud densities that increase with 
galactocentric radius as $\rho_c \propto r^{-2}$ (\S\ref{sec:vel_model}). If we assume that the initial
number density of atoms in a typical cloud is $n_0$ cm$^{-3}$ at the base of the
flow $r_{min}$, then the total cool gas mass between $r_{min} \sim 1$ kpc and $r=80$ kpc is 
\begin{eqnarray}
M_{CGM} (r<80)~\simeq ~3 \times 10^{10}~n_0 ~{\rm M_{\sun}}~, \\
\langle \dot{M}_{out} \rangle \simeq {\langle M_{CGM}\rangle \over t_{80}} \simeq 230~ n_0 {\rm M_{\sun} yr^{-1}}~. 
\end{eqnarray}
Note that this estimate does not attempt to account for hot or highly ionized gas that is also likely
to be associated with supernova-driven outflows (e.g., \citealt{strick09}). 

An estimate of $n_0$ follows from our earlier assertion that $R_{eff}$ for low-ion species
likely corresponds to a threshold N(\ion{H}{1})$\simgt 10^{17}$ cm$^{-2}$ (i.e., a Lyman Limit system). 
The neutral fraction for such \ion{H}{1} column density gas ionized by the metagalactic radiation
field is estimated to be $\sim 10^{-3}$ (e.g., \citealt{steidel90b}), so that the total H column
will be $\sim 10^{20}$ cm$^{-2}$. For a line of sight with $b \sim 80$ kpc, the corresponding average density
$n_H(80) \sim 4 \times 10^{-4}$ cm$^{-3}$. In the case of the aforementioned  $r^{-2}$ density dependence, 
\begin{eqnarray}
n_0 \sim \left[{R_{eff} \over r_{min}}\right]^2 ~  n_H(80) ~ \sim 1~{\rm cm}^{-3} ~.
\end{eqnarray}
Adopting $n_0=1$ cm$^{-3}$, cool gas in the CGM would account for at least 20\% of the total baryons
associated with a typical galaxy, with mass comparable to the total $M_{bar}$ (cold gas plus stars) 
in the central few kpc, which has 
$\langle M_{bar} \rangle \simeq 4\times 10^{10}$ M$_{\sun}$  at $z=2.3$ \citep{erb+06b}. 
If one assumes that the typical stellar mass $M_* \simeq 2 \times 10^{10}$ M$_{\sun}$ was formed
over the same timescale (i.e., $t_{80} \simeq 1.3 \times 10^8$ yr), then 
$\langle SFR \rangle \simeq 150$ M$_{\sun}$ yr$^{-1}$. Thus, a crude (but largely independent) calculation 
yields the same qualitative
result obtained with shell models in previous work: i.e., $\dot{M}_{out} \simgt \dot{M}_*$. 

In any case, it is difficult to escape the conclusion that the CGM contains a substantial fraction of
galactic baryons at $z \sim 2-3$, and that the gas-phase kinematics and geometry  points to a causal
connection between rapid star formation and the presence of large amounts of gas at large
galactocentric radius\footnote{See \citealt{menard09} for additional support for this causal connection.}. 

\section{Summary and Conclusions}
\label{sec:discussion}

We have used a relatively large sample of $1.9 \simlt z \simlt 2.6$ galaxies with accurate
measurements of systemic redshift $z_{sys}$ and reasonably high quality rest-frame far-UV
spectra to examine the relationship between galaxy properties and UV spectral morphology. Our
main focus has been on the kinematics and strength of 
interstellar absorption and \lya\ emission and their implications for the galaxy-scale outflows 
observed in all rapidly star-forming galaxies at high redshifts.  
Using this well-observed subset as a calibration, we then combined the rest far-UV galaxy spectra drawn from 
a much-larger parent sample with additional spatial information
for the same ensemble of galaxies provided by the spectra of background galaxies with small
angular separations. Using these joint constraints, we constructed simple models of the kinematics and geometry of 
the ``circumgalactic medium''-- the interface between star forming galaxies and the IGM. 
Our principal conclusions are as follows:

1. We have used the \Ha\ sample of $z\simeq 2-2.6$ galaxies together with their far-UV spectra to produce a revised
calibration that allows deriving systemic redshifts for star-forming galaxies 
from their far-UV spectral features (strong IS lines and \lya\ emission). 
In the absence of stellar photospheric
absorption lines or nebular emission lines in the rest-frame optical, 
the most accurate estimates of the systemic redshift
are derived from the centroids of strong IS absorption lines. 
Applying a shift of $\Delta v = +165$ \kms ($\Delta z = +0.0018$) at $z=2.3$ to a measured absorption redshift  
provides an estimate of a galaxy's systemic redshift accurate to $\pm 125$ \kms; 
redshifts measured from the \Ha\ emission
line (when available) have a precision of $\pm 60$ \kms. Both methods for estimating redshifts
have no significant systematic offset
relative to the redshift defined by stellar (photospheric) absorption features.  

2. Using only the \Ha\ sample of BX galaxies, we found mean velocity offsets for the centroids
of the strong IS lines and \lya\ emission line of $\dvis = -164\pm16$ \kms and $\dvla = +445\pm27$ 
\kms, respectively.  
We searched for significant correlations between the kinematics defined by the centroid velocities
of IS and \lya\ lines and other measured or inferred galaxy properties. Within the \Ha\ sample,
the only significant correlations found were between $\dvis$ and galaxy mass estimated from the sum of inferred
stellar and gas mass ($M_{bar}$) as well as  the independently estimated $M_{dyn}$. The sense of
the correlation is that
$\dvis$ is smaller (i.e., less
blue-shifted) in galaxies with larger masses.  

3. Despite the trend described in point 2 above, the velocity $|v_{max}|$ of the  maximum blue-shift  
observed in the absorption profiles  
is essentially identical for the sub-samples of
galaxies with ${\rm M_{bar}} > 3.7\times 10^{10}$ M$_{\sun}$ and ${\rm M_{bar}} < 3.7\times10^{10}$ M$_{\sun}$, 
with $|v_{max}| \simeq 800$ \kms. However, the higher-mass subset has both $\dvis$ and $\dvla$ shifted
toward positive velocities (i.e., more redshifted) by $\sim 200$ \kms. The differences
in the composite IS and \lya\ profiles can be explained by an additional component of
absorption at $v \simgt 0$ that appears to be absent in the lower-mass sub-sample. 
The extra  absorption, which has a peak near $v = 0$ and a centroid velocity of
$v\simeq +150$ \kms, could conceivably be a signature of infalling gas or stalled
winds falling back on the galaxy, but is most likely to be gas at small galactocentric
radii based on the line profiles. In any case, the velocity centroids of the IS lines, 
which are commonly used as a proxy for the ``wind velocity'' for high redshift
galaxy samples, are actually modulated 
almost entirely by gas that is not outflowing. 

4. The \lya\ emission profile, like $\langle \dvis \rangle$, is modulated by the covering fraction
(or apparent optical depth) of material near $v \simeq 0$. We show that, by using the information
on the covering fraction near $v \simeq 0$, the kinematics of blue-shifted gas,  and assuming spherical symmetry, the
behavior of \lya\ emission with respect to IS absorption can be reproduced using simple models. In
general, the red wing of \lya\ emission is produced by \lya\ photons scattered from outflowing gas
on the opposite side of the galaxy. The apparent redshift of the \lya\ centroid is a manifestation
of the fact that only photons scattering from material having a (redshifted) velocity large enough to
take the photons off the \lya\ resonance for any material between the last scattering and the observer
can escape in the observer's direction. 
The spectral morphologies of \lya\ emission and IS absorption are most easily understood if 
the highest velocity material (either redshifted or blue-shifted) is located at the largest distances
from the galaxy, i.e., that $v(r)$ is monotonically increasing with $r$ in a (roughly) spherically symmetric 
radial flow.  The same geometric picture works well to reproduce both the line-of-sight ($b=0$) absorption
profiles and the absorption line strength as a function of impact parameter (see point 6 below). 


5. We have demonstrated that the use of the far-UV spectra of galaxies within projected angular pairs (with 
discrepant redshifts) provides an opportunity to constrain the physical location of circumgalactic gas 
around typical galaxies. A set of 512 such galaxy pairs, on angular scales ranging from 
1\arcs\ to 15\arcs\ ($\simeq 8-125$ kpc), have been combined with the $b=0$ spectra of
the foreground galaxies in each pair. 
These are used together to measure each foreground galaxy's CGM along two independent 
lines of sight. Composite spectra stacked according to galactocentric impact parameter
allow us to measure the dependence of the line strength of several ionic species on impact parameter $b$ from
0-125 kpc (physical).  These lines are strongly saturated, so their strength is determined by a combination
of the covering fraction and the velocity spread in the absorbing gas.  
The fall-off of $W_0$ with $b$ implies $f_c(r) \propto r^{-\gamma}$, with $0.2 \le \gamma \le 0.6$, with
\lya\ and C{\sc IV} having smaller values of $\gamma$ compared to (e.g.) Si{\sc II} and Si{\sc IV}. 
For each observed species, the strength of absorption declines much more rapidly beginning
at $b \simeq 70-90$ kpc, with the exception of \lya\, which remains strong to $b \simeq 250$ kpc. 
Combining the known space density of the galaxies in our sample (i.e., including only 
those with apparent magnitude ${\cal R} \le 25.5$) with the average cross-section for 
absorption accounts for $\simeq 45$\% of all intergalactic \ion{C}{4}  
absorption with $W_0(1548) > 0.15$ \AA\, and $\simeq 70$\% of LLSs at $z \sim 2$ observed along QSO sightlines.

6. We have proposed a simple model constrained by a combination of $W_0(b)$ from the galaxy pairs
and the kinematics of IS species observed toward each galaxy's own stars ($b=0$).  The IS absorption line
shapes depend on a combination of $v_{out} (r)$, the radial dependence of outflow velocity which
in turn depends on the cloud acceleration $a(r)$ and $f_c(r,v)$, the radial dependence of the covering fraction, which
depends on cloud geometry and is inferred 
from $W_0(b)$. We show that a self-consistent model using parametrized forms of $a(r)$ 
and $f_c(r)$ can simultaneously 
reproduce the observed line shapes and strengths (apparent optical depth vs. velocity) as well as the observed
$W_0(b)$ relation from the galaxy-galaxy pair measurements. 
In the model, higher velocity gas is located at larger galactocentric
radii, and gas clouds are accelerated to $\simeq 800$ \kms before the absorption strength (i.e., covering
fraction)  is geometrically diluted.
There is little evidence in the line profiles for stalling or infalling wind 
material, and the transverse sightlines show metal-enriched gas to at least $\sim 125$ kpc\footnote{\cite{assp03},
and \cite{asp+05}
have argued that there is evidence that the metals affect regions of $\simeq 300$ kpc (proper) based on
\ion{C}{4}--galaxy cross-correlation, and that outflows would be expected to stall at approximately this distance due
to a number of separate considerations.} 
Taken together,
these suggest that high velocity outflows from $\simeq L*$ 
galaxies at $z \sim 2-3$ are able to deposit enriched material to very large radii. 
They also suggest that the observed metals are causally related to the observed galaxies, rather than
remnants of winds generated by other (earlier) galaxies deposited in the nearby volume.

7. The observed $v_{max} \simeq 800$ \kms\ that appears to be a general property of the galaxies in
our spectroscopic sample exceeds $v_{esc}(r_{vir})$ estimated for galaxy halo masses of $\simeq 9\times10^{11}$
M$_{\sun}$, the average halo mass expected given the observed space density and clustering of the $z \simeq 2$ galaxies.  
Similarly, the galactocentric radius at which the strength of low-ionization absorption line species begins
to decrease rapidly is similar to the estimated virial radius of $\simeq 80$ kpc for the same halo mass. 

8. We consider the observations in the context of ``cold accretion'' or ``cold flows'', which are believed
to be active over the same range of galactocentric radii and redshifts, for galaxies
in the same range of total mass as those in our sample ($4\times10^{11} 
\simlt M_{tot} \simlt 2\times 10^{12}$ M$_{\sun}$). There is possible evidence for the
presence of infalling gas in the far-UV spectra of galaxies with greater than the median baryonic mass
$M_{bar} = 4\times10^{10}$ M$_{\sun}$, while the lower-mass sub-sample shows no evidence for the expected
redshifted components of IS absorption lines.  The kinematics of \lya\ emission also seem inconsistent with 
the expectations for accreting material, while they are as expected in the context of the picture of
outflows we have presented.  We expect that the same gas seen in absorption to $b\simeq 50-100$ kpc
also acts as a scattering medium for escaping \lya\ photons initially produced in the galaxy \ion{H}{2}
regions. These \lya\ ``halos'' are probably a generic property of all high redshift LBG-like galaxies,
the more extreme of which have probably already been observed as ``\lya\ Blobs'' (\citep{steidel00,matsuda04}). More
typical LBGs likely exhibit this diffuse \lya\ emission, which is of low surface brightness and therefore
difficult to observe without extremely deep \lya\ images.    

9. The inferred mass of cool gas in the CGM (within $\sim 100$ kpc) is comparable to the
sum of cold gas and stars in the inner few kpc; together they account for $\simgt 50$\% of
the total baryonic mass ($\sim 1.5\times10^{11}$ M$_{\sun}$) associated with the average $M\sim 9\times 10^{11}$
M$_{\sun}$ dark matter halo.  
If the qualitative picture of the CGM we have proposed is correct, the wind material would still be 
traveling at $v_{out}  > v_{esc}(r_{vir})$ when it crosses the virial radius at $r \simeq 80$ kpc. Even if
$v_{out} \simeq v_{esc}$, the low-ionization wind material would
reach $r \sim 100$ kpc within $\sim 150$ Myr of the onset of the star formation episode if it is slowed only by
gravity and does not accumulate a large amount of swept-up ISM on its way. Since we have shown that
the CGM is outflowing and itself comprises a large baryonic reservoir, 
it is possible that in most directions there is little to impede outflows from propagating ballistically into the IGM.

The use of resolved background sources for studying the CGM in absorption 
is qualitatively different from using QSO sightlines. It seems clear that the multi-phase ISM observable in
a number of ionic species indicates differential changes in covering fraction of the outflowing gas as a function
of ionization level. This means that the radial behavior of the strength of saturated transitions is providing
information on the ``phase-space'' density of gas having physical conditions amenable to the presence of each
ion. Because absorption seen in the spectra of background {\it galaxies} averages over a $\simeq 2-3$ kpc region, 
as compared to background QSOs, whose apparent angular sizes are $\simeq 5$
orders of magnitude smaller, statistics such as IS line strength, covering fraction,
and velocity range vs. impact parameter, are much less subject to large variance along a single sightline. 

All of the galaxy spectra used in this paper were obtained as part of large surveys, and therefore have
low to moderate spectral resolution and S/N. It is possible with current generation telescopes and instruments
to obtain spectra of relatively high S/N and moderate resolution (e.g., \citealt{shapley06}) in 10-20
hours integration time (i.e., $\sim 10$ times longer than most in the current samples). However, spectra
similar to those currently possible only for strongly lensed galaxies ($R \sim 5000$, S/N$\simgt 30$) will
be routine at ${\cal R} \simeq 24-24.5$ using future 30m-class telescopes equipped with state of the art multi-object spectrometers
(\citealt{steidel+astro2010}). When such capability is realized, full observational access to the three-dimensional distribution of gas
near to and between galaxies will revolutionize the study of baryonic processes in the context of galaxy
formation.  The results described above are intended as an illustration of the application of rest-far-UV
galaxy spectra toward a better understanding of feedback processes and the CGM/IGM interface with 
the sites of galaxy formation. The feasibility of sharpening our understanding and interpretation with
higher quality data in the future is very encouraging.

\bigskip
\bigskip

This work has been supported by the US National Science Foundation through grants 
AST-0606912 and AST-0908805 (CCS), and by the David and Lucile Packard Foundation (AES, CCS). 
CCS acknowledges additional support
from the John D. and Catherine T. MacArthur Foundation.   
DKE is supported by the National Aeronautics and Space Administration under Award No. NAS7-03001 and the
California Institute of Technology. We would like to thank Juna Kollmeier and 
Joop Schaye for interesting and useful
conversations, and Patrick Hall, Martin Haehnelt, and the referee for comments that significantly 
improved the final version of the paper. Kurt Adelberger played a major role in the 
early days of the survey used in this paper;
his intellectual contributions have remained crucial though he has moved on to new challenges in the
``real'' world.  
Marc Kassis and the rest of the W.M. Keck Observatory staff keep the instruments
and telescopes running effectively, for which we are extremely grateful. 
We wish to extend thanks to those of Hawaiian ancestry on whose sacred mountain we are privileged
to be guests. 

\bibliographystyle{apj}
\bibliography{refs}

\end{document}